\documentclass[aps,prd,reprint,showpacs,showkeys,superscriptaddress,nofootinbib]{revtex4-1}

\usepackage{amsfonts,amssymb,mathrsfs,dsfont}
\usepackage{amsopn}
\usepackage{amsmath,amsthm}
\usepackage{enumitem}
\usepackage{graphicx}
\usepackage{tikz,braids}
\usetikzlibrary{arrows}
\usepackage{natbib} 
\setcitestyle{authoryear,square,comma,compress}
\renewcommand{\cite}{\citep}
\usepackage{hyperref}
\hypersetup{
   colorlinks,
   menucolor=black,
   linkcolor=black,
   citecolor=black,
   urlcolor=blue
}

\definecolor{darkgreen}{rgb}{0,0.5,0}
\definecolor{darkred}{cmyk}{0,1,1,0.4}

\makeatletter
\newcommand{\keyword}{\@dblarg\@keyword}
\def\@keyword[#1]#2{\textbf{#2}\index{#1}}
\newcommand{\fluxplane}[3]{
	\def\x{#1}
	\def\y{#2}
	\def\r{#3}
	\draw [thick,fill=black!5!white,opacity=0.5] (\x+0,\y+0) -- (\x+2,\y+2) -- (\x+10,\y+2) -- (\x+8,\y+0) -- (\x+0,\y+0);
	\draw [red] (\x+2.0,\y+1.0) [xscale=2.4,->] arc(250:100:.20);
	\draw [red] (\x+5.0,\y+1.5) [xscale=2.1,->] arc(100:210:.30);
	\draw [red] (\x+7.0,\y+0.9) [xscale=2.1,->] arc(-90:50:.30);
	\draw [fill,red!60!white] (\x+7.0,\y+0.9) 	ellipse (\r*0.14 and \r*0.07);
	\draw [fill,red!60!white] (\x+2.0,\y+1.0) 	ellipse (\r*0.14 and \r*0.07);
	\draw [fill,red!60!white] (\x+5.0,\y+1.5) 	ellipse (\r*0.14 and \r*0.07);
	\draw [fill] (\x+7.0,\y+0.9) 		ellipse (0.06 and 0.03);
	\draw [fill] (\x+2.0,\y+1.0) 		ellipse (0.06 and 0.03);
	\draw [fill] (\x+5.0,\y+1.5) 		ellipse (0.06 and 0.03);
}
\makeatother

\DeclareMathOperator{\curl}{\mathrm{curl}}
\DeclareMathOperator{\dist}{\mathrm{dist}}
\DeclareMathOperator{\spec}{\textrm{spec}}

\DeclareMathAlphabet{\mathpzc}{OT1}{pzc}{m}{it}

\newcommand\1{{\ensuremath {\mathds 1} }}
\newcommand{\C}{\mathbb{C}}

\newcommand{\N}{\mathbb{N}}

\newcommand{\bbS}{\mathbb{S}}
\newcommand{\R}{\mathbb{R}}
\newcommand{\Z}{\mathbb{Z}}
\newcommand{\bA}{\mathbf{A}}

\newcommand{\bB}{\mathbf{B}}
\newcommand{\be}{\mathbf{e}}
\newcommand{\bJ}{\mathbf{J}}
\newcommand{\bk}{\mathbf{k}}
\newcommand{\bL}{\mathbf{L}}

\newcommand{\bp}{\mathbf{p}}

\newcommand{\br}{\mathbf{r}}
\newcommand{\bx}{\mathbf{x}}
\newcommand{\bX}{\mathbf{X}}
\newcommand{\by}{\mathbf{y}}

\newcommand{\cC}{\mathcal{C}}
\newcommand{\cE}{\mathcal{E}}
\newcommand{\cF}{\mathcal{F}}

\newcommand{\cH}{\mathcal{H}}

\newcommand{\cS}{\mathcal{S}}
\newcommand{\cV}{\mathcal{V}}
\newcommand{\ha}{\hat{a}}
\newcommand{\hb}{\hat{b}}
\newcommand{\hH}{\hat{H}}
\newcommand{\hL}{\hat{L}}
\newcommand{\hT}{\hat{T}}
\newcommand{\hV}{\hat{V}}

\newcommand{\sU}{\mathrm{U}}

\newcommand{\sw}{\mathrm{w}}
\newcommand{\sx}{\mathrm{x}}
\newcommand{\sy}{\mathrm{y}}
\newcommand{\sz}{\mathrm{z}}

\newcommand{\vphi}{\varphi}

\newcommand{\sym}{\mathrm{sym}}
\newcommand{\asym}{\mathrm{asym}}

\newcommand{\CaTF}{C_{\mathrm{aTF}}}
\newcommand{\cEAF}{\cE^{\mathrm{af}}}
\newcommand{\cEGP}{\cE^{\mathrm{GP}}}
\newcommand{\cETF}{\cE^{\mathrm{TF}}}
\newcommand{\cEaTF}{\cE^{\mathrm{aTF}}}
\newcommand{\psiAF}{\psi^{\mathrm{af}}}
\newcommand{\rhoTF}{\varrho^{\mathrm{TF}}}
\newcommand{\rhoATF}{\varrho^{\mathrm{aTF}}}
\newcommand{\bloor}{{\lfloor \beta^{-1} \rfloor}}
\newcommand{\PsiLau}{\Psi^{\mathrm{Lau}}}
\newcommand{\PsiQH}{\Psi^{\mathrm{qh}}}
\newcommand{\cQH}{c^{\rm qh}}

\newcommand{\mat}[1]{
  \begin{bmatrix}
    #1
  \end{bmatrix}
}

\newcommand{\bDelta}{{\mbox{$\triangle$}\hspace{-7.3pt}\scalebox{0.8}{$\triangle$}}}

\theoremstyle{plain}

\theoremstyle{definition}

\theoremstyle{remark}

\newtheorem*{ack}{Acknowledgments}

\begin{document}

\title{Properties of 2D anyon gas}

\author{Douglas Lundholm}
\affiliation{Department of Mathematics, Uppsala University, Box 480, SE-751 06, Uppsala, Sweden}

\begin{abstract}
	An overview is given of the 2D many-anyon gas, including
	its definition (both for ideal and certain less-than-ideal 
	particles, as well
	as for abelian and nonabelian braid group representations),
	its corresponding known properties
	starting out from the intricate relationship between exchange and exclusion,
	as well as its emergence from bosons and/or fermions in 3D.\\
	\textit{\footnotesize 
	For the Encyclopedia of Condensed Matter Physics, 2e.
	Date: June, 2023}
\end{abstract}


\maketitle


\setlength\arraycolsep{2pt}
\def\arraystretch{1.2}

\subsection*{Key points/objectives}
\begin{itemize}[wide, labelwidth=!, labelindent=0pt]
\item How can the 2D anyon gas be defined physically and modelled mathematically?
\item What are the essential properties of the anyon gas?
	Does it admit some notion of exclusion statistics intermediate to bosons and fermions?
\item How can an anyon gas emerge in a realistic physical system consisting of 
	only bosons and/or fermions?
\end{itemize}

\section{Introduction} \label{sec:intro}

\keyword{Quantum statistics} refers to the symmetry and organizing principle assumed
by the components of a quantum system.
Primarily we think of particles, described by a joint state or wave function, 
which can fall
into certain symmetry classes due to the fundamental constraints on the
information that can be extracted from the system, such as those 
imposed by the uncertainty principle.
\keyword{Bosons} and \keyword{fermions} are ensembles of identical such particle subsystems 
characterized by symmetry, respectively antisymmetry, with respect to particle exchange.
In their ideal limits
they manifest the statistical state distributions defined by
Bose and Einstein, respectively Fermi and Dirac, 
with the latter subject to Pauli's exclusion principle.
In a sense, whereas bosons unify, fermions individualize.
\keyword{Anyons} are defined as identical quantum particles with an intermediate 
exchange symmetry characterized by representations of the braid group, 
which is the relevant group of continuous exchanges of particle configurations 
in 2D (two spatial dimensions).
An \keyword{anyon gas} is a quantum system consisting of a large number of such
particles subject to a fixed class of exchange symmetries,
which is dependent on the specific underlying kinematics and dynamics of
the quantum systems from which it emerged. 
The study of the behavior of the proper anyon gas
presents a significant challenge to mathematics and physics.
Our aim with this overview is to make these statements precise and to give an
updated status report on some of the many facets of the topic.

\subsection{Brief historic overview}

For completeness and to set our reference point for this overview, we give 
a very brief account on the origins of the anyon gas, referring to
\cite{BieLieSimWil-90,Froehlich-90,LeiMyr-22,Goldin-22} for further background.

\subsubsection{Three perspectives on anyons}

The discovery of anyons and intermediate quantum statistics in two spatial 
dimensions may be attributed to three independent research groups, 
working from three distinctly different perspectives\footnote{%
These three groups were initially unaware of each other's earlier contributions 
and their initial works therefore lack the respective citations. 
There have subsequently been attempts to remedy this in the literature 
\cite{BieLieSimWil-90}.}.
Chronologically, these are (cf.\ Fig.~\ref{fig:perspectives}):

\begin{figure}
	\centering
	\scalebox{0.8}{%
	\begin{tikzpicture} 
		\draw [red!60!white] (0,0) 	ellipse (1.5 and 0.7);
		\node at (-4,0) {\large\bf algebraic};
		\node at (0,0) {\large\bf geometric};
		\node at (4,0) {\large\bf magnetic};
		\draw [arrows=->,thick] (-2.8,0) -- (-1.8,0);
		\draw [arrows=->,thick] (2.8,0.2) -- (1.8,0.2);
		\draw [arrows=->,thick,dashed] (1.8,-0.2) -- (2.8,-0.2);
		\node [darkred] at (0,1.5) {\small most general};
		\draw [arrows=->,darkred] (0,1.2) -- (0,0.8);
		\node [darkred] at (4,1.5) {\small most practical}; 
		\draw [arrows=->,darkred] (4,1.2) -- (4,0.4);
		\node [darkred] at (-4,1.5) {\small most computational};
		\draw [arrows=->,darkred] (-4,1.2) -- (-4,0.4);
		\node at (3,-0.7) {\small statistics transmutation};
	\end{tikzpicture}
	}
	\caption{Three different perspectives on anyons.}
	\label{fig:perspectives}
\end{figure}
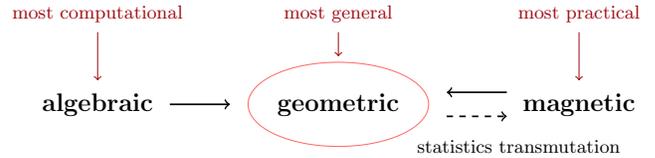

\begin{enumerate}[wide, labelwidth=!, labelindent=0pt]
\item {\bf Geometric:}
	In 1977, Leinaas and Myrheim 
	used geometric methods from gauge theory, fiber 
	bundles, and extensions of the kinetic energy operator via 
	boundary conditions (also known as \keyword{``Schr\"odinger quantization''}) 
	to arrive at the possibility for what they termed 
	\keyword{``intermediate statistics''}
	in 2D (as well as in 1D), and worked out some of its detailed consequences 
	for two particles \cite{LeiMyr-77}.
	They further pointed to the role of the fundamental group 
	of the configuration space for arbitrary 
	numbers of identical particles and the possibility for higher-dimensional fibers 
	(necessary for nonabelian representations, but they did not explicitly 
	mention the braid group or its representation theory in their work).

\item {\bf Algebraic:}
	In 1980-'81,
	starting out from the representation theory of the Lie algebra of local 
	currents on $\R^d$ and the group of diffeomorphisms 
	(akin to \keyword{``Heisenberg quantization''} 
	and Dirac's canonical quantization scheme
	\cite{Dirac-64}), 
	Goldin, Menikoff and Sharp 
	found a kinematical derivation of the possibilities for 
	quantum statistics with an unexpected dependence on the dimension $d$
	\cite{GolMenSha-80,GolMenSha-81}.
	In \cite{GolMenSha-81} 
	they also explored connections to the Aharonov--Bohm effect (in 3D), 
	while subsequently in 1983 they 
	highlighted the role of the one-dimensional 
	unitary representations of the braid group \cite{GolMenSha-83}, 
	and in 1985 
	pointed out that higher-dimensional braid group 
	representations induce inequivalent representations of the current algebra 
	\cite{GolMenSha-85}.
	Such identical particles were later termed \keyword{``plektons''} 
	(see, e.g., \cite{MunSch-95,DelFigTet-97,KorLanSch-99}) 
	or simply \keyword{``nonabelian anyons''} 
	(nowadays the more common term; cf.\ \cite{Nayak-etal-08}).
	It was also pointed out in \cite{GolMenSha-85}
	that nontrivial exchange phases 
	in 2D do not require indistinguishability; 
	the \emph{colored} braid group serves 
	as the fundamental group for configurations of \emph{distinguishable} particles.
	The general algebraic (and potentially nonabelian) approach to quantum 
	statistics was further developed in the field theory direction by 
	Tsuchiya and Kanie \cite{TsuKan-87,TsuKan-88}, Kohno \cite{Kohno-87}, 
	as well as Fr\"ohlich, Gabbiani, Kerler and Marchetti 
	\cite{Froehlich-88,FroMar-88,FroMar-89,Froehlich-90,FroGab-90,FroKer-93}, 
	and Fredenhagen, Rehren and Schroer \cite{FreRehSch-89}.

\item {\bf Magnetic:}
	In 1982,
	motivated by topological gauge theory examples in higher dimensions 
	(``$\theta$-vacua'' and ``dyons''), 
	Wilczek 
	considered a composite of a 2D magnetic flux and a charged particle, 
	coined for it the name \keyword{``anyon''},
	as well as predicted its fractional spin 
	and worked out the basic quantum mechanics for two such composite particles
	\cite{Wilczek-82a,Wilczek-82b}.
	The same line of investigation was continued by Wu in 1984, 
	who connected it to path integrals, extended it to arbitrary numbers 
	of particles, and independently and explicitly 
	invoked the braid group and its one-dimensional representations
	\cite{Wu-84a,Wu-84b}. 
	Further, Arovas, Schrieffer, Wilczek and Zee considered in 1985 the 
	statistical mechanics of a gas of such abelian anyons by means of its 
	lowest-order virial expansion 
	\cite{AroSchWilZee-85} (see also \cite{Dowker-85}).
\end{enumerate}

The first concrete physical application of intermediate/fractional 
statistics and anyons came from the magnetic perspective and concerned the 
\keyword{fractional quantum Hall effect} (FQHE) 
\cite{Stormer-99,Tsui-99,Laughlin-99}, 
in which the emergent fractionally 
charged quasiparticles/holes introduced by Laughlin 
were proposed by Halperin to have such properties 
\cite{Halperin-84}. 
Arovas, Schrieffer and Wilczek subsequently verified this \cite{AroSchWil-84}
(though under tacit assumptions of adiabaticity \cite{Forte-91}) 
by computing the corresponding Berry phase for quasiholes,
thereby converging the geometric and magnetic 
(and eventually algebraic)
perspectives.
Further,
the latent power of the algebraic perspective was brought to light
in the new millennium
as Kitaev proposed the usefulness of nonabelian anyons in topological
quantum computation \cite{Kitaev-03,FreKitLarWan-03,Kitaev-06}.
Most of the development of the field 
throughout this time span
has been covered in the books and reviews
\cite{Froehlich-90,Jackiw-90,Wilczek-90,Forte-92,IenLec-92,Lerda-92,CanJoh-94,Myrheim-99,DatMurVat-03,Khare-05,Ouvry-07,Stern-08,Nayak-etal-08}.

\subsubsection{Precursor ideas}

As is evident from the above, anyons and intermediate (``fractional'') 
quantum statistics are the convergence of many central ideas and concepts in 
mathematical physics. 
The former notion (anyon) is more closely tied into the concepts of 
\keyword{exchange phases} and symmetry classes of wave functions, 
while the latter (fractional statistics) to that of the 
\keyword{exclusion principle} and permissible probability distributions for particles. 
It has been---and still is---a nontrivial task to rigorously connect these two 
notions in the strictly intermediate case 
(see, e.g., the review \cite{CanJoh-94}). 
Exploration of intermediate statistics from the latter approach of exclusion 
with a finite occupation number of one-body states was considered by Gentile 
already in 1940-'42 \cite{Gentile-40,Gentile-42},
and---inspired by anyons---another, dimension-independent, 
approach allowing for \keyword{fractional exclusion statistics} 
was suggested by Haldane in 1991 \cite{Haldane-91}.

Ideas of charged particles encircling regions of magnetic flux were developed by 
Ehrenberg and Siday \cite{EhrSid-49}, Aharonov and Bohm \cite{AhaBoh-59}, 
and the more general concept of geometric phases by 
Pancharatnam \cite{Pancharatnam-56}, Berry \cite{Berry-84} and Simon \cite{Simon-83}. 
Methods of geometric quantization were developed by Souriau around 1967-'70, 
and he pointed out that if the configuration space is not simply connected 
then more than two different group characters could arise and this 
``would lead to prequantizations of a new type'' \cite{Souriau-70}. 
Around the same time, homotopy classes of Feynman paths brought another 
topological perspective on quantum statistics 
\cite{Schulman-68,LaiDeW-71,Dowker-72}
(cf., e.g., \cite{Mouchet-21}). 
For example, Laidlaw and DeWitt remarked that 
possibilities in two space dimensions
are not limited to bosons and fermions, 
but they did not elaborate further \cite{LaiDeW-71}.

Foundations for nonabelian anyons were laid much earlier in the theory of 
parastatistics (nonabelian representations of the permutation group) 
\cite{Green-53,MesGre-64}. 
Further, the central idea employed by Leinaas and Myrheim that equivalent 
configurations of indistinguishable particles need to be identified 
goes back all the way to Gibbs (cf.\ \cite{LeiMyr-77,Froehlich-90}).

The above shows that the essential ideas that were necessary for the 
discoveries of anyons were floating around at the time, and it is also 
of interest to remark that, while it took about half a century from the 
discovery of bosons and fermions to that of anyons, it took another half a 
century to see their experimental confirmation.

\subsubsection{Some remarks concerning one-dimensional anyons}

Although this article concerns anyons in the 2D setting, it is unavoidable 
to note the influence of ideas from generalized notions of quantum statistics 
in 1D (i.e. 1+1 spacetime dimensions). One of the first works in which 
generalized commutation relations for field operators can be found is 
Klaiber in 1968 \cite{Klaiber-68}, followed by Streater and Wilde \cite{StrWil-70},
and Fr\"ohlich \cite{Froehlich-76}. 
Transmutation of 1D hard-core bosons (studied classically by Tonks \cite{Tonks-36})
into fermions and vice versa was 
considered by Girardeau \cite{Girardeau-60}, 
who also noted topological and dimensional 
consequences for the configuration space \cite{Girardeau-65}. 
The Lieb--Liniger model \cite{LieLin-63} for point-interacting bosons in 1D 
constitutes the intermediate statistics found by Leinaas and Myrheim in 1D, 
however exchange phases (thus ``anyonic Lieb--Liniger'')
may also be added to that model \cite{Kundu-99}.
Leinaas and Myrheim later considered a different approach to statistics 
in 1D (and in higher dimensions) akin to ``Heisenberg quantization'' \cite{LeiMyr-93}, 
and found a relationship to the Calogero--Sutherland model 
\cite{Calogero-69a,Calogero-69b,Sutherland-71,Polychronakos-89}.
2D anyons that are forced into the lowest Landau level of a strong magnetic 
field may be understood using such 1D concepts \cite{Ouvry-07}.

\subsection{Key points/objectives}

We will in this overview 
focus on providing answers to the following key questions:
\begin{enumerate}
\item How can the 2D anyon gas be defined physically and modelled mathematically?
\item What are the essential properties of the anyon gas?
	Does it admit some notion of exclusion statistics intermediate to bosons and fermions?
\item How can an anyon gas emerge in a realistic physical system consisting of 
	only bosons and/or fermions?
\end{enumerate}

The first question is addressed in the first part of Section~\ref{sec:ideal} 
for the case of the ideal abelian anyon gas, 
and in the first part of Section~\ref{sec:nonideal} 
for the nonideal abelian gas.
Section~\ref{sec:nonabelian} introduces the nonabelian gas in brief.
The second question is the subject of the later parts of 
Sections~\ref{sec:ideal}, \ref{sec:nonideal} and \ref{sec:nonabelian}.
The third and final question is addressed in Section~\ref{sec:emergence}.
Conclusions and an outlook are given in Section~\ref{sec:conclusions}.

\section{The ideal anyon gas} \label{sec:ideal}

\subsection{Quantum statistics} \label{sec:quant-stat}

\begin{figure}
	\centering
	\scalebox{1.5}{%
	\begin{tikzpicture} 
		\draw [thick,darkred] (0,0) circle [radius=0.6];
		\draw [fill] (0.6,0) circle [radius=0.05];
		\draw [fill] (-0.6,0) circle [radius=0.05];
		\node [above right] at (0.6,-0.2) {\scalebox{0.7}{$+1$}};
		\node [above left] at  (-0.6,-0.2) {\scalebox{0.7}{$-1$}};
		\node [above right] at (-0.35,-0.2) {\scalebox{0.8}{$e^{i\alpha\pi}$}};
	\end{tikzpicture}
	}
	\caption{The circle of abelian anyons, 
	parametrized by the statistics parameter $\alpha$.
	Note that any choice of proper anyons 
	(i.e. neither bosons nor fermions) marks an additional point in this circle
	and thus breaks its orientation symmetry
	(mirror symmetry / complex conjugation).}
	\label{fig:circ}
\end{figure}
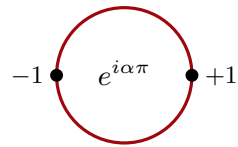

The possibility for anyons in 2D boils down to the peculiar geometry 
and topology of the plane as compared to spaces of higher dimensions. 
Consider for simplicity two particles at positions $\bx_1$
and $\bx_2$ in the Euclidean plane $\R^2$, with a wave function $\Psi$ encoding the
probability density $|\Psi(\bx_1,\bx_2)|^2$ of observing the particles at
these locations.
An exchange of the two positions admits then a change in phase:
$$
	\Psi(\bx_2,\bx_1)=e^{\pm i\theta}\Psi(\bx_1,\bx_2),
$$
where $\theta \in \R$. 
The sign is included above to specify the way in which the exchange is made,
and if 
there is \keyword{orientation symmetry} then we cannot distinguish between
clockwise and counterclockwise exchanges, so that
$$
	e^{i\theta} = e^{-i\theta}
	\quad \Leftrightarrow \quad
	e^{2i\theta} = 1,
$$
so $\theta/\pi \in \Z$ is either an even integer (bosons) 
or an odd integer (fermions).
However, if we relax the assumption on orientation symmetry and consider a
multivalued function for which we keep track of the 
number and orientation of elementary exchanges
(the winding number), then we can allow {\it any} phase $\theta$ (anyons) to appear here.
Conventionally, we write $\theta = \alpha\pi$ where $\alpha$ is known as the
\keyword{statistics parameter} and can be either $\alpha \in \R$
(defined modulo periodicity $2$) or
in a suitable interval $\alpha \in (-1,1]$ or $\alpha \in [0,2)$;
cf. Fig.~\ref{fig:circ}.

More generally, we may consider a wave function $\Psi$ for $N$ 
\keyword{distinct particles} in $d$-dimensional Euclidean space $\R^d$
and its probability density: 
$$
	|\Psi(\sx)|^2, \qquad \sx = (\bx_1,\bx_2,\ldots,\bx_N) \in \R^{dN} \setminus \bDelta_N,
$$
where, in order to indeed make them distinct we have removed the
(fat) \keyword{diagonal set} where any two or more particles overlap:
$$
	\bDelta_N := \{ \sx \in \R^{dN} : \bx_j = \bx_k \ \text{for some} \ j\neq k \}.
$$
Our reason for considering distinct particles is that we, 
for simplicity at this stage,
fix the particle number to $N$ and assume that number to be 
certain 
and conserved
(we shall return to possibilities of collisions on $\bDelta_N$ later).

In the case that the particles are \keyword{identical} 
and thus \keyword{indistinguishable}, then
we should reduce the configuration space further and ignore the (now artificial) labels:
$$
	X = \{\bx_1,\bx_2,\ldots,\bx_N\} \in \cC^N := (\R^{dN} \setminus \bDelta_N)/S_N,
$$
where the permutation group $S_N$ acts on the labels $j$ of $\sx = (\bx_j)_{j=1}^N$ 
while leaving the equivalence class (set) $X$ 
subject to this action fixed.
Indeed, our $N$-particle configuration space is 
identical to the manifold
$$
	\cC^N = \{ X = \{\bx_1,\bx_2,\ldots,\bx_N\} \subset \R^d : |X|=N \}
$$
of all $N$-point subsets of $\R^d$.
If we now consider exchanges of particles then we can no longer act by permutations
since it has lost meaning on configurations $X$ of identical particles, 
however what we can consider instead
is the {\it continuous} exchange of positions in $\cC^N$.
Namely, among the continuous paths from a point $X \in \cC^N$ to another point
$Y \in \cC^N$ there are the \emph{loops}, 
starting and ending at the same point
$X=Y$, which we hereby take to represent 
\keyword{continuous physical exchanges} of identical particles \cite{LeiMyr-77}. 
Given such \keyword{exchange loops} $\sigma$ 
(continuous functions $[0,1] \to \cC^N$ of a parameter, s.t.\ $\sigma(0)=\sigma(1)$), 
we may then consider lifting any wave function $\Psi$ which has been 
assigned 
a value on a configuration $X \in \cC^N$
to a multivalued function\footnote{More concretely, 
a section of a bundle with base space $\cC^N$, 
or a $\rho$-equivariant function on the covering space of $\cC^N$;
see e.g.\ \cite{Dowker-85,Froehlich-90,MunSch-95,DelFigTet-97,Myrheim-99,Lundholm-17,LunQva-20}.
The range/fiber is initially $\C$.}
upon which the loops can act.
Let us denote the action of a loop $\sigma$ on $\Psi$ at the point $X$ 
by $\Psi(\sigma.X)$.
Then we demand the \keyword{exchange conditions} 
\begin{equation} \label{eq:Psi-equivariance}
	\Psi(\sigma.X) = \rho(\sigma) \Psi(X),
\end{equation}
where $|\rho(\sigma)| = 1$, so that
\begin{equation} \label{eq:Psi-prob}
	|\Psi(\sigma.X)|^2 = |\Psi(X)|^2
\end{equation}
unambiguously defines a  
probability density at $X \in \cC^N$.
Further, $\rho(1) = 1$ for the trivial (constant) loop $\sigma=1$, 
and if we compose two loops $\sigma_1$ and $\sigma_2$ into a new loop 
$\sigma_1\sigma_2$ then we demand the compatibility of the action,
$$
	\rho(\sigma_1\sigma_2) = \rho(\sigma_1)\rho(\sigma_2).
$$
In other words, $\rho$ is a group homomorphism 
and thus a \keyword{representation} of the group of exchange loops at $X$ 
with composition.

In comparison,
if we consider a single particle, $\cC^{N=1} = \R^d$, 
charged and subject to magnetism, then the action of a path should be understood 
as the (parallel) transport of the charge along the path,
and different loops $\sigma$ may then give rise to different 
\keyword{holonomies} or phases given by the flux enclosed by the loop:
$$
	\rho(\sigma) = e^{i\Phi_\sigma}, \qquad
	\Phi_\sigma = \oint_\sigma \bA \cdot d\br = \int_{\text{interior of $\sigma$}} B(\bx) \,d\bx,
$$
where $\curl \bA = B$ is the magnetic field 
(viewed as a (pseudo)scalar field in $d=2$ and appropriately projected in $d \ge 3$).
However, for $N\ge 1$ and in the case that there is no external field
so that $\rho(\sigma)$ 
should depend only on the topology 
of the loop $\sigma$, i.e.\ if we require that $\rho(\sigma) = \rho(\sigma')$
for any two topologically equivalent loops $\sigma \sim \sigma'$
(connected by some continuous deformation),
then we consider loops only up to homotopy and
the relevant group of (equivalence classes of) loops is the \keyword{fundamental group} 
of the configuration space
\cite{GolMenSha-83,Wu-84a,GolMenSha-85}:
$$
	\sigma \in 
	\pi_1(\cC^N) = \begin{cases}
		\{1\}, & d=1,\\
		B_N, &d=2,\\
		S_N, &d \ge 3,
	\end{cases}
$$
where $B_N$ denotes the \keyword{braid group}, defined below.
The group is trivial\footnote{%
Interesting statistics can still emerge as a choice of boundary condition 
on $\bDelta_N$ \cite{LeiMyr-77}.}
in one dimension because particles then cannot 
exchange continuously without colliding, i.e.\ hitting the diagonal set $\bDelta_N$.

Now, if $\Psi$ takes values in $\C$, then $\rho(\sigma)$ is a phase 
--- the {\bf exchange phase} corresponding to the exchange loop $\sigma$ --- 
and we may identify different \keyword{exchange quantum statistics} as symmetry classes
of $\Psi$ defined by $\rho$ in \eqref{eq:Psi-equivariance}:
\begin{description}[wide, labelwidth=!, labelindent=0pt]
\item[\keyword{Bosons}] $\rho(\sigma)=+1$, the trivial representation.
	Wave functions $\Psi$ extend symmetrically from $\cC^N$ to $\R^{dN}$,\footnote{%
	$L^2$ denotes the Hilbert space of square-integrable functions
	with inner product $\langle\Psi,\Phi\rangle = \int \overline{\Psi(\sx)}\Phi(\sx) \,d\sx$.}
	$$
		L^2_\sym(\R^{dN}) := \{ \Psi \in L^2 :
		\Psi(\sigma.\sx) = \Psi(\sx),\ \sigma \in S_N \}
	$$
	(the usual action of permutations, and
	with any missing data on $\bDelta_N$ to be filled in, 
	depending on our choice of Hamiltonian).
	For example, particles may be independent and identically distributed:
	$\Psi_0 = \otimes^N \psi_0\ \in L^2_\sym$,
	$$
		\Psi_0(\sx) = (\otimes^N \psi_0)(\sx) = \prod_{j=1}^N \psi_0(\bx_j),
	$$
	corresponding to Bose--Einstein condensation in the single one-body state 
	$\psi_0 \in L^2(\R^d)$.
	
\item[\keyword{Fermions}] $\rho(\sigma)=\mathrm{sign}(\sigma) \in \{-1,+1\}$,
	the signed representation on $S_N$ (signature of permutation).
	In this case wave functions $\Psi$ extend \emph{anti}symmetrically from $\cC^N$ to $\R^{dN}$,
	$$
		L^2_\asym(\R^{dN}) := \{ \Psi \in L^2 \!:\!
		\Psi(\sigma.\sx) = \mathrm{sign}(\sigma)\Psi(\sx), \,\sigma \in S_N \}
	$$
	(in this case it is natural to require that $\Psi$ vanishes on $\bDelta_N$).
	We apparently obtain determinantal correlations and the \keyword{Pauli principle}, 
	since functions in $L^2_\asym$ are spanned by \keyword{Slater determinants}, such as
	$\Psi_0 = \psi_0 \wedge \psi_1 \wedge \ldots \wedge \psi_{N-1}$, 
	where
	$$
		\bigwedge_{k=0}^{N-1} \psi_k (\bx_1,\ldots,\bx_N)
		:= \det \mat{\psi_0(\bx_1)& \cdots & \psi_0(\bx_N)\\
					\vdots & & \vdots \\
					\psi_{N-1}(\bx_1)& \cdots & \psi_{N-1}(\bx_N)}.
	$$
	
\item[\keyword{Anyons}] in $d=2$ the most general $\rho$ 
	is a unitary representation of $B_N$.
	Proper anyons then correspond to those $\rho$ which do not factor through $S_N$,
	i.e. for which orientation symmetry is broken.
	Let us denote by $L^2_\rho$ 
	the Hilbert space of square-integrable multivalued
	wave functions $\Psi$ based on the configuration space $\cC^N$ 
	and subject to the exchange conditions \eqref{eq:Psi-equivariance}.
	A central question is, whether we can anticipate 
	\keyword{intermediate or fractional exclusion statistics} in $L^2_\rho$?
	Actually, it turns out that the information at hand is not 
	sufficient to decide this,
	and even the above conclusions for bosons and fermions can be misleading,
	as we will see below.
\end{description}

\subsection{The braid group} \label{sec:braid-group}

For $d=2$, the group of continuous loops in $\cC^N$ modulo homotopy is
conveniently visualized as a group of $N$-particle world lines that evolve in 
$2+1$-dimensional spacetime,
subject to composition via some arbitrary but fixed reference configuration $X_0 \in \cC^N$.
The resulting group, denoted $B_N$, is the {\bf braid group on $N$ strands}
\cite{Artin-25,Artin-47,Birman-74}:
\begin{align*}
	B_N = \Bigl\langle \sigma_1,\ldots,\sigma_{N-1} \ : \ 
		&\sigma_j \sigma_{j+1} \sigma_j = \sigma_{j+1} \sigma_j \sigma_{j+1}, \\ 
		&\sigma_j \sigma_k = \sigma_k \sigma_j
		\Bigr\rangle^{j,k=1\ldots N-1}_{k \neq j\pm 1} 
\end{align*}
That is, the group generated by $N-1$ elementary braids
$$
	\sigma_j: \quad
	\begin{tikzpicture}[scale=0.4,font=\footnotesize,anchor=mid,baseline={([yshift=-.5ex]current bounding box.center)}]
		\braid[number of strands=7] s_4^{-1};
		\node at (1, -2.0) {$1$};
		\node at (2, -2.0) {$2$};
		\node at (3, -2.0) {$\ldots$};
		\node at (4, -2.0) {$j$};
		\node at (6, -2.0) {$\ldots$};
		\node at (7, -2.0) {$N$};
	\end{tikzpicture}
	\qquad
	\sigma_j^{-1}: \quad
	\begin{tikzpicture}[scale=0.4,font=\footnotesize,anchor=mid,baseline={([yshift=-.5ex]current bounding box.center)}]
		\braid[number of strands=7] s_4;
		\node at (1, -2.0) {$1$};
		\node at (2, -2.0) {$2$};
		\node at (3, -2.0) {$\ldots$};
		\node at (4, -2.0) {$j$};
		\node at (6, -2.0) {$\ldots$};
		\node at (7, -2.0) {$N$};
	\end{tikzpicture}
$$
which are composed by stacking one on top of another.
These elementary braids are subject to two sets of topological relations;
one of which are known as the \keyword{Yang-Baxter relations}, and the other
implement the commutativity of independent braids.
In $B_4$ we have, e.g.,
\begin{eqnarray*}&
	\begin{tikzpicture}[scale=0.4,font=\footnotesize,anchor=mid,baseline={([yshift=-.5ex]current bounding box.center)}]
		\braid[number of strands=4] s_1^{-1} s_2^{-1} s_1^{-1};
	\end{tikzpicture}
	\ = \ 
	\begin{tikzpicture}[scale=0.4,font=\footnotesize,anchor=mid,baseline={([yshift=-.5ex]current bounding box.center)}]
		\braid[number of strands=4] s_2^{-1} s_1^{-1} s_2^{-1};
	\end{tikzpicture}
	\qquad
	\qquad
	\begin{tikzpicture}[scale=0.4,font=\footnotesize,anchor=mid,baseline={([yshift=-.5ex]current bounding box.center)}]
		\braid[number of strands=4] s_3^{-1} s_1^{-1};
	\end{tikzpicture}
	\ = \ 
	\begin{tikzpicture}[scale=0.4,font=\footnotesize,anchor=mid,baseline={([yshift=-.5ex]current bounding box.center)}]
		\braid[number of strands=4] s_1^{-1} s_3^{-1};
	\end{tikzpicture}
	\\&
	\sigma_1\sigma_2\sigma_1 = \sigma_2\sigma_1\sigma_2
	\qquad\qquad\qquad
	\sigma_3 \sigma_1 = \sigma_1 \sigma_3
\end{eqnarray*}
If we add a third set of relations $\sigma_j^2 = 1 \ \forall j$,
or equivalently, $\sigma_j = \sigma_j^{-1}$, 
i.e.\ clockwise and counterclockwise exchanges
cannot be distinguished, then we obtain the {\bf permutation group} $S_N$.
This will be the case for $d \ge 3$ since an elementary exchange may be rotated out of the plane
(continuously and homotopically).
From now on we shall stick to the anyonic case $d=2$.

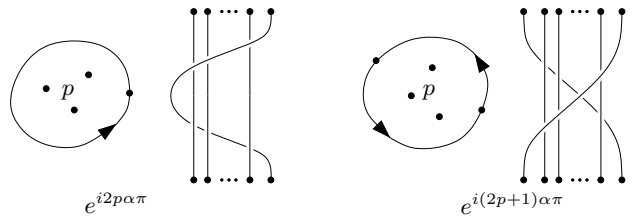
\begin{figure}
	\centering
	\begin{tikzpicture} 
		\node [below right] at (0,0) {\scalebox{0.7}{\includegraphics{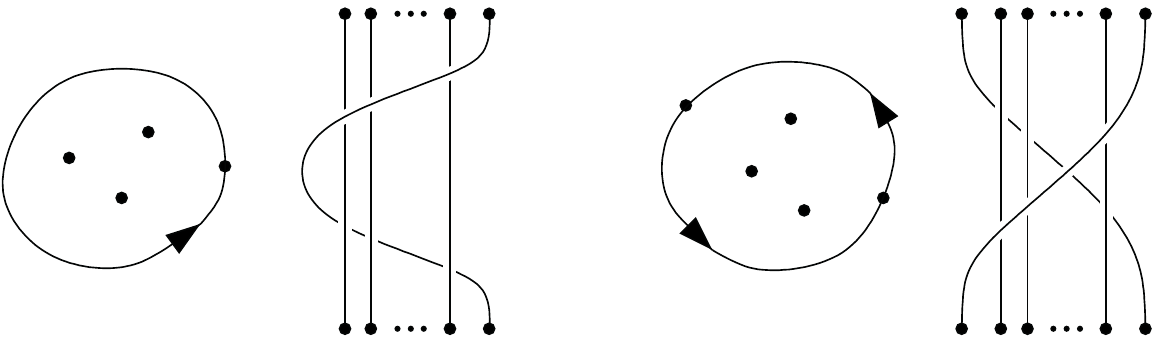}}};
		\node [below right] at (1,-2.5) {$e^{i2p\alpha\pi}$};
		\node [below right] at (6,-2.5) {$e^{i(2p+1)\alpha\pi}$};
		\node [above right] at (0.7,-1.5) {$p$};
		\node [above right] at (5.5,-1.5) {$p$};
	\end{tikzpicture}
	\caption{Exchange loops for a single resp.\ a pair of particles with $p$ other particles
	enclosed, and corresponding braid diagrams and phases.
	From \cite{LunSol-13a}.}
	\label{fig:loops}
\end{figure}

If one looks for irreducible abelian unitary representations, 
i.e.\ \keyword{exchange phases}
$$
	\rho\colon B_N \to \sU(1) = \{ z \in \C : |z|=1 \},
$$
then the Yang-Baxter relations single out a unique phase:
\begin{equation} \label{eq:exch-phase}
	\rho(\sigma_j) = e^{i\alpha\pi}, \qquad \text{for all $j=1,2,\ldots,N-1$};
\end{equation}
cf.\ Figs.~\ref{fig:circ}-\ref{fig:loops}.
The only representations that factor through $S_N$
require either $\alpha=0$ ($\alpha \in 2\Z$) or $\alpha=1$ ($\alpha \in 2\Z+1$),
thus corresponding to bosons or fermions.
We will return to the nonabelian case, and representations with higher rank
than $1$, in Section~\ref{sec:nonabelian}.

\subsection{Kinetic energy} \label{sec:kinetic-energy}

Above we considered wave functions $\Psi \in L^2_\rho$ 
that implement an exchange symmetry \eqref{eq:Psi-equivariance}
by choice of a representation $\rho$ of $B_N$. However, it turns out that this is only one part of
the data that specifies an anyon model. 
The other, equally essential, part is the choice
of a Hamiltonian or kinetic energy operator.
Namely, in order to be able to consider continuous exchanges of particles
we also need to resolve space and time at the appropriate energy scale.

Our starting point is the nonrelativistic\footnote{We remain nonrelativistic
throughout this overview and likewise we will ignore spin;
see however \cite{Jackiw-90,Froehlich-09,Mund-09}.} 
free kinetic energy for $N$ particles with mass $m$:
\begin{equation} \label{eq:T-free}
	T = \frac{1}{2m} \sum_{j=1}^N \bp_j^2,
\end{equation}
where $\bp_j \in \R^2$ is the momentum of the particle with position $\bx_j \in \R^2$.
If the particles are all distinguishable then the canonical quantization 
of this expression is obtained with momentum operators 
$\hat{\bp}_j = -i\hbar\nabla_{\bx_j}$ canonically conjugate 
to the position operators $\hat{\bx}_j$, i.e.
\begin{equation} \label{eq:T-dist}
	\hT_{\rm dist} := \frac{\hbar^2}{2m} \sum_{j=1}^N (-i\nabla_{\bx_j})^2,
\end{equation}
acting on (suitably regular) $\Psi \in L^2(\R^{2N})$.

In our case of identical anyons we may consider the particles to be 
\emph{locally} distinguishable.
Namely, at any fixed point $X \in \cC^N$
we may attach arbitrary labels to them:
$$
	\cC^N \ni X = \{\bx_1,\ldots,\bx_N\} 
	\ \leftrightarrow \ 
	(\bx_1,\ldots,\bx_N) \in \R^{2N} \setminus \bDelta_N,
$$
and we can keep track of these labels along any paths from $X$ in $\cC^N$
as long as they cannot exchange nontrivially, 
i.e.\ as long as any two possible paths do not
compose to a topologically nontrivial loop.
So in order to define free anyons, the idea is then that 
we should keep the kinetic energy operator 
as above whenever we
consider wave functions $\Psi$ supported on topologically trivial subsets of $\cC^N$
(i.e.\ not encircling any of the holes in the manifold),
which then can anyway be treated as subsets of the configuration space 
$\R^{2N} \setminus \bDelta_N$
of distinguishable and distinct particles.
Additional information must then be added to $\hT_{\rm dist}$ 
whenever we pass between different such subsets
of $\cC^N$ so as to complete topologically nontrivial loops 
$\sigma \in \pi_1(\cC^N) = B_N$, $\sigma \neq 1$.
One way to implement this concretely is to treat the exchange condition \eqref{eq:Psi-equivariance}
as a topological boundary condition\footnote{%
Technically, $\Psi$ is defined as a complex-valued function on the covering space of $\cC^N$
subject to the $\rho$-equivariance condition \eqref{eq:Psi-equivariance},
which is equivalent to regarding $\Psi$ as a section of a complex line bundle over 
$\cC^N$ with its geometry
specified by $\rho\colon B_N \to \sU(1)$ 
(locally flat, but globally curved by the holonomies 
specified by $\rho(\sigma)$ for $\sigma \in B_N$); see, e.g.,
\cite{Dowker-85,MunSch-95,LunQva-20}.}
that $\Psi$ must satisfy
in order to be in the domain of the operator, which we denote by
\begin{equation} \label{eq:T-rho}
	\hT_\rho := \frac{\hbar^2}{2m} \sum_{j=1}^N (-i\nabla_{\bx_j}^{(\rho)})^2,
\end{equation}
or simply by $\hT_\alpha$ if $\rho(\sigma_j) = e^{i\alpha\pi}$.
For convenience we also denote the Hilbert space $L^2_\rho$ by $L^2_\alpha$ in this case.
Again, we stress that $\hT_\alpha$ acts \emph{locally} as the standard Laplacian 
$\hT_{\rm dist} = \frac{\hbar^2}{2m}\sum_{j=1}^N(-\Delta_{\bx_j})$ 
on (suitably regular) $\Psi \in L^2_\alpha$, 
thus \emph{independently} of $\alpha$, 
but \emph{globally} we have also the exchange conditions to account for, such as
\keyword{pair exchange}
\begin{multline} \label{eq:Psi-exch}
	\Psi(\bx_1,\ldots,\bx_k,\ldots,\bx_j,\ldots,\bx_N) \\
	= e^{i(2p+1)\alpha\pi} \Psi(\bx_1,\ldots,\bx_j,\ldots,\bx_k,\ldots,\bx_N),
\end{multline}
where $p$ is the number of other particles enclosed
in the exchange loop as $\bx_j$ and $\bx_k$ are exchanged once in a 
counterclockwise manner 
(as well as other conditions; cf. Fig.~\ref{fig:loops}).

As an example, consider the two-particle case $N=2$ 
in center-of-mass and relative coordinates
$\bX = \frac{1}{2}(\bx_1+\bx_2)$, $\br = \bx_1-\bx_2$. 
Then the kinetic energy separates,
\begin{equation} \label{eq:T2-separate}
	\hT_\alpha = \frac{\hbar^2}{4m}(-\Delta_\bX) + \frac{\hbar^2}{m}(-\Delta_\br),
\end{equation}
and further, expressed in polar coordinates $(r=|\br|,\vphi)$:
\begin{equation} \label{eq:T2-polar}
	{-}\Delta_\br = -\partial_r^2 - \frac{1}{r}\partial_r - \frac{1}{r^2}\partial_\vphi^2.
\end{equation}
These operators act on functions $\Psi(\bX,r,\vphi)$ 
(here with its proper multivaluedness on $\cC^2$ incorporated 
in the winding of $\vphi \in \R$ over multiples of $\pi$) 
such that
\begin{equation} \label{eq:T2-bc}
	\Psi(\bX,r,\vphi+\pi) = e^{i\alpha\pi} \Psi(\bX,r,\vphi), \quad r > 0,
\end{equation}
corresponding to the exchange condition \eqref{eq:Psi-exch}
(all braids in $B_2$ are generated by this simple exchange $\sigma_1$).
The manifold $\cC^2$ may be parametrized by
the set in $\R^4$ given by the half-space
$$
	\Omega = \{\bX \in \R^2, \ r > 0, \ 0 \le \vphi < \pi\},
$$
and its boundary 
$$
	\partial\Omega = \{\bX \in \R^2, \ r = 0 \ \text{or} \ \vphi=0 \ \text{or} \ \vphi=\pi \}
$$
relates the two points $(r>0,\vphi=0)$ and $(r>0,\vphi=\pi)$
topologically to the same point in $\cC^2$.
An equivalent description is therefore 
\eqref{eq:T2-separate}-\eqref{eq:T2-polar} defined on $L^2(\Omega)$
and subject to the boundary condition
$$
	\Psi(\bX,r,\vphi=\pi) = e^{i\alpha\pi} \Psi(\bX,r,\vphi=0), \quad r > 0.
$$

The operator $\hT_\rho$ as defined above is formally hermitian but not self-adjoint 
(and not even necessarily \emph{essentially} self-adjoint) 
since we have still not 
supplied any information at the diagonal set $\bDelta_N$, which cuts out 
holes of codimension 2 in the manifold $\cC^N$ (for $N=2$ this set is 
$\bDelta_2 = \{\bX \in \R^2, r=0\}$).
There is however a canonical choice of a self-adjoint realization 
(known in the mathematical literature as the \keyword{Friedrichs extension}),
obtained by considering the expectation value of the kinetic energy in 
its formal quadratic form expression\footnote{This
is more appropriately defined on $\cC^N$ and observed to extend unambiguously 
to the covering space \cite{LunQva-20}.}
\begin{equation} \label{eq:T-form}
	\langle \Psi, \hT_\rho \Psi \rangle 
	= \frac{\hbar^2}{2m} \int_{\R^{2N}} \sum_{j=1}^N |\nabla_{\bx_j}^{(\rho)}\Psi|^2 d\sx \ \ge 0
\end{equation}
initially on (multivalued) wave functions $\Psi$ 
that are smooth and compactly supported away
from $\bDelta_N$ and subject to the exchange conditions 
\eqref{eq:Psi-exch} resp.\ \eqref{eq:Psi-equivariance}.\footnote{%
There is a natural extension (closure) of the energy form \eqref{eq:T-form}
to the largest subspace 
of the Hilbert space $L^2_\rho$ 
upon which the energy is defined and finite, 
which is also the largest domain of the momentum operator $-i\nabla^{(\rho)}$.
Denoting its adjoint $(-i\nabla^{(\rho)})^*$,
one then defines 
$\hT_\rho := \frac{\hbar^2}{2m}(-i\nabla^{(\rho)})^* (-i\nabla^{(\rho)})$,
canonically self-adjoint on a corresponding smaller subspace of $L^2_\rho$.
One might object that this approach 
seems to assume a \keyword{hard-core condition} on the anyons
since functions are initially taken to vanish close to $\bDelta_N$, 
however
for bosons and fermions this eventually yields exactly the standard 
free kinetic energy $\hT_{\rm asym/sym}$
(with the same expression as $\hT_{\rm dist}$)
on the (Sobolev) spaces of anti/symmetric 
square-integrable functions with both one and two partial derivatives,
that are also square-integrable.
It is in fact a peculiarity of dimensions $d\ge 2$ that any extra conditions
at the diagonals disappear in this procedure, 
so that our assumption on distinct particles is justified a posteriori; 
cf \cite{BorSor-92,LunSol-14}. 
This can also be understood by means of a more precise 
version of the Hardy inequality \eqref{eq:Hardy} discussed below \cite{LunQva-20}.}
For $N=2$ the energy form \eqref{eq:T-form} is
\begin{multline} \label{eq:T-form-2}
	\langle \Psi, \hT_\alpha \Psi \rangle 
	= \frac{\hbar^2}{m} \int_{\bX \in \R^2} \int_{r=0}^\infty \int_{\vphi=0}^{2\pi} \Bigl( 
	 \frac{1}{4} |\nabla_{\bX}\Psi|^2 \\
	+ |\partial_r\Psi|^2 + \frac{1}{r^2}|\partial_\vphi\Psi|^2 \Bigr) d\vphi \, rdr \, d\bX,
\end{multline}
again subject to the exchange condition \eqref{eq:T2-bc}.
In general then, this specific choice of a self-adjoint operator 
$\hT_\rho$ resp. $\hT_\alpha$
is what we take to mean precisely by \keyword{\emph{free}}
(noninteracting, as compared to point-interacting; see below) 
\keyword{ideal anyons},
by analogy with bosons and fermions.

\subsection{Statistics transmutation in 2D} \label{sec:stat-transm}

So far we have worked exclusively in the geometric perspective of anyons,
but will now be able to make the connection to the magnetic perspective.

Since we have fixed our attention on the plane $\R^2$, 
it is occasionally convenient to switch to complex notation:
$$
	\R^{2N} \ni \sx = (\bx_1,\ldots,\bx_N) 
	\ \leftrightarrow \ 
	\sz = (z_1,z_2,\ldots,z_N) \in \C^N, 
$$
where the real (imaginary) part in $\C$ is canonically identified with the 
first (second) coordinate in $\R^2$.

Now, consider either a bosonic or a fermionic wave function 
$\Psi \in L^2_{\sym/\asym}(\C^N)$
restricted to the subset of distinct coordinates $\sz \in \C^N \setminus \bDelta_N$
and make there the gauge transformation $\Psi = U\tilde{\Psi}$, where
\begin{equation} \label{eq:U-transmute}
	U(\sz) := \prod_{j<k} \frac{z_j-z_k}{|z_j-z_k|} 
	= \exp \left( i\sum_{j<k} \arg(z_j-z_k) \right),
\end{equation}
the product of all relative phases of pairs of particles.
One may verify that this expression is well defined and single valued on 
$\C^N \setminus \bDelta_N$, unitary $|U(\sz)|=1$,
and antisymmetric w.r.t.\ permutations of labels in $\sz=(z_j)_{j=1}^N$, 
so that if $\Psi$ is symmetric then $\tilde\Psi = U^{-1}\Psi$
is antisymmetric, and vice versa.
Thus, modulo the smaller set of diagonals $\bDelta_N$ where it is not defined, 
the unitary transformation by $U$ achieves a
\keyword{statistics transmutation}
$L^2_\sym \leftrightarrow L^2_\asym$ between bosons and fermions in 2D.
This perhaps seems confusing, 
however, recall that the kinetic energy is equally important in defining
the exchange statistics.
Indeed, if we consider the momentum then
this transmutation comes at the cost of a \keyword{gauge potential}:
\begin{equation} \label{eq:gauge-transf-bf}
	-i\nabla_{\bx_j} \Psi = U(-i\nabla_{\bx_j} + \bA_j)\tilde{\Psi},
\end{equation}
where we have defined the vector potentials
\begin{equation} \label{eq:A-ideal}
	\bA_j(\sx) := -i U^{-1}\nabla_{\bx_j}U 
	= \sum_{k \neq j} \frac{(\bx_j-\bx_k)^\perp}{|\bx_j-\bx_k|^2},
\end{equation}
for $j=1,\ldots,N$, and we denoted a $\pi/2$ rotation in $\R^2$
$$
	\bx = (x,y)
	\quad \mapsto \quad
	\bx^\perp = (-y,x)
$$
(or $z \mapsto iz$).
In other words, if we had initially
$$
	\hT_\asym = \frac{\hbar^2}{2m} \sum_{j=1}^N (-i\nabla_{\bx_j})^2
	\quad \text{acting on $L^2_{\asym}$},
$$
then we now obtain
$$
	\hT_{\sym\to\asym} := \frac{\hbar^2}{2m} \sum_{j=1}^N \left( -i\nabla_{\bx_j} + \bA_j \right)^2
	\quad \text{acting on $L^2_{\sym}$},
$$
so that indeed fermions may be represented as bosons, 
although with an additional magnetic interaction.
The magnetic field seen by particle $j$,
\begin{equation} \label{eq:B-ideal}
	\curl_{\bx_j} \bA_j = 2\pi \sum_{k \neq j} \delta(\bx_j-\bx_k),
\end{equation}
corresponds to the attachment of \keyword{Aharonov--Bohm point fluxes} 
of magnitude $2\pi$ to each of the other particles.\ 
The set where this field is singular is precisely $\bDelta_N$.

Similarly, if we fix a statistics parameter $\alpha \in \R$ and consider
$\Psi = U^\alpha \tilde\Psi$
(now typically a singular gauge transformation involving multivalued functions 
over $\C^N \setminus \bDelta_N$) then
\begin{equation} \label{eq:gauge-transf-anyon}
	-i\nabla_{\bx_j} \Psi = U^\alpha(-i\nabla_{\bx_j} + \alpha\bA_j)\tilde{\Psi},
\end{equation}
and thus if we originally had the anyonic kinetic energy
$$
	\hT_\alpha = \frac{\hbar^2}{2m} \sum_{j=1}^N (-i\nabla_{\bx_j})^2
	\quad \text{acting on $\Psi \in L^2_\alpha$},
$$
then 
we obtain an equivalent description in terms of bosons:
$$
	\hT_{\sym\to\alpha} := \frac{\hbar^2}{2m} \sum_{j=1}^N \left( -i\nabla_{\bx_j} + \alpha \bA_j \right)^2,
	\ \text{acting on $\tilde\Psi \in L^2_{\sym}$}.
$$
If we instead consider the transformation
$\Psi = U^{\alpha-1}\tilde\Psi$ where $\tilde\Psi \in L^2_\asym$,
then we arrive at a description in terms of fermions:
$$
	\hT_{\asym\to\alpha} := \frac{\hbar^2}{2m} \sum_{j=1}^N \left( -i\nabla_{\bx_j} + (\alpha-1) \bA_j \right)^2,
$$
acting on $L^2_{\asym}$.
The magnetic field of $\alpha\bA_j$ 
corresponds to the attachment of Aharonov--Bohm fluxes of magnitude $2\pi\alpha$ 
to each of the other particles.
In this way we have the freedom to treat anyons as either bosons or fermions
with topological magnetic interactions.
This tool can certainly be beneficial as we can trade a simple
operator (the free kinetic energy) but on a complicated geometry 
(multivalued functions with appropriate exchange conditions),
for a complicated operator (magnetic interactions) but on a simple geometry
(symmetric or antisymmetric functions on $\R^{2N}$).
The former choice is usually referred to as the \keyword{anyon gauge}
while the latter is the \keyword{magnetic gauge},
and this connection also brings us from an idealized world of 
quantum statistics of point particles and closer
to the various possibilities for emergent statistics transmutation phenomena.

The procedure we referred to above to ensure the self-adjointness of the 
kinetic energy operator by extending the energy form
(Friedrichs extension) also applies in this magnetic gauge formulation of 
the problem, 
so that indeed the operators $\hT_{\sym/\asym\to\alpha}$ 
are self-adjoint on appropriate
domains in $L^2_{\sym/\asym}$ where the energy is finite \cite{LunSol-14}.
Further, we remark that multiples of the unitary multiplication operator
$$
	U^2\colon L^2_{\sym/\asym/\alpha} \to L^2_{\sym/\asym/(\alpha+2)}
$$
may be used, together with complex conjugation symmetry, 
to obtain that the free ideal
anyonic kinetic energies are unitarily equivalent w.r.t.\ the transformations
\begin{equation} \label{eq:ideal-symmetries}
	\alpha \mapsto \alpha + 2n, \ n \in \Z
	\quad \text{and} \quad
	\alpha \mapsto -\alpha.
\end{equation}

\subsection{Exchange vs. exclusion} \label{sec:exch-excl}

Now that we have managed to define more or less concrete but still
hypothetical models for ideal abelian anyons, 
we come to the next 
question: How do such anyons actually behave? 
If we are ever to observe them in nature then we need to know their behavior.
So, are they more like bosons or more like fermions, or something in between,
like the terms ``intermediate statistics'' or ``fractional statistics'' suggest?
In other words, can we in any way relate the intermediate (braid)
\keyword{exchange statistics} encoded into anyons 
to some intermediate \keyword{exclusion statistics} 
concerning the occupation numbers 
of various states as seen from a one-body perspective. 
Of course, since the magnetic gauge picture clarifies 
that anyons are (possibly) at least as complicated as strongly interacting systems
of bosons or fermions, it is a priori not at all obvious whether this is possible.
The issue of exchange vs.\ exclusion was referred to as ``$\alpha$ to $\beta$'' 
by Canright and Johnson in their review \cite{CanJoh-94} on the matter.

\begin{figure}
	\centering
	\includegraphics[scale=0.9]{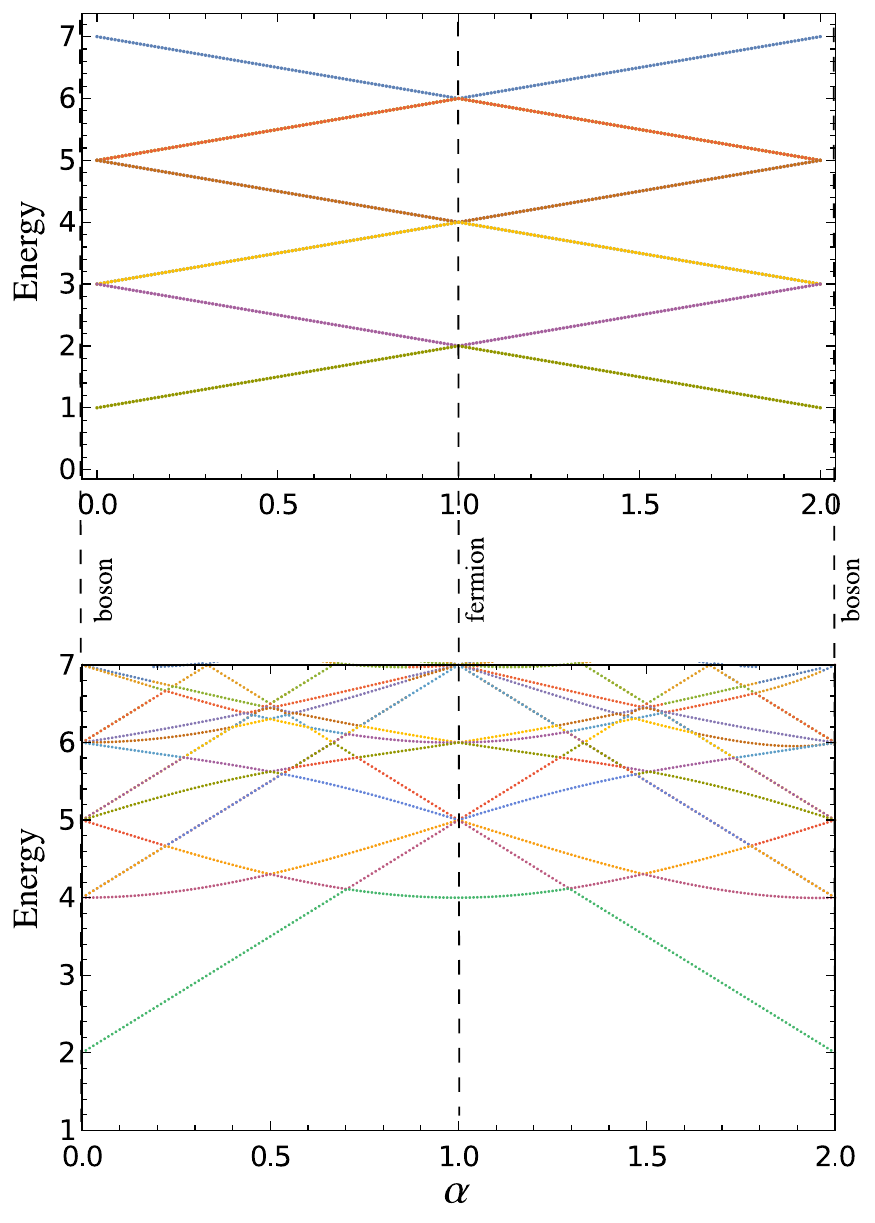}
	\caption{Energy spectra for $N=2$ (upper) and $N=3$ (lower) anyons
	in a harmonic trap (minus the center of mass energy 1, and in suitable units). 
	From \cite{Yakaboylu-etal-19}; 
	cf.\ also \cite{LeiMyr-77,MurLawBraBha-91,SpoVerZah-91}.}
	\label{fig:spectrum-2-3}
\end{figure}

Only in cases of very few particles, $N=2,3,4$, are 
energy spectra for ideal anyons well understood,
because they have then been computed 
analytically 
(for $N=2$ \cite{LeiMyr-77,Wilczek-82b,AroSchWilZee-85},
also extending to a collective \cite{DatMur-93,DatMurVat-03} 
part of the spectrum for all $N$ \cite{Wu-84b,Chou-91a}) 
or numerically 
(the remaining part of the spectrum \cite{MurLawBraBha-91,SpoVerZah-91,SpoVerZah-92}),
in a few situations.

In order to lift the degeneracy of the spectrum and facilitate comparison,
it is convenient to confine the particles in a harmonic trapping potential 
$V(\bx) = \frac{1}{2}m\omega^2|\bx|^2$.
For $N$ anyons in the magnetic gauge description w.r.t.\ bosons
we thus consider the Hamiltonian operator 
$$
	\hH_N = \hT_{\sym\to\alpha} + \hV
	= \sum_{j=1}^N \left[
		\frac{\hbar^2}{2m} \bigl( -i\nabla_{\bx_j} + \alpha \bA_j \bigr)^2 
		+ V(\bx_j)
	\right],
$$
and its corresponding ground-state energy (g.s.e.)
\begin{equation} \label{eq:E-harmonic}
	E_N := \inf \textup{spec} \,\hH_N 
	= \inf_{0 \neq \Psi \in L^2_\sym} \langle \hH_N \rangle_\Psi,
\end{equation}
where
$\langle\hat{A}\rangle_\Psi := \langle\Psi,\hat{A}\Psi\rangle / \langle\Psi,\Psi\rangle$
denotes the expectation of any operator $\hat{A}$ in a state $\Psi \neq 0$.
The spectrum of the 2D harmonic oscillator $\hH_{N=1}$ is well known:
$$
	\spec \hH_1 = \{ \hbar\omega n \}_{n=1,2,3,\ldots} \ \text{(each with multiplicity $n$)},
$$
and for bosons one thus obtains $N$ times the lowest eigenvalue, 
$E_N(\alpha=0) = \hbar\omega N$, 
while for (spinless) fermions, due to the Pauli principle,
we must sum over the $N$ first eigenvalues according to their multiplicity:
$E_N(\alpha=1) \sim \frac{\sqrt{8}}{3} \hbar\omega N^{3/2}$
as $N \to \infty$.
Further, because of rotation symmetry, 
the Hamiltonian commutes with the total angular momentum operator
$$
	\hL/\hbar 
	= -i\sum_{j=1}^N \left[ x_j\partial_{y_j} - y_j\partial_{x_j} \right]
	= \sum_{j=1}^N \left[ z_j \frac{\partial}{\partial z_j} - \bar{z}_j \frac{\partial}{\partial \bar{z}_j} \right],
$$
and its eigenvalues $L \in \Z$ can therefore be used to separate the energy spectrum further.

The spectrum of $\hH_N$ for $N=2$ and $N=3$ is shown in Fig.~\ref{fig:spectrum-2-3},
and while in the former case it manifests a strictly linear interpolation
between bosons and fermions, \cite{LeiMyr-77}
\begin{equation} \label{eq:E2-harmonic}
	E_2(\alpha) = \hbar\omega\left(2 + \min_{q \in \Z}|2q+\alpha| \right),
\end{equation}
in the latter case there are also nonlinearly
interpolating states at various $L$, and such a state with $L \neq 0$ 
comes down from an excited level at $\alpha=0$ to reach the 
Fermi energy at $\alpha=1$.
Similar features are also seen in the numerical spectrum for $N=4$ \cite{SpoVerZah-92}
and expected to occur for all $N \ge 3$ \cite{ChiSen-92}.

\begin{figure}
	\centering
	\includegraphics[scale=0.15]{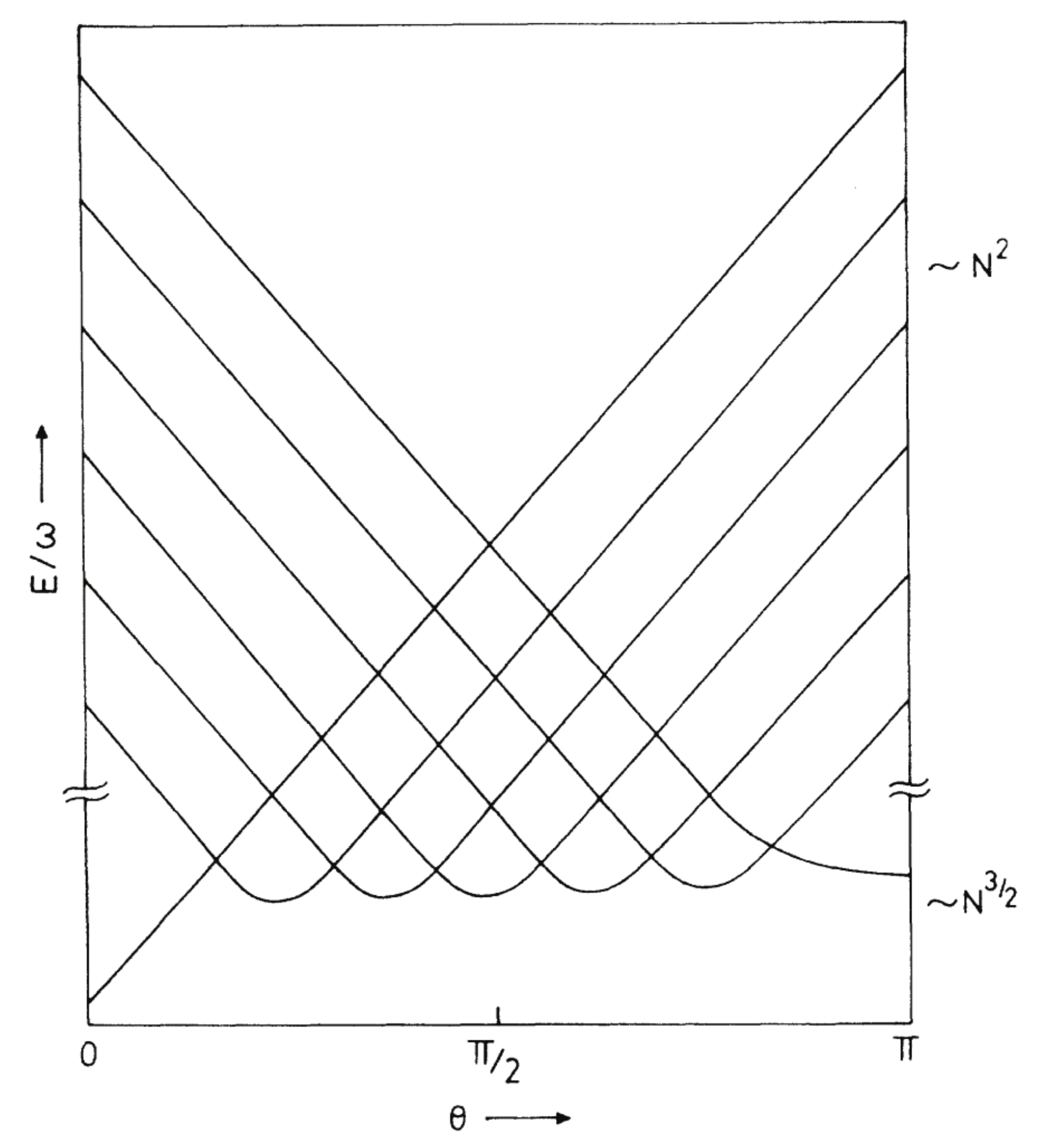}
	\includegraphics[scale=0.15]{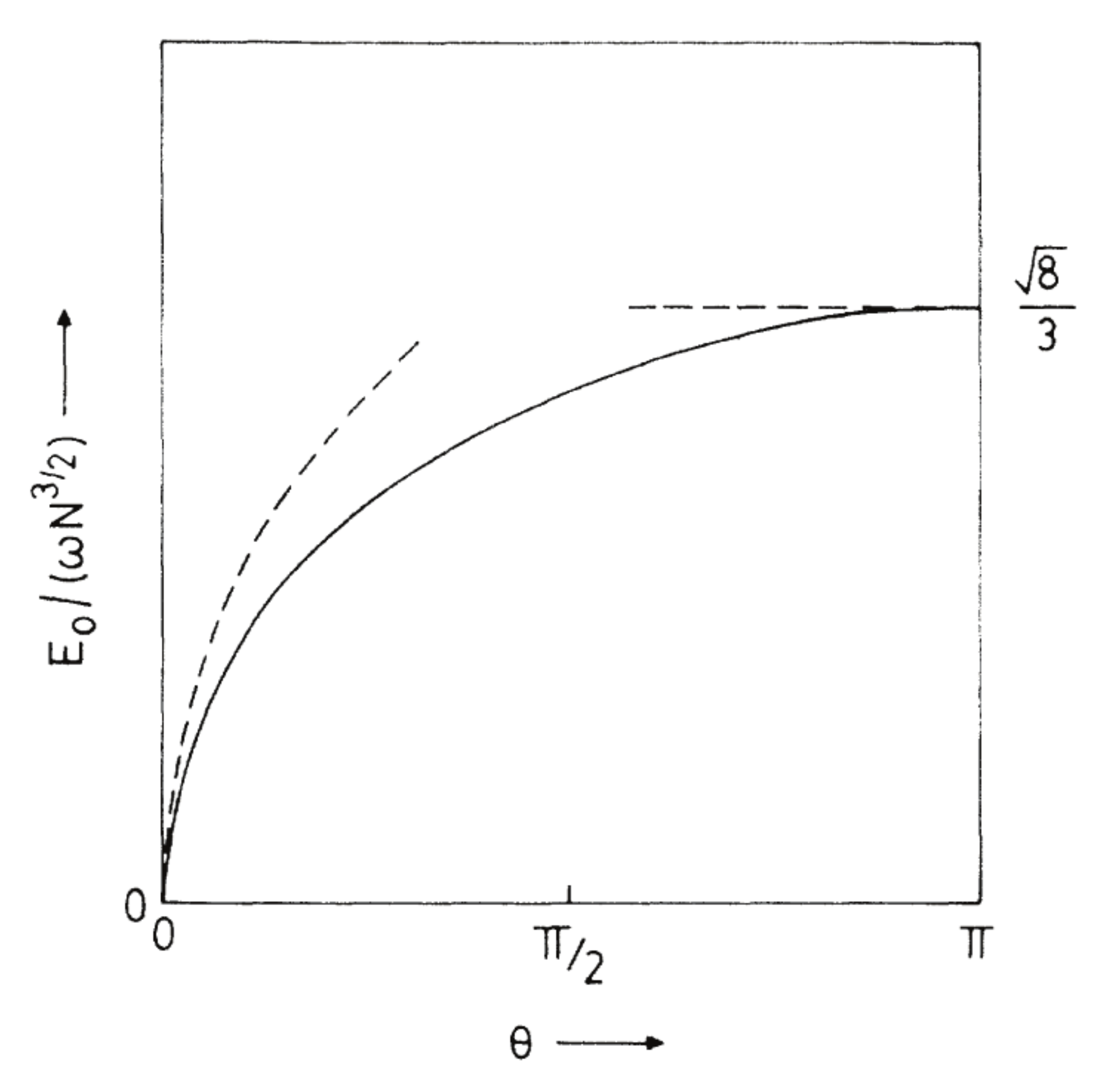}
	\caption{Schematic energy spectrum for $N \gg 1$ (left) and 
	estimates for the g.s.e.\ near bosons and fermions (right) 
	for anyons with statistics $\alpha=\theta/\pi$
	in a harmonic trap. From \cite{ChiSen-92}.}
	\label{fig:spectrum-N}
\end{figure}

Rough upper bounds for $E_N$ at any $\alpha$ and $N$ 
are obtained simply by considering trial states
which localize the particles on separate domains in $\R^2$, so that their exchange 
statistics is not seen 
(in the magnetic gauge picture one may then gauge away all
magnetic interactions, while in the anyon gauge picture
one then stays within a domain of distinguishability).
By balancing the localization energy and the spread into the potential
one then finds that $E_N(\alpha) \lesssim \hbar\omega N^{3/2}$, 
i.e.\ the energy is at most at the order of fermions.
Note however that the fermionic energy does not provide a strict upper bound, 
as seen for $N=3$ where $E_3(\alpha \approx 0.7) > E_3(\alpha=1)$.

On the other hand,
it is also useful to note that the bosonic energy, 
for which the g.s.\ $\Psi_0$ is a Gaussian at the origin $\sx=0$,
always gives a rigorous lower bound:
\begin{equation} \label{eq:boson-bound}
	E_N(\alpha) \ge E_N(0) = \hbar\omega N,
\end{equation}
since the kinetic energy satisfies a \keyword{diamagnetic inequality} \cite{LunSol-14}
\begin{equation} \label{eq:diamag-ineq}
	\langle \hT_\alpha \rangle_\Psi \ge \langle \hT_\sym \rangle_{|\Psi|}.
\end{equation}
This is observed by factoring $\Psi$ into its modulus and phase,
and dropping the phase contribution to the momentum $-i\nabla\Psi$.

Chitra and Sen \cite{ChiSen-92} provided a more interesting lower bound to 
the spectrum for any $N$, extending \eqref{eq:E2-harmonic}:
for an arbitrary state $\Psi$ with angular momentum $L$, it holds
\begin{equation} \label{eq:CS-bound}
	\langle \hH_N \rangle_\Psi 
	\ \ge \ 
	\hbar\omega \left( N + \left|L + \alpha \frac{N(N-1)}{2} \right| \right).
\end{equation}
Combined with our a priori upper bound, this 
implies that in the ground state
the angular momentum needs to be close to
\begin{equation} \label{eq:opt-ang-mom}
	L \approx -\alpha \binom{N}{2}, \ \text{since} \ 
	\left|L + \alpha \binom{N}{2}\right| \lesssim O(N^{3/2}),
\end{equation}
so that on the average there needs to be an angular momentum
$-\alpha$ for every pair of particles in order to cancel the 
shifted inherent/anyonic angular momentum 
(the number of attached units of magnetic flux per particle).
This also means that as $N$ grows large there will be a large number $O(N^{1/2})$
of level crossings in the g.s.\ between different integer angular momenta, 
because any fixed angular momentum $L$ can only be good on a scale 
$\Delta\alpha = O(N^{-1/2})$.
The situation is sketched in Fig.~\ref{fig:spectrum-N}, 
where at fixed $N \gg 1$ one expects to find curves corresponding 
to families of states at fixed $L$, 
that are smoothly varying with $\alpha$, 
to eventually come down to an energy $O(N^{3/2})$ on a $O(N^{1/2}$)-neighborhood 
around a suitable minimum $\alpha \approx \alpha_0(L)$,
and then to again shoot upwards in energy with a slope $\gtrsim O(N^2)$.

Chitra and Sen went on to estimate 
(again, see Fig.~\ref{fig:spectrum-N})
the enveloping g.s.e.\ curve close to bosons
(suggesting $E_N/(\hbar\omega N^{3/2}) \approx \frac{7\sqrt{3}}{9} \sqrt{\alpha}$) 
and to fermions 
(suggesting $E_N/(\hbar\omega N^{3/2}) \approx \frac{\sqrt{8}}{3}$),
but left it as an open problem whether there are nonanalyticities
in the limiting function
\begin{equation} \label{eq:harm-energy}
	h(\alpha) := \liminf_{N \to \infty} E_N(\alpha)/(\hbar\omega N^{3/2}),
\end{equation}
concluding that ``This is an interesting but difficult question to answer.''

\subsection{Towards density functionals for anyons} \label{sec:functionals}

The few-particle spectra led to some initial results on the thermodynamics
of the anyon gas, by considering the high-temperature, low-density virial
expansions \cite{AroSchWilZee-85,Dowker-85} 
(see \cite{Myrheim-99,Khare-05,ManTroMus-13a} for review).
In the low-temperature regime,
in order to make the many-anyon ground state problem more manageable,
we could wish for a limiting effective description 
with a relatively limited number of degrees of freedom that are 
sufficient to describe the collective state of the system at $N \to \infty$.
Such a state might converge towards the minimizer of a suitable functional, 
i.e. (taking some suitable scale $a>0$ for normalization):
\begin{eqnarray*}
	&E_N/N^a \xrightarrow{N \to \infty} \text{minimum of $\cE[\psi]$ or $\cE[\varrho]$, and}\\
	&\text{some marginal of g.s.\ $\Psi_0$ of $\hH_N$} \xrightarrow{N \to \infty} \text{minimizer of $\cE$},
\end{eqnarray*}
where $\cE$ is typically an energy functional of either one-body states 
$\psi \in L^2(\R^2)$
or of corresponding probability densities $\varrho = |\psi|^2 \in L^1(\R^2)$
(with mass $1$ or $N$).

For example, for 2D fermions in a trapping potential $V$, 
one has the {\bf Thomas--Fermi approximation} \cite{Thomas-27,Fermi-27}:
\begin{equation} \label{eq:TF-approx}
	E_N = 
	\inf_{0 \neq \Psi \in L^2_\asym} \langle \hT_\asym + \hV \rangle_\Psi
	\approx \inf_{\varrho \ge 0 : \int_{\R^2}\varrho=N} \cETF[\varrho],
\end{equation}
where the density $\varrho = \rhoTF \approx \varrho_{\Psi_0}$ 
(one-body marginal of g.s.) 
minimizes the \keyword{Thomas--Fermi (TF) functional}
\begin{equation} \label{eq:TF-func}
	\cETF[\varrho] := \int_{\R^2} \left[ 
		2\pi\varrho(\bx)^2 + V(\bx)\varrho(\bx) \right] d\bx.
\end{equation}
({\bf Note:} {\it To simplify our discussion, 
we shall now and in the remainder 
of this article assume that the physical constant $\hbar^2/(2m)=1$,
i.e. $T=\bp^2$, by an appropriate choice of units $\hbar=2m=c=e=1$.})
The kinetic energy density is here given by the energy per unit area 
of the \keyword{homogeneous 2D Fermi gas} at density $\varrho(\bx)$, 
obtained e.g.\ by confining $N$ fermions in a box of area $L \times L$, 
summing the one-body eigenvalues, and taking the thermodynamic limit:
$$
	E_N/L^2 \xrightarrow{N,L \to \infty} 2\pi \bar\varrho^2, \qquad 
	\text{at fixed} \ \bar\varrho = N/L^2.
$$

On the other hand, for bosons with sufficiently weak \emph{scalar} interactions 
one has the \keyword{Gross--Pitaevskii (GP) functional} \cite{Gross-61,Pitaevskii-61}:
\begin{equation} \label{eq:GP-func}
	\cEGP[\psi] := \int_{\R ^2} \Bigl[ 
	\bigl|({-i}\nabla + \bA_{\rm ext})\psi\bigr|^2 + V|\psi|^2 + g|\psi| ^4  \Bigr],
\end{equation}
describing approximate condensation into the state $\psi \in L^2(\R^2)$
as $E_N/N \to$ minimum of $\cEGP$.
In regimes where the interaction strength $g \in \R$ 
(related to the scattering length of the interaction potential)
dominates over the kinetic term one may neglect that term 
and keep only the last two terms,
again leaving a functional of $\varrho=|\psi|^2$ of the Thomas--Fermi type
with $g \leftrightarrow 2\pi N$
(this is only a formal analogy at the level of functionals, 
since the two mechanisms of the approximations are very different).

In view of the above, could one then suggest the following interpolating
functional for $0 < \alpha \le 1$,
describing an \keyword{``average-field'' approximation} for anyons 
\begin{equation} \label{eq:cf-func}
	\cEAF[\varrho] \approx
	\int_{\R^2} \Bigl[ 
		2\pi \alpha \varrho(\bx)^2 + V(\bx)\varrho(\bx) \Bigr] d\bx
	\quad ?
\end{equation}
Indeed, in a mean-field\footnote{%
We differentiate between ``average-field'' and ``mean-field'', partly to distinguish the
anyonic context, but also because of more technical reasons 
that will be clarified below.}
(\keyword{Hartree}) ansatz for the ground state
in the bosonic representation, consider:
\begin{equation} \label{eq:Hartree}
	\Psi(\sx) = \psi(\bx_1) \psi(\bx_2) \ldots \psi(\bx_N),
\end{equation}
and we may think of a single particle $\psi \in L^2(\R^2)$ in the magnetic field 
of the others (identically distributed):
\begin{equation} \label{eq:avg-field}
	B(\bx) \approx 2\pi\alpha (N-1)|\psi(\bx)|^2 
	\approx 2\pi\alpha \varrho_\Psi(\bx).
\end{equation}
If the density $\varrho_\Psi=N|\psi|^2$ is approximately constant on an area 
$L \times L$ 
where $N$ bosons condense into the lowest Landau level 
of the field $B$ (cf.\ \eqref{eq:H-Landau} below), 
then one indeed obtains an energy per unit area
\begin{equation} \label{eq:avg-energy}
	E_N/L^2 \approx B N/L^2 \approx 2\pi\alpha \bar\varrho^2, 
	\qquad \bar\varrho = N/L^2,
\end{equation}
thus suggesting \eqref{eq:cf-func}.
For the ideal anyon gas this argument is not rigorous, 
because one cannot apply such an ansatz in the singular magnetic field
(it is not in the appropriate domain of $\hT_\alpha$).
This difficulty can be overcome using regularization however, 
and we return to it in Section~\ref{sec:almost-bosonic} in conjunction 
with the extended anyon gas.
Thus, after accepting some refinements to the ``average-field'' approach in that context,
it has been successfully applied 
in the limits $\alpha \to 0$ and $\alpha \to 1$ as $N \to \infty$.

To be precise, if we wish to refer to the approximation as above with an 
almost constant field (on the scale of a trap)
then it is perhaps more appropriate to call it a 
\keyword{``constant-field'' approximation}.
We also note that if \eqref{eq:cf-func} would be a valid description 
for proper anyons
then it suggests that for the harmonic oscillator problem
\begin{align} 
	E_N &\stackrel{?}{\approx}
		\inf_{\substack{\varrho \ge 0 \\ \int\varrho = N}} \int_{\R^2} 
		\Biggl[ \frac{\hbar^2\pi\alpha}{m} \varrho(\bx)^2 + 
		\frac{m\omega^2}{2}|\bx|^2 \varrho(\bx)
		\Biggr] d\bx \nonumber\\
		&= \frac{\sqrt{8}}{3} \sqrt{\alpha} \,\hbar\omega N^{3/2},
		\qquad 0 < \alpha \le 1.
		\label{eq:avg-field-hosc}
\end{align}

Early applications of variants of \keyword{density functional theory} (DFT) 
to the anyon context were made by Chitra and Sen, 
who proposed the refinement as in Fig.~\ref{fig:spectrum-N} due to the
bosons' hard core \cite{ChiSen-92},
and by Li, Bhaduri and Murthy who considered an extension of TF theory
to excited states in the spectrum \cite{LiBhaMur-92}.
Again, only some aspects have been rigorously justified,
in the limits close to bosons and fermions.

\subsection{Local exclusion principle and degeneracy pressure for the ideal anyon gas} \label{sec:local-exclusion}

An approach that has been successful for an increased qualitative understanding of
the genuine ideal anyon gas at arbitrary $\alpha$ is to take a strictly 
\emph{local} route to exclusion via the wider 
concept of \keyword{statistical repulsion}.
Namely, one may note that 
statistical repulsion manifests itself in the anyon gas
in (at least) three ways:
\begin{enumerate}[wide, labelwidth=!, labelindent=0pt]
\item {\bf Effective \emph{scalar} pairwise repulsion.}
	The exchange conditions \eqref{eq:Psi-exch} for each pair of particles 
	transpire to yield a pair repulsion between particles.
	Its simplest form can be stated as a lower bound 
	(known in mathematics as a many-particle \keyword{Hardy inequality} 
	\cite{HofLapTid-08,LunSol-13a}) 
	for the kinetic energy: for any $N$,
	\begin{equation} \label{eq:Hardy}
		\hT_\alpha \ \ge \ \frac{4\alpha_N^2}{N} \sum_{j < k} \frac{1}{|\bx_j-\bx_k|^2},
	\end{equation}
	interpreted in the sense of forms or expectation values, and
	where the so-called \keyword{$N$-fractionality of $\alpha$}, defined
	\begin{align*}
		\alpha_N 
		&:= \min_{p=0,1,\ldots,N-2} \min_{q \in \Z} |(2p+1)\alpha - 2q|,
	\end{align*}
	is the arcwise distance (in units of $\pi$)
	on the unit circle from the set 
	$\{e^{i(2p+1)\alpha\pi}\}_{p=0,1,\ldots,N-2}$
	of all possible pair-exchange phases 
	to the point $+1$ of bosons.
	Taking the limit as $N \to \infty$ one finds (see Fig.~\ref{fig:popcorn})
	\begin{equation} \label{eq:fractionality}
		\alpha_{N \to \infty}
			= \left\{ \begin{array}{ll}
			\frac{1}{\nu}, & 
			\text{if $\alpha = \frac{\mu}{\nu}$ \emph{odd}-numerator reduced rational,}\\
			0, & \text{otherwise,}
			\end{array}\right.
	\end{equation}
	an odd variant of Thomae's ``popcorn function''.
	
\begin{figure}
	\centering
	\scalebox{1.1}{%
	\begin{tikzpicture}
		\node [above right] at (0,0) {\includegraphics[scale=0.65, trim=0.4cm 0cm 0cm 0cm]{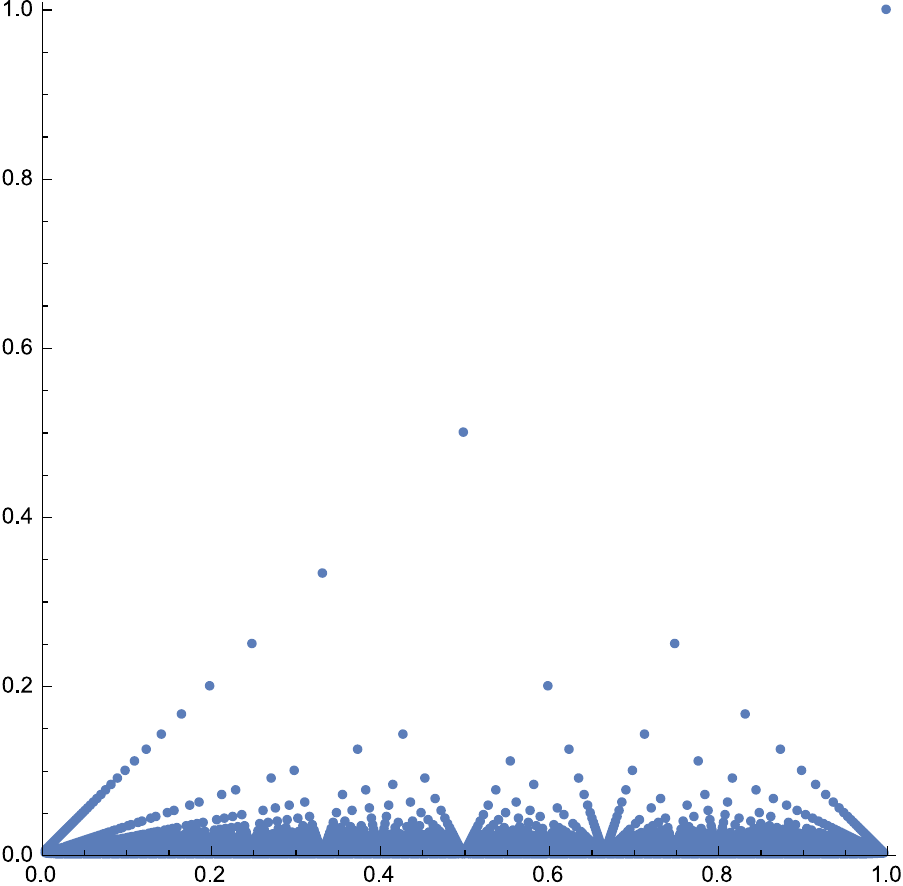}};
		\node [above right] at (5.70,0) {\scalebox{0.8}{$\alpha$\hspace{-20pt}}};
		\node [above right] at (-0.1,5.95) {\scalebox{0.8}{$\alpha_{N \to \infty}$}};
	\end{tikzpicture}
	}
	\caption{%
	$\infty$-fractionality of $\alpha$ given by the odd popcorn function 
	\eqref{eq:fractionality} \cite{LunSol-13a,Lundholm-16}.}
	\label{fig:popcorn}
\end{figure}

	The origin of the scalar lower bound \eqref{eq:Hardy} lies in the effective 
	angular momentum barrier of each pair of particles at distance $r>0$:
	\begin{equation} \label{eq:statistical-repulsion}
		V_{\textrm{stat}}(r) = |(2p+1)\alpha - 2q|^2\frac{1}{r^2} 
		\ \ge \ \frac{\alpha_N^2}{r^2}.
	\end{equation}
	Namely, in an exchange of two particles, as in \eqref{eq:Psi-exch}
	where $p$ other particles happen to be ensnared,
	an anyonic phase shift $e^{i(2p+1)\alpha\pi} \neq 1$ 
	leaves a residual angular momentum according to \eqref{eq:statistical-repulsion}.
	Heuristically, we can anticipate this effect 
	in the two-particle energy form \eqref{eq:T-form-2},
	where the quantization condition \eqref{eq:T2-bc} in the angular variable $\vphi$
	is appropriately modified to account for the presence of any extra particles,
	assuming all other position variables are fixed and
	the relative angular variable undergoes a full loop $\varphi \to \varphi+\pi$
	in configuration space.
	The anyonic phase then shifts the 
	relative angular momentum of the pair,
	which however is quantized over the even integers due to the half-circle symmetry
	for identical particles.\footnote{%
	The operator $\hat{\ell}=-i\partial_\vphi$ on $[0,\pi]$ 
	with twisted boundary conditions $\psi(\pi)=e^{i\alpha\pi}\psi(0)$
	has eigenvalues $\{\alpha-2q\}_{q \in \Z}$.
	Therefore the minimum of $\hat{\ell}^2$ is $\min_{q \in \Z}|\alpha-2q|^2 = \alpha_2^2$.
	Compare also \eqref{eq:E2-harmonic}.}

	We expect the repulsive barrier \eqref{eq:statistical-repulsion} 
	to be stronger than the r.h.s.\ on average as the number $p$ fluctuates.
	Indeed,
	the coupling constant of the pair-interaction potential on the r.h.s.\ of \eqref{eq:Hardy}
	can be strengthened to a nearest-neighbor version,
	which only depends on the 2-fractionality $\alpha_2$ (periodized $\alpha$), 
	if it is at the same time weakened by yet another factor of $N$ 
	\cite{LunQva-20,RouYan-23b}.
	Anyway, the repulsion offered by the potential 
	is strong enough to imply that $\Psi \to 0$ at $\bDelta_N$
	if $\alpha_2 \neq 0$, and can therefore be viewed as a
	\keyword{generalized Pauli principle} for anyons 
	\cite{LunSol-13b,Lundholm-16,LunQva-20}
	(other variants were considered in \cite{FroMar-88,FroMar-89,GolSha-96,GolMaj-04}).

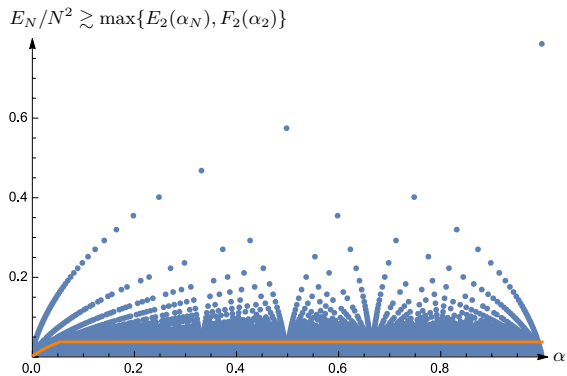
\begin{figure}
	\centering
	\begin{tikzpicture}
		\node [above right] at (0,0) {\includegraphics[scale=0.79]{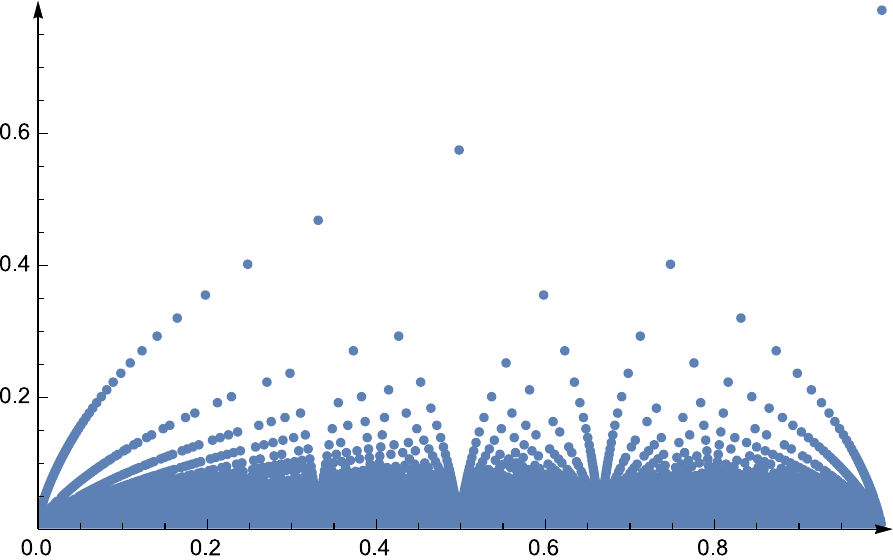}};
		\node [above right] at (7.23,.18) {\scalebox{0.8}{$\alpha$\hspace{-2pt}}};
		\node [above right] at (0,4.6) {\scalebox{0.8}{$E_N/N^2 \gtrsim \max\{E_2(\alpha_N),F_2(\alpha_2)\}$}};
		\node [above right] at (0.28,0.23) {\scalebox{6.8}[5.2]{\begin{tikzpicture}
			\draw[very thin,color=orange,domain=0:0.054,samples=50] plot (\x,{0.25*3.8*\x/sqrt(1+16.5*\x)});
			\draw[very thin,color=orange,domain=0.052:1] plot (\x,{0.25*0.147});
			\end{tikzpicture}}};
	\end{tikzpicture}
	\caption{Lower bounds to the homogeneous ideal anyon gas energy 
	\cite{LarLun-16,LunSei-17}.
	The blue points are numerical lower bounds for 
	$E_2(\alpha) = 4\pi\alpha + O(\alpha^{4/3})$
	at $\alpha=\alpha_{N \to \infty}$, while the orange curve describes
	$F_2(\alpha_2) = \frac{1}{4}\min\{ E_2(\alpha_2),0.147 \}$, correspondingly.
	}
	\label{fig:ideal}
\end{figure}
			
\item {\bf Local exclusion principle.} 
	The bound \eqref{eq:Hardy} (with its various refinements 
	\cite{LunSol-13a,LarLun-16,LunQva-20})
	can also be applied locally on subsets of the plane
	and is strong enough to yield a linear bound 
	$E_N \gtrsim N-1$
	for the local (Neumann) energy \cite{LunSol-13b,LunSei-17}.
	Again, this holds modulo a (positive) constant that either depends on $\alpha_N$,
	or a typically weaker one (still positive) that only depends on $\alpha_2$
	(cf.\ Fig.~\ref{fig:ideal}).
	
	For comparison, take e.g.\ the unit square $Q=[0,1]^2$ and
	consider fermions which there satisfy the bound
	\begin{equation} \label{eq:E-N-Neumann-Fermi}
		E_N(\alpha=1,Q) \ge \pi^2 (N-1) \qquad \text{for}\ N\ge 1,
	\end{equation}
	since each added particle is orthogonal to the ground state 
	(the constant function) and thus adds at least the energy $E_2=\pi^2$ of a
	first-excited state $\psi(x,y) \propto \cos(\pi x)$ or $\cos(\pi y)$.
	
	For anyons, we denote the N-particle \keyword{Neumann energy on $Q$} by 
	\begin{equation} \label{eq:E-N-Neumann}
		E_N(\alpha,Q) := 
		\inf_{0 \neq \Psi \in L^2_\sym(Q^N)} \langle \hT_{\sym\to\alpha} \rangle_\Psi.
	\end{equation}
	It is known that \cite{LarLun-16,LunSei-17}
	\begin{equation} \label{eq:E-2-Neumann}
		E_2(\alpha,Q) = 4\pi\alpha_2 + O(\alpha_2^{4/3})
		\qquad\text{for any}\ \alpha,
	\end{equation}
	and furthermore, all higher-particle energies can be bounded uniformly
	in terms of this energy, or a suitable approximation\footnote{%
	$\tilde{E}_2(\alpha) \approx 2\pi (j_{\alpha_2}')^2 \ge 4\pi\alpha_2$, 
	where $j_\alpha'$ denotes the 
	first zero of the derivative of the Bessel function $J_\alpha$ 
	\cite{LarLun-16}.}
	$\tilde{E}_2$ to it: 
	for $N \ge 2$,
	\begin{equation} \label{eq:E-2-approx}
		E_N \ge (N-1)\tilde{E}_2(\alpha_N),
	\end{equation}
	as well as
	\begin{equation} \label{eq:E-N-4}
		E_N / N \ge \frac{1}{4}\min \{E_2,E_3,E_4\} \ge 
		\frac{1}{4}\min\{ E_2(\alpha_2),0.147 \}.
	\end{equation}
	This r.h.s.\ we denote by $F_2(\alpha_2)$; cf.\ Fig.~\ref{fig:ideal}.
	An interpretation of the above is that 
	the nearest-neighbor energy $E_2$ is a relatively good approximation
	at every scale (at least as a lower bound)
	due to a balancing of uncertainty and exclusion in the gas.
	
\item {\bf Degeneracy pressure.} 
	Lastly, a combination of the above local exclusion principle and the
	uncertainty principle (also in a local formulation) yields a convenient
	measure of the degeneracy pressure in the anyon gas.
	
	Compare the TF functional \eqref{eq:TF-func} which shows that the effect 
	of the exclusion principle for fermions is a self-energy for the density 
	that penalizes localization
	and thus forces a spread of the profile into the more expensive regions 
	of the trapping potential.
	Indeed, the ideal Fermi gas satisfies the following kinetic energy inequality
	(known in mathematics as a \keyword{Lieb--Thirring inequality} \cite{LieThi-75})
	which captures this type of degeneracy pressure:
	for any number $N$ of particles and any 
	$N$-fermion state $\Psi \in L^2_\asym(\R^{2N})$,
	\begin{equation} \label{eq:LT-fermions}
		\langle \hT_\asym \rangle_\Psi 
		\gtrsim 
		\int_{\R^2} \varrho_\Psi(\bx)^2 \,d\bx
	\end{equation}
	(modulo a universal constant\footnote{%
	Deciding on the best constant 
	in this inequality 
	is a long-standing open problem, 
	but it is conjectured \cite{LieThi-76} to be 
	given by that of 
	its restriction to $N=1$
	(a Sobolev constant)
	and is strictly smaller than the TF constant $2\pi$.
	See \cite{SeiSol-23} for recent progress.}),
	where $\varrho_\Psi$ is the one-body density associated to $\Psi$,
	obtained by marginalizing 
	$N-1$ particles.
	Inequalities of this type have been 
	useful to prove qualitative properties of large fermionic systems such as 
	their thermodynamic stability w.r.t.\ Coulomb interaction with a mixture of charges
	\cite{DysLen-67,LieThi-75,LieSei-09}.
	In the case that $\nu$ species of fermions or spin states are considered,
	the constant in \eqref{eq:LT-fermions} is weakened by a factor $1/\nu$.
	
	The degeneracy pressure for anyons is measured by a similar inequality 
	\cite{LunSol-13a,LunSol-13b} (see \cite{LunQva-20} for review):
	for any number $N$ of particles and any $N$-anyon state $\Psi \in L^2_\alpha$,
	\begin{equation} \label{eq:LT-anyons}
		\langle \hT_\alpha \rangle_\Psi \gtrsim \alpha_2 \int_{\R^2} \varrho_\Psi(\bx)^2 \,d\bx
	\end{equation}
	(again modulo a universal constant).
	Its strength thus depends to leading order on the 
	2-fractionality,
	i.e.\ the nearest-neighbor exchange statistics $\alpha$.
	It then follows, like for fermions, that charged anyonic systems 
	(except bosons)
	with Coulomb interaction are also stable \cite{LunSol-14,LunSei-17}.
\end{enumerate}

Applied to the harmonic trap \eqref{eq:E-harmonic}, 
the degeneracy pressure \eqref{eq:LT-anyons} yields the bound
\begin{equation} \label{eq:harm-LT-bound}
	E_N(\alpha,\text{harm.osc.}) \gtrsim \sqrt{\alpha_2} \,\hbar\omega N^{3/2},
\end{equation}
which indeed matches the constant-field approximation \eqref{eq:avg-field-hosc},
and Fig.~\ref{fig:spectrum-N},
up to the value of the universal constant.

\begin{figure}
	\centering
	\includegraphics[scale=0.4]{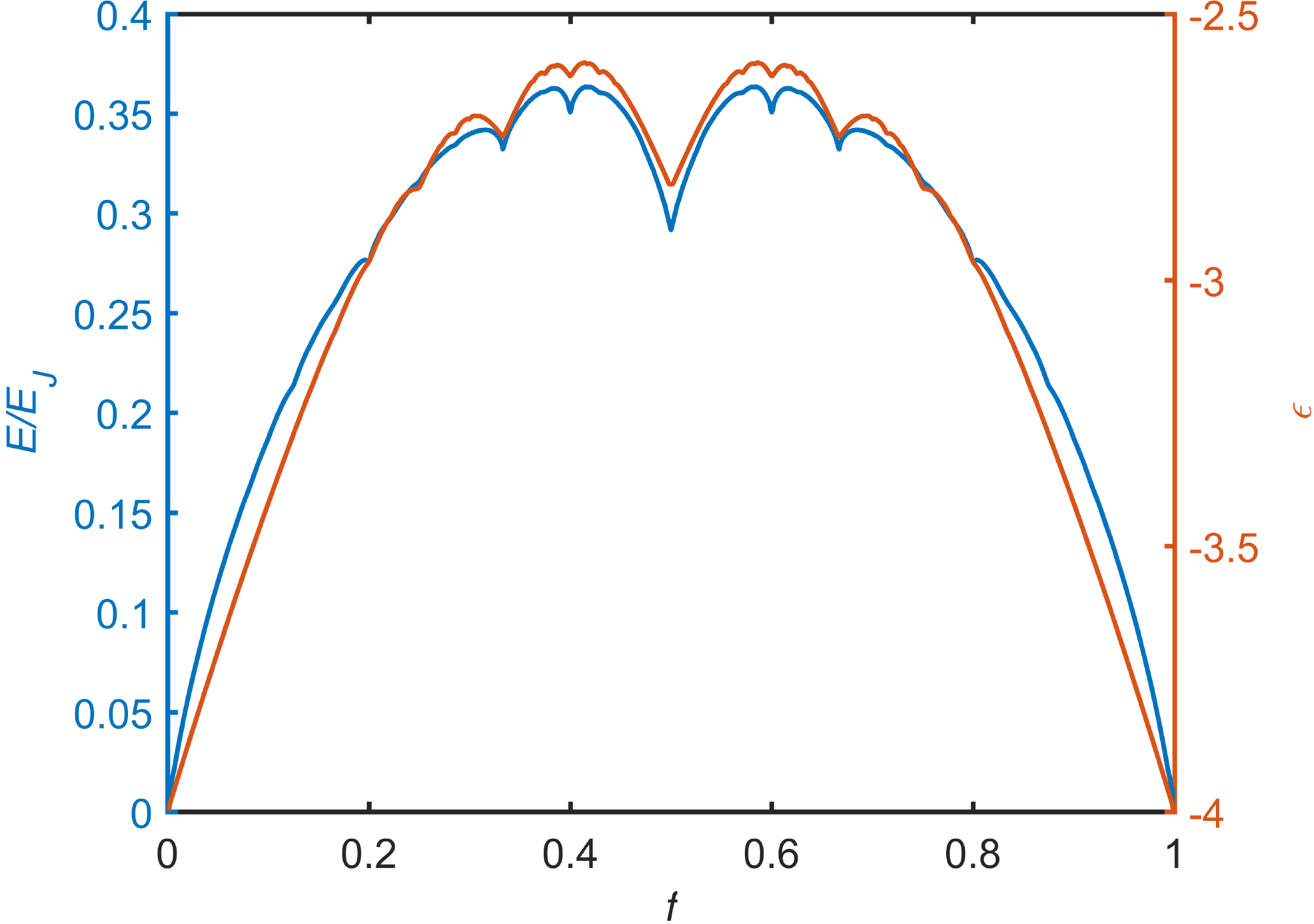}
	\caption{Universal energy phenomena in 2D, from \cite{Lankhorst-etal-18}: 
	lowest energy branch of a $100 \times 100$ square Josephson junction array (left/blue)
	resp.\ the Hofstadter butterfly (right/orange), 
	which also is proportional to the critical temperature of a superconducting
	square network. (For comparison, $\alpha \sim 2f$.)}
	\label{fig:universal}
\end{figure}

On a disk geometry $D$ with unit area, the analysis of Chitra and Sen 
again suggests a large number of level crossings as $N \to \infty$, 
and thus possibly nonanalyticities in the limiting g.s.e.
per particle and unit density of the 
\keyword{homogeneous ideal anyon gas}\footnote{%
Although it has not been rigorously shown, we may
expect the thermodynamic limit to be independent of the shape of the domain
(and otherwise take the infimum over reasonable shapes).}
\begin{equation} \label{eq:homo-energy}
	e(\alpha) := \liminf_{N \to \infty} E_N(\alpha,D)/N^2.
\end{equation}
Most details about this function are still open, but
from a further analysis of the homogeneous ideal anyon gas 
on the unit square $Q$ (dividing it into smaller squares),
one can at least derive the rigorous uniform bounds (cf.\ Fig.~\ref{fig:ideal})
\cite{LunSei-17}
\begin{equation} \label{eq:homo-bounds}
	\frac{1}{4} \max\{\tilde{E}_2(\alpha_N),F_2(\alpha_2)\} 
	\le e(\alpha) \le 2\pi^2,
\end{equation}
and
\begin{equation} \label{eq:homo-bounds-lin}
	\alpha_2 \lesssim e(\alpha) \lesssim \alpha_2,
\end{equation}
modulo lower and upper universal constants (certainly not optimal).
Subject to the assumption that $e(\alpha)$ indeed has a fractal structure 
which captures universal aspects of planar geometry,
a conjecture on its possible form can be given as in Fig.~\ref{fig:universal};
cf.\ \cite{Lankhorst-etal-18}.

\subsection{Some states of particular interest} \label{sec:trial-states}

\begin{figure}
	\centering
	\scalebox{0.96}{\begin{tikzpicture}[>=stealth']
		\def\x{0}
		\def\y{0}
		\def\c{red}
		\coordinate (p1) at (0+\x,0+\y);
		\coordinate (p2) at (0+\x,1+\y);
		\coordinate (p3) at (1+\x,1+\y);
		\coordinate (p4) at (1+\x,0+\y);
		\draw [fill,\c] (p1) circle [radius = 0.04];
		\draw [fill,\c] (p2) circle [radius = 0.04];
		\draw [fill,\c] (p3) circle [radius = 0.04];
		\draw [fill,\c] (p4) circle [radius = 0.04];
		\draw [\c] (p1) -- (p2) -- (p3) -- (p4) -- (p1);
		\draw [\c] (p1) -- (p3);
		\draw [\c] (p2) -- (p4);

		\def\x{2}
		\def\y{0}
		\def\c{green}
		\coordinate (p1) at (0+\x,0+\y);
		\coordinate (p2) at (0+\x,1+\y);
		\coordinate (p3) at (1+\x,1+\y);
		\coordinate (p4) at (1+\x,0+\y);
		\draw [fill,\c] (p1) circle [radius = 0.04];
		\draw [fill,\c] (p2) circle [radius = 0.04];
		\draw [fill,\c] (p3) circle [radius = 0.04];
		\draw [fill,\c] (p4) circle [radius = 0.04];
		\draw [\c] (p1) -- (p2) -- (p3) -- (p4) -- (p1);
		\draw [\c] (p1) -- (p3);
		\draw [\c] (p2) -- (p4);
		
		\def\x{1}
		\def\y{2}
		\def\c{blue}
		\coordinate (p1) at (0+\x,0+\y);
		\coordinate (p2) at (0+\x,1+\y);
		\coordinate (p3) at (1+\x,1+\y);
		\coordinate (p4) at (1+\x,0+\y);
		\draw [fill,\c] (p1) circle [radius = 0.04];
		\draw [fill,\c] (p2) circle [radius = 0.04];
		\draw [fill,\c] (p3) circle [radius = 0.04];
		\draw [fill,\c] (p4) circle [radius = 0.04];
		\draw [\c] (p1) -- (p2) -- (p3) -- (p4) -- (p1);
		\draw [\c] (p1) -- (p3);
		\draw [\c] (p2) -- (p4);
		
		\node [below right] at (0,0) {\scalebox{0.9}{$\cV_1$}};
		\node [below right] at (2,0) {\scalebox{0.9}{$\cV_2$}};
		\node [below right] at (1,2) {\scalebox{0.9}{$\cV_3$}};
		\node [below right] at (0.95,0.2) {\scalebox{0.9}{$j$}};
		
		\node [above right] at (3.6,1.2) {$\Rightarrow$};

		\def\x{5}
		\def\y{0}
		\draw [arrows=->,thick,] 	(3.2+\x,0.3+\y) -- (3.7+\x,0.95+\y);
		\draw [arrows=->,thick,] 	(3.2+\x,0.3+\y) -- (2.7+\x,-0.3+\y);
		\def\c{red}
		\coordinate (p1) at (0.3+\x,0.2+\y);
		\coordinate (p2) at (0.0+\x,3.2+\y);
		\coordinate (p3) at (2.6+\x,2.9+\y);
		\coordinate (p4) at (3.2+\x,0.3+\y);
		\draw [fill,\c] (p1) circle [radius = 0.04];
		\draw [fill,\c] (p2) circle [radius = 0.04];
		\draw [fill,\c] (p3) circle [radius = 0.04];
		\draw [fill,\c] (p4) circle [radius = 0.04];
		\draw [\c] (p1) -- (p2) -- (p3) -- (p4) -- (p1);
		\draw [\c] (p1) -- (p3);
		\draw [\c] (p2) -- (p4);
		\def\c{green}
		\coordinate (p1) at (0.1+\x,0.1+\y);
		\coordinate (p2) at (0.3+\x,2.8+\y);
		\coordinate (p3) at (2.4+\x,3.2+\y);
		\coordinate (p4) at (2.8+\x,0.1+\y);
		\draw [fill,\c] (p1) circle [radius = 0.04];
		\draw [fill,\c] (p2) circle [radius = 0.04];
		\draw [fill,\c] (p3) circle [radius = 0.04];
		\draw [fill,\c] (p4) circle [radius = 0.04];
		\draw [\c] (p1) -- (p2) -- (p3) -- (p4) -- (p1);
		\draw [\c] (p1) -- (p3);
		\draw [\c] (p2) -- (p4);
		\def\c{blue}
		\coordinate (p1) at (0.2+\x,0.5+\y);
		\coordinate (p2) at (0.1+\x,3.0+\y);
		\coordinate (p3) at (3.2+\x,3.1+\y);
		\coordinate (p4) at (2.6+\x,0.4+\y);
		\draw [fill,\c] (p1) circle [radius = 0.04];
		\draw [fill,\c] (p2) circle [radius = 0.04];
		\draw [fill,\c] (p3) circle [radius = 0.04];
		\draw [fill,\c] (p4) circle [radius = 0.04];
		\draw [\c] (p1) -- (p2) -- (p3) -- (p4) -- (p1);
		\draw [\c] (p1) -- (p3);
		\draw [\c] (p2) -- (p4);

		\draw [dashed] (0.2+\x,0.35+\y) circle [radius = 0.50];
		\draw [dashed] (0.1+\x,3.0+\y) circle [radius = 0.50];
		\draw [dashed] (2.8+\x,3.15+\y) circle [radius = 0.50];
		\draw [dashed] (2.85+\x,0.3+\y) circle [radius = 0.50];

		\node [below right] at (3.3+\x,0.5+\y) {\scalebox{0.9}{$\bx_j$}};
		\node [below right] at (3.3+\x,1.45+\y) {\scalebox{0.9}{$\alpha\bA_j$}};
		\node [below right] at (2.2+\x,-0.1+\y) {\scalebox{0.9}{$\bJ_j$}};
		\node [below left] at (-0.2+\x,2.8+\y) {\scalebox{0.9}{$\cV_1^*$}};
	\end{tikzpicture}}
	\caption{From \cite{Lundholm-16}. 
		A coloring of $N=12$ particles with
		$\nu=3$ colors into $K=4$ 3-clusters $\cV_q^*$. 
		Each colored edge $(j,k) \in \cE_q$ corresponds
		to one unit $-\mu$ of pairwise angular momentum.
		Also shown is the contribution to the magnetic potential
		$\alpha\bA_j$ and the current $\bJ_j$ of particle $j$ due solely 
		to the 3-cluster $\cV_1^*$.}
	\label{fig:clustering}
\end{figure}
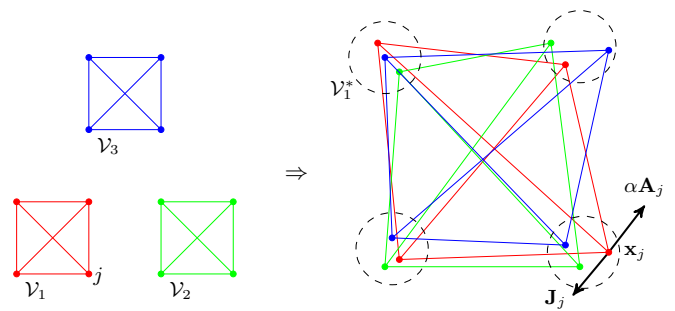

We saw above that there is a nontrivial and 
delicate balance between exclusion and uncertainty
in the ground state energy.
Certain types of states have been proposed \cite{LunSol-13b,Lundholm-16}
to minimize the kinetic energy 
$\hT_{\sym\to\alpha}$ at particular values of the statistics parameter: 
in the case of $\alpha = \mu/\nu \in [0,1]$ being 
an \emph{even}-numerator reduced fraction, 
and for $N = \nu K$, $K=1,2,\ldots$, a suitable sequence of particle numbers, let
\begin{align} \label{eq:trial-even}
	\Psi_\alpha &:= \prod_{j<k} |z_{jk}|^{-\alpha} 
		\,\cS\left[ \prod_{q=1}^\nu 
		\prod_{(j,k) \in \cE_q} (\bar{z}_{jk})^\mu 
		\right] \prod_{l=1}^N \psi_0(\bx_l), 
\end{align}
while for \emph{odd} numerators $\mu$, let
\begin{equation} \label{eq:trial-odd}
	\Psi_\alpha := \prod_{j<k} |z_{jk}|^{-\alpha} 
		\,\cS\left[ \prod_{q=1}^\nu 
		\prod_{(j,k) \in \cE_q} (\bar{z}_{jk})^\mu 
		\bigwedge_{k=0}^{K-1} \psi_k \,(\bx_{l \in \cV_q})
		\right].
\end{equation}
Here $z_{jk} := z_j - z_k$ are the pairwise relative complex coordinates
and we have grouped, or `colored', the particles into $\nu$ different colors 
where $G_q = (\cV_q,\cE_q)$ denotes the complete graph 
over each such group of $|\cV_q| = K$ vertices=particles 
(see Fig.~\ref{fig:clustering}).
The symmetrization $\cS$ over all the particles then amounts to 
symmetrization over all such colorings,
and can be viewed as passing from a set of distinguishable particles (by color)
to indistinguishable (cf.\ \cite{RegGoeJol-08}).
The $\psi_k$, $k=0,1,2,\ldots$, are the 
eigenstates (ordered by increasing energy) 
of the corresponding one-body Hamiltonian $\hH_1$.
The states $\Psi_\alpha$ are moderately singular 
and so we should actually consider
$\Psi = \Phi \Psi_\alpha \in L^2_{\textup{sym}}(\R^{2N})$
with $\Phi$ a suitably chosen regularization factor to allow
inclusion into the domain of $\hT_{\sym\to\alpha}$.

Some of the reasons why these states appear to be of interest 
in the many-anyon problem include:
\begin{itemize}[wide, labelwidth=!, labelindent=0pt]
\item They have the right average angular momentum \eqref{eq:opt-ang-mom}
	and are (for harmonic trap and certain $K$) 
	homogeneous with a degree suitable to 
	optimize the bound
	\eqref{eq:CS-bound}.
\item They enjoy certain clustering properties that are beneficial to minimize
	the pair repulsion $V_{\rm stat}(r)$ on large scales $r$.
	States of the form \eqref{eq:trial-even} for $\mu=2$, $\nu=3,5,\ldots$
	are known in the FQHE 
	context
	as Read--Rezayi states and are known to 
	form clusters of $\nu$ particles bound to $\mu$ vortices
	\cite{ReaRez-99,CapGeoTod-01},
	while \eqref{eq:trial-odd} for $\mu=1$, $\nu=2$ correspond to 
	Dyson--Moore--Read/BCS/Pfaffian states subject to pairing \cite{Dyson-67,MooRea-91}.
	Further, excitations on top of these states appear to have nonabelian
	character \cite{Stern-08,Nayak-etal-08,BonGurNay-11}.
\item They manifest certain structural stabilities with the smallness of
	$1 \le \mu \le \nu$, 
	similar to the Laughlin states of the FQHE 
	and indicative of a scale of robustness.
\item Some of the states \eqref{eq:trial-even} are exact but 
	moderately singular/generalized
	low-energy eigenfunctions of the Hamiltonian with singular boundary conditions
	(cf.\ point interactions below, Sec.~\ref{sec:point-int}) 
	\cite{Chou-91a,MurLawBhaDat-92}.
\end{itemize}

\section{The nonideal anyon gas} \label{sec:nonideal}

The ``average(/constant)-field'' approach, suggesting \eqref{eq:cf-func},
leads to difficulties when
applied to strictly ideal and pointlike particles, 
and it is both realistic and helpful to
instead consider extended particles as well as other nonideal models of
anyon gases which can more easily
incorporate the average effects of exchange phases or flux.
Heuristic
approaches to ensure averaged influence of the statistics include a 
``self-consistency argument'' 
wherein a typical cyclotron orbit of the generated field ought to contain
a large number of particles, 
suggesting a better approximation if \cite{CheWilWitHal-89,Trugenberger-92b}
\begin{equation} \label{eq:self-consistency}
	\alpha_2 \ll 1 \qquad \text{or} \qquad
	(1-\alpha_2)^2 \ll 1
\end{equation}
(i.e.\ slightly better around fermions due to Pauli repulsion),
as well as a similar orbital argument but based instead on the scattering angles 
of combined Aharonov--Bohm and hard-disk interaction \cite{CaeMac-94},
again suggesting the conditions \eqref{eq:self-consistency}.

\subsection{The extended anyon gas} \label{sec:extended}

In the \keyword{extended anyon gas}, we work in the magnetic perspective 
and replace the point magnetic fluxes of ideal anyons by extended fluxes.
For simplicity 
(and not completely ad hoc from the perspective of emergent models) 
we may consider the flux to be constantly spread over a disk-shaped
magnetic field attached to each particle, with radius $R>0$. 
We then replace the anyon magnetic potentials \eqref{eq:A-ideal} 
by
\begin{equation} \label{eq:A-extended}
	\bA_j^R(\sx) := \sum_{k \neq j} \frac{(\bx_j-\bx_k)^\perp}{|\bx_j-\bx_k|_R^2},
	\qquad |\bx|_R := \max\{|\bx|,R\},
\end{equation}
so that, 
if $\1_{D(\by,R)}$ 
denotes the indicator function on a disk of radius $R$ centered at $\by$,
the field is
$$
	\curl_{\bx_j} \alpha\bA_j^R 
	= 2\pi\alpha \sum_{k \neq j} \frac{\1_{D(\bx_k,R)}(\bx_j)}{\pi R^2}
	\ \xrightarrow{R \to 0} \ 
	2\pi\alpha \sum_{k \neq j} \delta_{\bx_k} (\bx_j).
$$
The kinetic energy of the extended anyon gas with bosons as reference is then
$$
	\hT_{\sym\to\alpha}^R := \frac{\hbar^2}{2m} \sum_{j=1}^N \left( -i\nabla_{\bx_j} + \alpha \bA_j^R \right)^2,
$$
acting on $L^2_\sym$,
while if we use fermions as reference then
$$
	\hT_{\asym\to\alpha}^R := \frac{\hbar^2}{2m} \sum_{j=1}^N \left( -i\nabla_{\bx_j} + (\alpha-1) \bA_j^R \right)^2,
$$
acting on $L^2_\asym$.

There is now an additional natural dimensionless parameter in the problem
that can be defined as the ratio of the size of
the magnetic flux disk to the average interparticle distance, 
\begin{equation} \label{eq:filling-ratio}
	\bar\gamma := R\bar\varrho^{1/2}.
\end{equation}
This density parameter has been
called the \keyword{magnetic filling ratio} in 
\cite{Trugenberger-92,Trugenberger-92b,LarLun-16}.
The limit $\bar\gamma=0$ corresponds to ideal anyons,
while $\bar\gamma \sim 1$ is the ``smearing'' limit when particles typically 
begin to overlap, 
and $\bar\gamma \to \infty$ would describe the constant-field limit.

At any fixed choice of the two parameters $\alpha \in \R$, $\bar\gamma \ge 0$,
the energy per particle and unit of density of the 
\keyword{homogeneous extended anyon gas}
may thusly be defined via the thermodynamic limit
\begin{equation} \label{eq:ext-gas-bound}
	e_{\substack{\sym/\\\asym}}(\alpha,\bar\gamma) :=
	\liminf_{\substack{N, L \,\to\, \infty \\ N/L^2 = \bar\varrho}} 
	\frac{E_N(\alpha,R,LQ)}{\bar\varrho N},
\end{equation}
where $E_N(\alpha,R,LQ)$ denotes the g.s.\ energy of the respective
operator $\hT_{\sym/\asym\to\alpha}^R$ on $L^2_{\sym/\asym}(LQ^N)$.
A corresponding TF-type functional for a \keyword{local-density approximation} 
within a trapping potential $V$ is then
\begin{equation} \label{eq:ext-func} 
	\cE^{\alpha,R}_{\substack{\sym/\\\asym}}[\varrho] := \int_{\R^2} \left[ 
		e_{\substack{\sym/\\\asym}}\bigl( \alpha, R\varrho(\bx)^{1/2} \bigr) \varrho(\bx)^2
		+ V(\bx)\varrho(\bx) \right] d\bx,
\end{equation}
with $\int_{\R^2} \varrho = N$.

The qualitative analysis of local exclusion effects 
of the ideal gas discussed in Section~\ref{sec:local-exclusion} 
has been extended to the extended gas, both in the bosonic 
and the fermionic references.
Up to constants, one finds the qualitative bounds 
\cite{LarLun-16}
$$
	e_\sym(\alpha, \bar\gamma) \gtrsim \begin{cases} 
		|{\ln \bar\gamma}|^{-1} + \alpha_\infty, &\bar\gamma \ll 1 \ \text{($\alpha$ fixed)}, \\
		|\alpha|, &\bar\gamma \gtrsim 1,
		\end{cases}
$$
where the first regime is a combination of a dilute 2D Bose gas \cite{Schick-71,LieYng-01}
and the strongest of the two lower bounds \eqref{eq:homo-bounds} 
for the ideal gas at odd numerators,
$\tilde{E}_2(\alpha_\infty) \gtrsim \alpha_\infty$,
while the second regime indeed captures the constant-field energy \eqref{eq:cf-func}
for arbitrarily large $\alpha$.
The full regime $\bar\gamma \lesssim 1$ 
is more difficult to describe,
since it covers both soft-core and hard-core behavior
(see Sec.~\ref{sec:other-int} below),
and we should expect to replace $\alpha_\infty$ with $\alpha_2$ in a suitable limit.
Trugenberger \cite{Trugenberger-92b} suggested that this 
gas with fixed 
extension parameter $R>0$ will at low temperatures prefer to adjust its density to 
$\bar\gamma \gtrsim 1$,
where the average-field approximation is good and the gas becomes a superfluid,
while for high temperatures the density drops, $\bar\gamma \ll 1$,
the average-field loses validity and superfluidity is lost.

In the fermionic reference \cite{GirRou-22}
$$
	e_\asym(\alpha,\bar\gamma) \gtrsim \begin{cases}
		\alpha_2, 
			&\quad \bar\gamma \lesssim 1,\\
		1, 
			&\quad \bar\gamma \gtrsim 1.
	\end{cases}
$$
In this case the inherent Pauli principle improves the dependence
to nearest neighbor 
due to its scale independence.
Note that both energies $e_{\sym/\asym}$ 
must be periodic in $\alpha$ at $\bar\gamma=0$, 
but not at any finite $\bar\gamma>0$, 
which therefore requires an interesting interpolation between the regimes.

\subsection{Almost-bosonic anyons} \label{sec:almost-bosonic}

The average-field approach \eqref{eq:cf-func}
can be made fully rigorous in the limit $N \to \infty$ close to bosons:
$$
	\alpha = \beta/N \to 0, \qquad \text{at fixed}\ \beta \in \R,
$$
in which the interaction may indeed be treated in a mean-field sense
if anyons are also extended/smeared out to a finite size $R>0$, as above.
It is even possible to
simultaneously take $R \sim N^{-\eta} \to 0$ at some sufficiently slow rate $\eta$
(an ``almost-ideal'' limit).
The idea is then as a final step to take $\beta=\alpha N$ (total flux) large 
(then ``less bosonic'')
and study the appropriate ground-state energy functionals in this limit.
Because of this particular order of limits, it is not completely clear 
whether this indeed describes the ideal anyon gas as defined above, 
or perhaps some nonideal model (see Sec.~\ref{sec:other-int}). 
We shall simply refer to it as \keyword{``almost-bosonic'' anyons}.

In any case, the first steps above yield the seemingly
\emph{correct} {\bf average-field functional} w.r.t.\ bosons 
for any fixed $\beta \in \R$ \cite{LunRou-15}:
\begin{equation} \label{eq:af-func}
	\cEAF_\beta[\psi] = \int_{\R^2} \left[
		\left| (-i\nabla + \beta\bA[|\psi|^2]) \psi\right|^2
		+ V |\psi|^2 \right]. 
\end{equation}
The magnetic potential $\bA[\varrho=|\psi|^2]$ 
is chosen such that it generates exactly the field $2\pi\varrho$:
\begin{equation} \label{eq:self-field}
	\curl \beta\bA[\varrho](\bx) = 2\pi\beta\varrho(\bx),
	\qquad \int_{\R^2} \varrho = 1.
\end{equation}
Thus, the \emph{scalar} self-interaction of the GP model \eqref{eq:GP-func} 
is here replaced with a \emph{magnetic} self-interaction.
This produces a peculiar new balancing problem, namely in order to
lower its energy in the self-generated magnetic field, 
as the coupling $\beta$ (total number of flux units) grows
there needs to be an increasing phase and angular momentum in 
$\psi = |\psi|e^{i\phi}$
to cancel this magnetic field:\footnote{%
Due to the reasons described below,
the cancellation is not complete, but both terms in \eqref{eq:mod-phase-decomp}
contribute an energy of order $\beta$.}
\begin{equation} \label{eq:mod-phase-decomp}
	\bigl|(-i\nabla + \beta\bA)\psi\bigr|^2 
	= \bigl|\nabla|\psi|\bigr|^2 + \bigl|\nabla\phi + \beta\bA\bigr|^2|\psi|^2.
\end{equation}
Further, in order to preserve regularity, any nontrivial such phase circulation 
must appear in the form of quantized vortices. 
On a typical disk $D(\bx,R)$, the phase circulation
\begin{equation} \label{eq:phase-circ}
	\Phi(\bx,R) := \oint_{\partial D(\bx,R)} \beta \bA \cdot d\br
	\approx -\int_{D(\bx,R)} \curl \nabla \phi \,d\by
\end{equation}
then needs to minimize the residual angular momentum,
\begin{equation} \label{eq:phase-circ-bound}
	\int_0^{2\pi} \bigl|\partial_\vphi \phi + \beta A_\vphi\bigr|^2 \,d\vphi
	\ge \min_{q \in \Z} \bigl|q-\Phi(\bx,R)\bigr|^2.
\end{equation}
However, the amount of flux to be cancelled
by each vortex also depends on the typical value of the density $\varrho$ there,
which therefore sets a varying length scale for vortex formation \cite{CorLunRou-16}.
It is thus expected (see also \cite{CheWilWitHal-89} for an early conjecture)
and indeed confirmed in numerical studies \cite{CorDubLunRou-19,Girardot-21}
that there is the formation of an Abrikosov-like vortex lattice.
However, unlike in the Abrikosov setting with homogeneous rotation/field,
the density of vortices here follows the average density profile in the trap $V$,
which for $\beta \gg 1$ turns out to be given again by the minimizer of an
effective {\bf Thomas--Fermi-type functional} \cite{CorLunRou-16}: 
\begin{equation} \label{eq:aTF-func}
	\cEaTF_\beta[\varrho]
	:= \int_{\R^2} \Bigl[ \CaTF |\beta| \varrho(\bx)^2 + V(\bx)\varrho(\bx) \Bigr] \,d\bx.
\end{equation}

\begin{figure}
	\centering
	\includegraphics[scale=0.8]{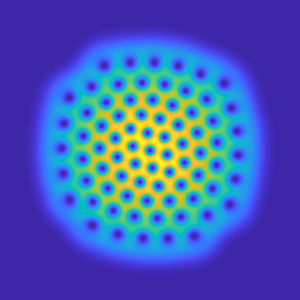}
	\includegraphics[scale=0.8]{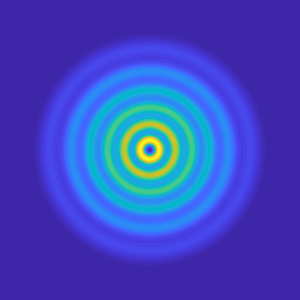}\\
	\includegraphics[scale=0.4]{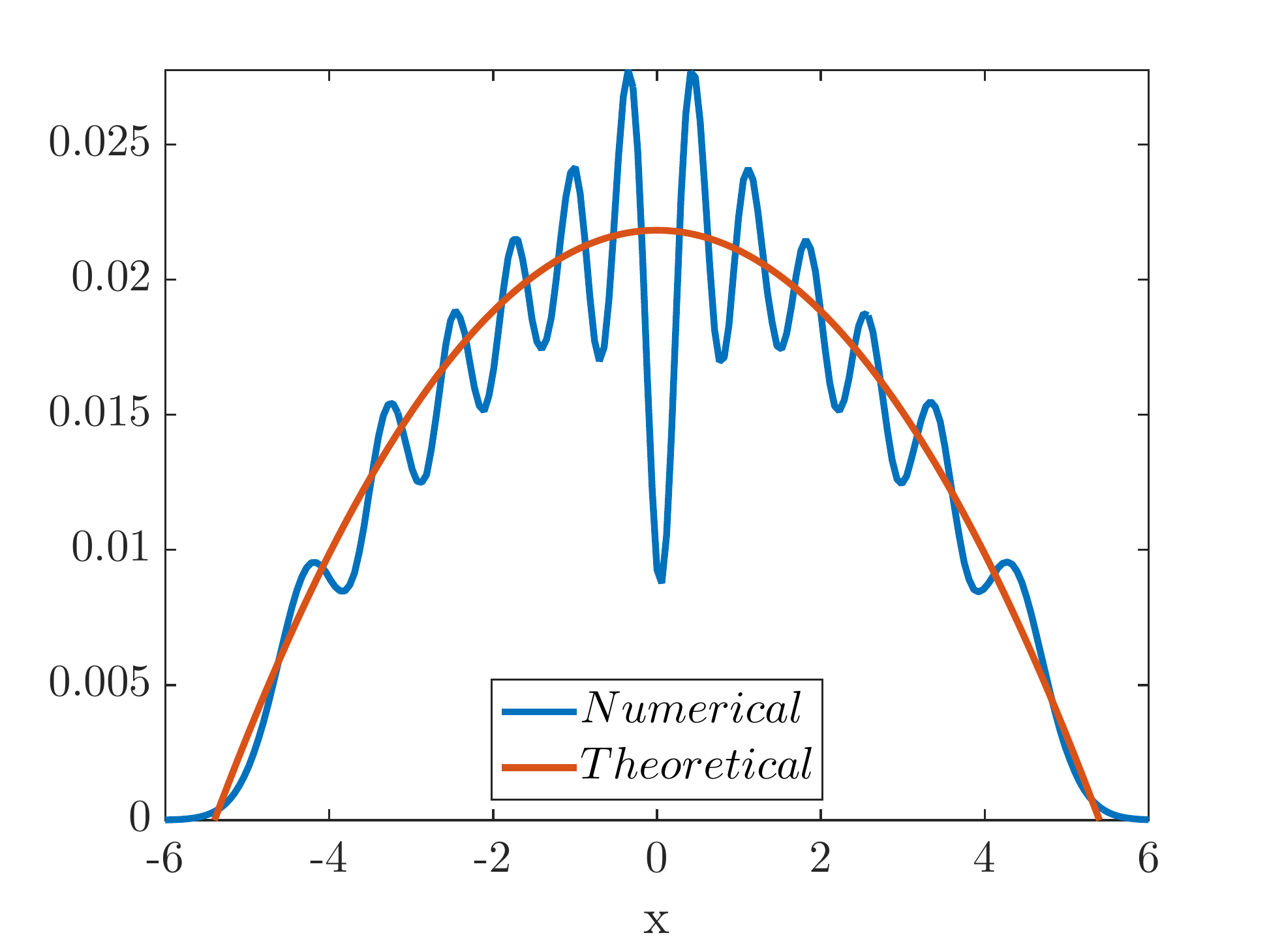}
	\caption{Numerical simulation of the ground-state density 
	$|\psiAF_\beta|^2$ in the average-field functional
	\eqref{eq:af-func} for $\beta=90$ in a harmonic trap,
	averaged over rotations and then compared with the exact minimizer 
	$\rhoATF_\beta$ 
	of \eqref{eq:aTF-func}. From \cite{CorDubLunRou-19}.}
	\label{fig:af-bosons}
\end{figure}

Indeed, numerical simulations \cite{CorDubLunRou-19,Girardot-21} 
(see Fig.~\ref{fig:af-bosons})
confirm the theoretical prediction of 
\cite{CorLunRou-16,CorLunRou-proc-17}
that the g.s.\ $\psiAF_\beta$ of \eqref{eq:af-func} 
exhibits an approximately triangular vortex lattice distribution 
with scales set by 
the TF profile (minimizer) $\rhoATF_\beta$ of \eqref{eq:aTF-func}, 
and it is further estimated numerically that \cite{CorDubLunRou-19}
\begin{equation} \label{eq:CaTF-estimate}
	\CaTF \approx 4\pi^{3/2}/3 
	\approx 1.18 \times 2\pi.
\end{equation}
The fact that this constant is strictly bigger than the fermionic TF constant 
and constant-field prediction\footnote{%
Also in comparison to Fig.~\ref{fig:spectrum-N}, Chitra and Sen,
who estimated a substantially larger factor \cite{ChiSen-92}, 
assumed rotational symmetry for the density to simplify their
functional, plus added a hard-core delta 
regularization that itself amounts to the constant-field energy 
$2\pi|\beta|\varrho^2$.}
$2\pi$ can be understood from the microscopic lattice inhomogeneity of
$|\psiAF_\beta|^2$ as compared to its average density $\rhoATF_\beta$,
which effectively raises the energy by a factor
which is at least as big as the Abrikosov constant 
($\approx 1.1596$; cf.\ \cite{AftBlaNie-06a,AftBlaNie-06b}).

\subsection{Almost-fermionic anyons} \label{sec:almost-fermionic}

A similar regularization approach can be taken close to fermions: 
$$
	\alpha = 1 - \beta/\sqrt{N} \to 1, \qquad \beta, \hbar\ \text{fixed},
$$
for which the leading terms of the kinetic energy can be compared with the Fermi gas 
$E_N(\alpha=1) \sim 2\pi N^2$, $N \to \infty$.
By rescaling the parameters, 
it is equivalent to study a semiclassical mean-field limit
$$
	\alpha = 1 - \beta/N \to 1, \qquad \hbar \sim N^{-1/2} \to 0.
$$
Also here we can allow ``almost-ideal'' or ``virtually'' extended anyons, 
$0 < R \sim N^{-\eta} \to 0$, if the rate $\eta$ is small enough\footnote{%
The rate is better than for bosons because of the Pauli principle; 
cf. \eqref{eq:self-consistency}.}.
	
This \keyword{almost-fermionic} approach 
takes us back to the actual TF 
functional for fermions 
\eqref{eq:TF-func},
which remains correct for any finite $\beta$ in this limit 
\cite{GirRou-21,Girardot-21}.
More precisely, one obtains a semiclassical {\bf Vlasov functional}:
$$
	\cE^{\rm Vla}_\beta[\mu] := \frac{1}{(2\pi)^2} \int_{\R^4} \bigl| \bp + \beta\bA[\varrho] \bigr|^2 \mu(\bx,\bp) \,d\bx d\bp
	+ \int_{\R^2} V\varrho \,d\bx
$$
for $0 \le \mu(\bx,\bp) \le 1$ a measure on phase space $\R^4$ 
which is subject to the Pauli principle and
normalized $\int_{\R^4}\mu = (2\pi)^2$.
The minimizing measure $\mu$ of this functional is the corresponding filled Fermi sea:
\begin{equation} \label{eq:mu-Vla}
	\mu(\bx,\bp) = \1\left\{ |\bp + \beta\bA[\varrho](\bx)|^2 \le 4\pi\varrho(\bx) \right\},
\end{equation}
with \keyword{spatial density} $\varrho=\rhoTF$ \emph{independent} of $\beta$:
$$
	\varrho(\bx) = \frac{1}{(2\pi)^2} \int_{\R^2} \mu(\bx,\bp) \,d\bp 
	= (4\pi)^{-1}(\lambda^{\rm TF} - V(\bx))_+.
$$
The \keyword{momentum density} $t=t^{\rm TF}_\beta$,
$$
	t(\bp) = \frac{1}{(2\pi)^2} \int_{\R^2} \mu(\bx,\bp) \,d\bx,
$$
\emph{does} depend on $\beta$ however, in a nonlocal way.
The energy $\cE^{\rm Vla}_\beta[\mu]=\cETF[\rhoTF]$ 
is locally a constant at $\alpha \approx 1$ in this approximation, thus
confirming this aspect of Fig.~\ref{fig:spectrum-N}.

The almost-fermionic regime was also studied numerically 
by Girardot et al.\ in \cite{Girardot-21}
using a \keyword{Hartree--Fock} approach, i.e.\ on Slater determinants, 
and with a self-generated field $B = 2\pi\beta\varrho/N$
(total flux $2\pi\beta$),
showing good agreements to the above.
In fact, it can be interpreted as a multicomponent version of the bosonic
functional \eqref{eq:af-func} on orthogonal states $\psi_k$, $k=1,2,\ldots,N$.
These are then able to complement each other's densities
so as to fill in any holes left by vortices,
and thus eventually smoothen out the overall density profile:
\begin{equation} \label{eq:af-density-approx}
	\varrho(\bx) = \sum_{k=1}^{N} |\psi_k(\bx)|^2 \approx \rhoTF(\bx),
\end{equation}
thereby validating the ``constant-field'' (local density) 
approximation in this regime when $\beta$ is not too large.

\subsection{Magnetic TF theory} \label{sec:MTF}

For higher ratios of the magnetic field to the density one may expect \cite{Girardot-21}
a \keyword{magnetic Thomas-Fermi theory} \cite{McEuen-etal-92,LieSolYng-95} 
to become valid.
Namely, for a field $B(\bx)$ which is approximately constant 
on the scale of the trapping potential $V$, 
we can think of distributing the density
$$
	\varrho(\bx) = \sum_{n=0}^\infty \varrho_n(\bx)
$$
locally into the \keyword{Landau levels} (LLs) $\cH_n$ of the corresponding 
one-body \keyword{Landau Hamiltonian}\footnote{%
In our units the cyclotron frequency is $\omega_c = 2|B|$.}
\begin{equation} \label{eq:H-Landau}
	\hH_1^{\rm Lan} = (-i\nabla_\bx + B\bx^\perp/2)^2 
	= \sum_{n=0}^\infty |B|(2n+1) \1_{\cH_n}.
\end{equation}
The degeneracy (per unit area) of each level is 
the number of flux units (per unit area),
$d_B := |B|/(2\pi)$.
Thus, defining the corresponding expected local magnetic energy 
at a given density $\varrho$,
$$
	j_B(\varrho) := \sum_{n=0}^\infty |B|(2n+1) \varrho_n,
	\qquad 0 \le \varrho_n \le d_B,
$$
the \keyword{magnetic TF functional} 
for $N$ fermions in a fixed \emph{external} field $B > 0$ is then
\begin{equation} \label{eq:MTF-func}
	\cE^{\rm mTF}[\varrho] := \int_{\R^2} \left[ j_B(\varrho(\bx)) + V(\bx)\varrho(\bx) \right] d\bx,
	\quad \int_{\R^2} \varrho = N.
\end{equation}
At fixed $B$ and $\varrho$, in order to minimize $j_B(\varrho)$
we fill all $N_B := \lfloor \varrho/d_B \rfloor$ 
lower Landau levels maximally:
$$
	\varrho_n = \begin{cases}
	d_B, &0 \le n \le N_B-1,\\
	\varrho - N_Bd_B, &n = N_B,\\
	0, &n \ge N_B + 1,
	\end{cases}
$$
so that in the limit
$B \to 0$, implying $d_B \to 0$, 
\begin{multline*}
	j_B(\varrho) \approx \sum_{n=0}^{N_B-1} |B|(2n+1)d_B = |B| N_B^2 d_B \\
	\approx |B|\varrho^2/d_B = 2\pi \varrho^2,
\end{multline*}
returning us to the usual TF for the free Fermi gas.
On the other hand, if $B$ is large enough that $d_B \ge \varrho$,
then all particles will fit into the lowest Landau level (LLL):
\begin{equation} \label{eq:MTF-LLL}
	j_B(\varrho) = |B|\varrho.
\end{equation}

Now, considering a sufficiently extended anyon gas and
formally taking the anyonic (average) magnetic field 
$B(\bx) = 2\pi\beta\varrho(\bx)$, 
where $\beta=1-\alpha \in [0,1]$ 
is the number of flux units per particle attached to fermions,
we then obtain a \keyword{magnetic TF self-interaction functional}
for anyons:
\begin{equation} \label{eq:aMTF-func}
	\cE^{\rm amTF}_\beta[\varrho] := \int_{\R^2} \left[ 
		j_{B=2\pi\beta\varrho(\bx)}(\varrho(\bx)) 
		+ V(\bx)\varrho(\bx) \right] d\bx.
\end{equation}
At any point $\bx \in \R^2$ where $\varrho(\bx) \neq 0$ 
let us consider there the fractions of density 
$f_n := \varrho_n(\bx)/\varrho(\bx)$ in the LLs:
$$
	\frac{j_B(\varrho)}{2\pi\varrho^2} 
	= \sum_{n=0}^\infty \beta(2n+1) f_n,
	\qquad 0 \le f_n \le \beta \le 1,
$$
which again is minimized by filling the lowest
$N_B = \bloor$ levels:
$$
	f_n = \begin{cases}
	\beta, &0 \le n \le \bloor - 1,\\
	1 - \beta\bloor, &n = \bloor,\\
	0, &n \ge \bloor + 1.
	\end{cases}
$$
Hence, our minimum is
$$ 
	\frac{j_B(\varrho)}{2\pi\varrho^2}
	= \sum_{n=0}^{\bloor-1} (2n+1) \beta^2 + \beta(2\bloor + 1)(1-\beta\bloor), 
$$ 
which simplifies to
$$
	\frac{j_B(\varrho)}{2\pi\varrho^2} - 1
	= \beta^2(1 - \{\beta^{-1}\})\{\beta^{-1}\} =: M(\beta) \in [0,\beta^2/4),
$$
where $\{\beta^{-1}\} = \beta^{-1} - \bloor \in [0,1)$ 
is the fractional part of $\beta^{-1}$.
Therefore our proposed functional for the g.s.\ 
(a pointwise lower bound to \eqref{eq:aMTF-func})
takes the form
\begin{equation} \label{eq:red-aMTF-func}
	\tilde\cE^{\rm amTF}_\beta[\varrho] := \int_{\R^2} \left[ 
		2\pi\varrho(\bx)^2 \left(1+M(\beta)\right) 
		+ V(\bx)\varrho(\bx) \right] d\bx,
\end{equation}
thus refining 
both the ``constant-field'' approximation \eqref{eq:cf-func} around fermions
as well as the above almost-fermionic approximation.

\begin{figure}
	\centering
	\includegraphics[scale=0.82]{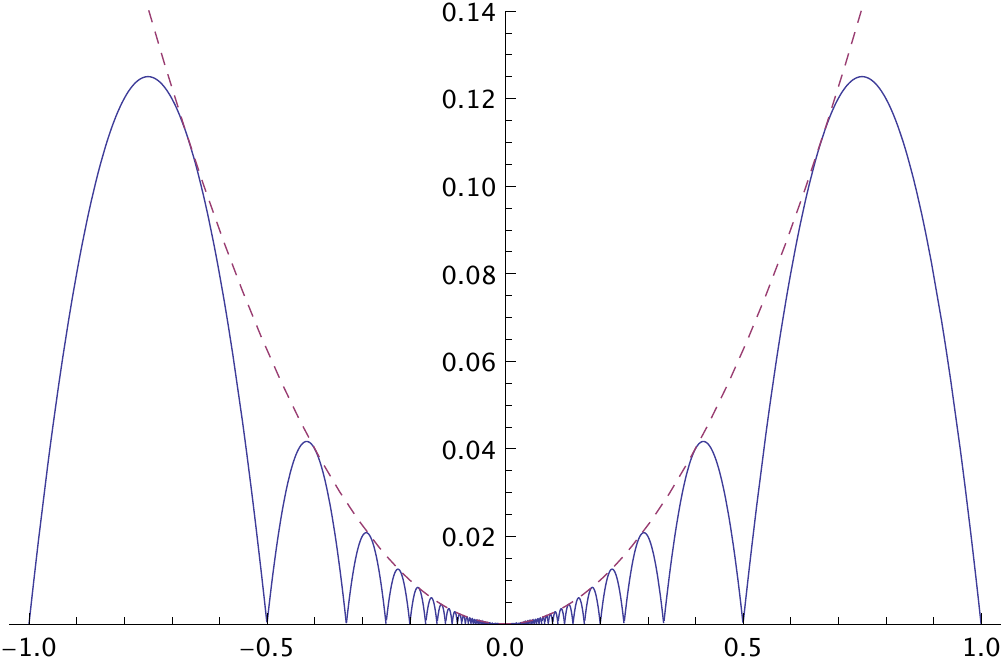}
	\caption{The factor 
	$M(\beta) = \beta^2(1 - \{\beta^{-1}\})\{\beta^{-1}\} \le \beta^2/4$ 
	(r.h.s.\ dashed) 
	in the magnetic TF functional \eqref{eq:red-aMTF-func}
	as a function of the relative 
	statistics parameter 
	$\beta = 1-\alpha$ around fermions $\beta=0$.}
	\label{fig:MTF}
\end{figure}

The corresponding energy landscape (see Fig.~\ref{fig:MTF}) 
is now indicative of \keyword{stability} at the Fermi point $\beta=0$.
A possible interpretation for the gas is that the discretization of the Fermi sea
into Landau levels imposes a more rigid structure in phase space,
i.e., as the attached flux $\beta$ per particle increases,
particles come from plane-wave states which homogenize easily in space
and then need to start organizing into the respective cyclotron orbits
with decreasing radii as the field increases.
Thus the Fermi sea becomes less fluid and more crystalline,
with the most synergetic 
situation 
when a whole number of levels are approximately filled 
($\{\beta^{-1}\} \approx 0$ or $1$)
so adjustments to the density can be accommodated at all levels in parallel
while keeping the fractions $f_n$ constant, 
and least synergetic when half a level is filled 
($\{\beta^{-1}\} \approx 1/2$)
and a larger fraction $\beta\{\beta^{-1}\}$ of 
particles 
need to organize their motion into that specific LL.
However, we cannot expect this picture to be valid all the way to bosons at
$\beta=1$ (compare also \eqref{eq:self-consistency})
because there will also be correlations across Landau levels,
leading to effective attractions, clustering and condensation,
and for example, if the LLL is filled and there are only few particles 
in the first-excited LL then one might consider the statistics transmutation 
situation discussed below where the excited particles would 
effectively rather be treated as bosons than fermions.

An analysis at these seemingly beneficial fractions $\alpha = 1 - 1/n$
was conducted by Chen et al. \cite{CheWilWitHal-89} in the homogeneous setting
(building on \cite{FetHanLau-89}),
where they indeed observed characteristics of \keyword{superfluidity}.
For example, the energy there increases linearly when a small external magnetic 
field is applied in either direction, which is interpreted as a Meissner effect.
Recently another approach to a DFT for anyons, 
via Kohn--Sham theory with neglected exchange correlation 
(then similar to self-consistent Hartree--Fock), 
was taken by Hu et al.\ to study these effects \cite{Hu-etal-21}. 
Their numerics agree well with the semiclassical TF 
approximation close to fermions and show significant deviations 
only when the magnetic field is strong enough that particles 
in the bulk occupy only the lowest few LLs.
For trapped systems they also observe a boundary effect 
which can compensate for the otherwise necessary fractional
filling of LLs, but which vanishes in the thermodynamic limit.
Indeed, 
due to the many parallels to the FQHE,
Chen et al.\ suggested that 
``the physics of the anyon gas at general values of $\theta$ is likely
to be quite rich and to depend quite strongly on `number-theoretic' properties
of $\theta$.''

\subsection{Point-interacting anyons} \label{sec:point-int}

Introducing
\keyword{point or contact interactions} 
corresponds to specifying the behavior of wave functions
at the diagonal (contact) set $\bDelta_N$,
i.e.\ imposing boundary conditions there 
so as to render the Hamiltonian self-adjoint,
while allowing a different behavior than that of the canonical/free choice 
(Friedrichs; cf. Sec.~\ref{sec:kinetic-energy}).
This is equivalent to considering \emph{all} of the possible boundary conditions,
or \keyword{self-adjoint extensions}, of the operator $\hT_\alpha$ 
starting from its initial definition on wave functions vanishing 
at $\bDelta_N$ and having finite expected energy in the standard sense 
\eqref{eq:T-form}.
For general $N$ this is difficult even for distinguishable particles,
but one may reduce this problem significantly by adding assumptions 
such as translation and scale covariance.
Our present understanding of general point-interacting 
(and pointlike, i.e.\ non-extended) anyons 
is currently rather limited
(cf.\ \cite{DelFigTet-97}),
however for $N=2$ anyons the situation is now well understood
\cite{Grundberg-etal-91,ManTar-91,BorSor-92,MurLawBhaDat-92,CouNogPer-92,CorOdd-18,Oddis-20},
and extends the corresponding analysis of the standard non-magnetic two-body 
point-interaction problem in various dimensions 
(see, e.g., \cite{Albeverio-etal-05}).

For $\alpha_2 \neq 1$ (non-fermions) the problem admits 
a circle of different possibilities for point interaction, 
all of which are attractive and correspond to allowing wave functions with
singularities at the diagonals.
One simple way to understand this is via the form \eqref{eq:T-form-2}, 
which in the s-wave (radially symmetric) with $\Psi = r^{\pm\alpha_2}\phi(r)$
turns out to be equivalent to 
\cite{ManTar-91,Oddis-20}
\begin{equation} \label{eq:T-point}
	\int_0^\infty |\phi'(r)|^2 r^{d_\alpha-1} dr,
\end{equation}
for a parameter giving an effective \keyword{``intermediate dimension''}
\begin{equation} \label{eq:eff-dim}
	2 \le d_\alpha := 2 + 2\alpha_2 \le 4.
\end{equation}
Thus, for bosons the problem is indeed two-dimensional 
and has a circle of extensions, 
where a unique one $\hT_\sym$ (Friedrichs, $\Psi \sim 1$ as $r \to 0$) 
is nonnegative.
For fermions it is more akin to four dimensions 
where the corresponding bosonic problem is essentially self-adjoint, 
i.e.\ has a unique extension $\hT_\asym$ (Friedrichs, $\Psi \sim r$).
For proper anyons the situation is intermediate and actually 
closest akin to three dimensions.
In this case there is again a circle of extensions, 
but the circle now splits into two halves (cf.\ Fig.~\ref{fig:circ}): 
in one half one has negative extensions and a scale of increasingly 
negative bound states,
and in the other a scale of decreasingly nonnegative 
(i.e.\ increasingly attractive) extensions.
The largest (least attractive / ``free'') such extension is the 
usual free kinetic energy $\hT_\alpha = \hT_\alpha^{\rm F}$ 
(Friedrichs, $\Psi \sim r^{\alpha_2}$),
while the smallest $\hT_\alpha^{\rm K}$
(most attractive, $\Psi \sim r^{-\alpha_2}$,
and known as the \keyword{Krein extension} \cite{AloSim-80})
is also singled out as 
scale covariant while all the other intermediate extensions have scale.

If these observations 
extend to $N \ge 3$ particles then it is reasonable
to expect such \keyword{``kreinyons''} to be of interest in physics, and in fact
may even correspond to some of the moderately singular ``noninterpolating'' 
exact eigenstates \cite{MurLawBhaDat-92} 
found in the harmonic oscillator problem (cf. Section~\ref{sec:trial-states}).
Other relevant connections of anyons to geometry 
and quantum mechanics on cone-like manifolds 
were reviewed in \cite{Jackiw-90}.

\subsection{Other interactions and fields} \label{sec:other-int}

One can obviously also include other (scalar) interactions into 
the magnetic models for anyons,
as well as an external magnetic field (coupled by a charge $-q$), e.g.
\begin{multline} \label{eq:H-interacting}
	\hH_N := \frac{\hbar^2}{2m} \sum_{j=1}^N \left( -i\nabla_{\bx_j} + q\bA_{\rm ext} + \alpha \bA_j^R \right)^2\\
	+ \sum_{j=1}^N V(\bx_j) + \sum_{j < k} W(\bx_j-\bx_k).
\end{multline}
Reasonable external fields have been incorporated into some parts of
the average-field theory \cite{Girardot-20,GirRou-21},
as well as point interactions \cite{Oddis-20,CorFer-21},
and interacting anyons are certainly relevant in emergent models, 
as we will see below.
Coulomb interactions and the thermodynamic stability of 
``anyonic matter'' was addressed in 
\cite{LunSol-14,LunSei-17,Lundholm-17}.
We note that since scalar interactions can show features akin to statistics,
such as effective TF-type functionals and degeneracy pressure 
\cite{LieSeiSolYng-05,LunPorSol-15,LunNamPor-16,Lundholm-17,LarLunNam-19}, 
it is important to isolate the features that are due to 
the actual quantum statistics (cf.\ \cite{LamLunRou-22}).

Another way to view the effect of the magnetic flux disks in the average-field
approximation is as an effective scalar potential 
\begin{equation} \label{eq:W-effective}
	W(\bx_j-\bx_k) = \frac{2\pi|\alpha|}{\pi R^2} \1_{D(\bx_k,R)}(\bx_j)
\end{equation}
added to anyons 
(in the bosonic reference this is indeed correct as a lower bound \cite{LarLun-16}).
One may then consider the combination of parameters
\begin{equation} \label{eq:int-strength}
	\frac{|\alpha|}{\bar\varrho R^2} = |\alpha|\bar\gamma^{-2},
\end{equation}
i.e.\ the height of the potential compared to the average density,
as a dimensionless measure of interaction strength,
and for fixed $\alpha$ and $\bar\gamma \ll 1$ 
this then corresponds to a hard-core interaction, 
while if $|\alpha|\bar\gamma^{-2} \ll 1$,
such as if $\sqrt{|\alpha|} \ll \bar\gamma$ 
where either side may be taken to grow or decline,
then $W$ instead corresponds to a weak soft-core interaction.
With the convergence rates $\eta$ currently allowed in the derivations 
of almost-bosonic and almost-fermionic models 
($0 < \eta < 1/4$ \cite{Girardot-20} resp.\ $0 < \eta < 1/3$ \cite{GirRou-22}), 
we find that on a fixed homogeneous trap,
$\bar\gamma \sim N^{1/2-\eta} \to \infty$
and $|\alpha|\bar\gamma^{-2} \sim N^{2\eta-2} \to 0$ 
resp. $|1-\alpha|\bar\gamma^{-2} \sim N^{2\eta-3/2} \to 0$ 
as $N \to \infty$,
that is a combined high-density and weak soft-core limit,
indeed beneficial to the average-field approximation.

\subsection{Lowest Landau level (LLL) anyons} \label{sec:LLL}

If the anyons are ideal but in a strong external constant magnetic field, 
then all anyons fall into the LLL (cf.\ \eqref{eq:MTF-LLL}), 
and the model actually becomes exactly solvable.
The reason is that their wave function becomes (anti)analytic (modulo a Gaussian factor)
and corresponds to changing the Jastrow factors in \eqref{eq:trial-even}-\eqref{eq:trial-odd}
from $|z_{jk}|^{-\alpha}$ to $(\bar z_{jk})^{\alpha}$, $\alpha \in [0,1]$.
It results in a description in terms of 
\keyword{fractional exclusion statistics} (cf.\ \cite{Haldane-91})
with one-body occupation numbers (angular momenta) shifted by $\alpha$,
and a relation to Calogero--Sutherland models.
We refer to \cite{Ouvry-07} for an overview of this approach.

\section{The nonabelian anyon gas} \label{sec:nonabelian}

Let us now generalize our context further and consider an ideal
\keyword{$N$-anyon wave function} as \emph{locally}
a map\footnote{Again, one way to concretize this is to regard the
multivalued function $\Psi$ as a $\rho$-equivariant function on the
covering space of $\cC^N$, or as a section of a vector bundle $E \to \cC^N$
with fiber $\cF$ \cite{LunQva-20}.} 
\begin{equation} \label{eq:nonab-exch}
	\Psi\colon \cC^N \to \cF,
	\quad \text{subject to} \quad
	\Psi(\sigma.X) = \rho(\sigma) \Psi(X),
\end{equation}
where the \keyword{fiber} $\cF$ 
is a Hilbert space of `internal' (non-spatial) degrees of freedom
on which $B_N$ acts unitarily\footnote{%
We can even let go of the unitarity condition on reps if one is willing to
consider multivalued probability distributions 
(cf. \cite{AbrBra-11}).}:
$$
	\rho\colon B_N \to \sU(\cF).
$$
Now $\rho(\sigma)$ is an \keyword{exchange operator} on $\cF$
corresponding to the braid $\sigma \in B_N$.
There may be other observables of the considered system acting in $\cF$, 
but for simplicity we here assume that it is finite dimensional, 
i.e. $\cF = \C^D$ for some $D \in \N$, and $\sU(\C^D) = \sU(D)$ 
is the group of $D \times D$ unitary matrices.
We then identify three possibilities for these representations (`reps'):
\begin{enumerate}[wide, labelwidth=!, labelindent=0pt]
\item \keyword{Irreducible abelian}: $\cF = \C$, 
	which is the case we treated above, with elementary exchange phase
	$$
		\rho(\sigma_j) = e^{i\alpha\pi} \qquad \text{for all $j$,}
	$$
	for a fixed statistics parameter $\alpha$ of abelian anyons.
\item
	\keyword{Reducible abelian}: $\cF = \C^D$, $D>1$,
	where all elementary exchange operators $\rho(\sigma_j)$ 
	commute and thus are simultaneously diagonalizable:\footnote{%
	We write $A \cong B$ if two matrices or operators are similar, i.e. unitarily equivalent.}
	$$
		\rho(\sigma_j) \cong \mathrm{diag}(e^{i\alpha(1)\pi},\ldots,e^{i\alpha(D)\pi}) 
		\qquad \text{for all $j$,}
	$$
	for statistics parameters $\alpha(k)$ corresponding to different sectors
	of particles that do not interact quantum statistically.
\item
	\keyword{Nonabelian}: $\cF = \C^D$, $D>1$,
	where not all elementary exchange operators commute:
	$$
		\rho(\sigma_j)\rho(\sigma_k) \neq \rho(\sigma_k)\rho(\sigma_j)
		\qquad \text{for some $j \neq k$.}
	$$
	Actually, it turns out that 
	\begin{equation} \label{eq:rank-growth}
		D \ge N-2 \qquad \text{if} \ \ N \ge 7,
	\end{equation}
	because otherwise the rep is necessarily abelian
	(this is a theorem in the rep theory of the braid group \cite{Formanek-96}; 
	cf. \cite{Weinberger-15}, \cite[Theorem~3.13]{LunQva-20}).
	Furthermore, if $N \ge 7$ and $D=N-2$ or $D=N-1$ then the rep is either abelian
	or, up to an abelian factor, the nonabelian (reduced) Burau representation.
	For any other reps we need $D \ge N$.
\end{enumerate}

Because of this last remark, we cannot simply fix the fiber Hilbert space $\cF$
or the rank of the representation $D = \dim \cF$
and then study growing numbers $N$ of anyons, 
because they will then eventually abelianize.
Further,
in the nonabelian case it is, 
as far as this author is aware,
not certain whether
statistics transmutation is available, 
i.e.\ whether one may always pass from the geometric
to the magnetic perspective 
(this is a question in topology; 
see \cite{LunQva-20,MacSaw-19} for further discussion).
Some models have magnetic formulations however \cite{Kohno-87,Wen-91,LeeOh-94},
but for simplicity in this overview 
we will remain in the geometric perspective
and furthermore pull in a small sample of results from the 
\keyword{algebraic perspective} 
as well, where certain reps can be systematically and consistently constructed
from elementary \keyword{fusion and braiding} rules, and
exchange operators can be computed explicitly.
The topic is quite complex and extends far outside the scope of this overview;
instead we refer to 
\cite{Beer-etal-18,LunQva-20,Stern-08,Nayak-etal-08,LahPac-17,MasMizNit-23} 
for more detail on the algebraic and diagrammatic aspects of anyon models
(and note that the 
representation theory of the braid group 
is a still incomplete and very active field of mathematics).
Anyons with nonabelian braid group representations
are generally called \keyword{``nonabelions''} or \keyword{``plektons''}.

A \keyword{many-anyon model} is a sequence of fiber Hilbert spaces $\cF_N$,
$D_N = \dim\cF_N$,
and braid group representations
\begin{equation} \label{eq:rho-N}
	\rho_N\colon B_N \to \sU(\cF_N),
	\qquad N=1,2,3,\ldots,
\end{equation}
together with a sequence of kinetic energy operators 
\begin{equation} \label{eq:T-nonab}
	\hT_{\rho_N} = \frac{\hbar^2}{2m} \sum_{j=1}^N (-i\nabla_{\bx_j}^{(\rho_N)})^2
\end{equation}
which implement these reps (by means of parallel transport) 
as the corresponding exchange operators in \eqref{eq:nonab-exch}.\footnote{%
One might also require a compatibility between reps $\rho_N$ 
w.r.t.\ the embedding sequence
$B_{N-1} \hookrightarrow B_N$ of braid groups, which may be nontrivial
(see \cite{RowWan-12,DelRowWan-16}).}

The free kinetic energy operator $\hT_{\rho_N}$
is again locally the same as the free Laplacian $\hT_{\rm dist}$, 
supplemented with holonomies
$\rho_N(\sigma) \in \sU(D_N)$ w.r.t.\ any nontrivial exchange loops $\sigma \in B_N$.
Instead of the abelian pair-exchange conditions \eqref{eq:Psi-exch},
one now has
\begin{multline} \label{eq:Psi-exch-nonab}
	\Psi(\bx_1,\ldots,\bx_k,\ldots,\bx_j,\ldots,\bx_N) \\
	= U_{N,p} \Psi(\bx_1,\ldots,\bx_j,\ldots,\bx_k,\ldots,\bx_N),
\end{multline}
with \keyword{pair-exchange operators}\footnote{%
Defined up to similarity; note e.g.\ that 
$\rho(\sigma_j) \cong \rho(\sigma_k)$
with similarities depending on $j$ and $k$ \cite{LunQva-20}.}
\begin{equation} \label{eq:pair-exch-ops}
	U_{N,p} \cong \rho_N(\sigma_1 \sigma_2 \ldots \sigma_p \sigma_{p+1} \sigma_p \ldots \sigma_2 \sigma_1)
\end{equation}
for any counterclockwise two-particle exchange loop 
in which $p \le N-2$ other particles are enclosed,
corresponding to the second figure in Fig.~\ref{fig:loops}.
We denote their eigenvalues by
$$
	\spec U_{N,p} = \{ e^{i\pi\gamma_k} \}_{k=1,2,\ldots,D_N},
$$
where we normalize the full range of \keyword{statistics parameters}
$\gamma_k = \gamma_k(N,p)$ in the interval $(-1,1]$.

Nonabelions are still
harder to treat as perturbations close to bosons or fermions, since typically
the eigenvalues of $U_{N,p}$ are separated from (or coincide with) $\pm 1$.
Also, because of a lack of general magnetic formulations,
it is even more unclear what we should mean with an extended gas of nonabelions,
except as derived from emergent models and Berry connections (see below).
However,
concerning the qualitative features and local exclusion properties 
for the ideal gas,
somewhat analogous results as for the abelian gas may be obtained 
\cite{LunQva-20}, as follows.

The \keyword{$N$-fractionality} is now replaced by
\begin{equation} \label{eq:fractionality-nonab}
	\alpha_{N,n} := \min_{p=0,1,\ldots,n-2} \beta_{N,p},
	\qquad n \le N,
\end{equation}
where the \keyword{pair-exchange parameter} 
with $p \le N-2$ particles enclosed,
$$
	\beta_{N,p} := \min_{k=1,2,\ldots,D_N} |\gamma_k(N,p)|
	= \dist\bigl(\spec(U_{N,p}),+1\bigr),
$$
is again the 
arcwise distance (in units of $\pi$) from the spectrum of $U_{N,p}$ to $+1$.
Then, the statistical repulsion as given both by the Hardy inequality \eqref{eq:Hardy} 
(and its variants, depending on $\alpha_{N,n}$), 
the local exclusion principle \eqref{eq:E-2-approx}-\eqref{eq:E-N-4},
as well as the degeneracy pressure \eqref{eq:LT-anyons}, 
generalize to the ideal nonabelion gas:
\begin{equation} \label{eq:LT-nonab}
	\langle \hT_{\rho_N} \rangle_\Psi \gtrsim \alpha_{N,2} \int_{\R^2} \varrho_\Psi(\bx)^2 \,d\bx.
\end{equation}
Thus, it is again the nearest-neighbor pair exchange 
(in the rep $\rho_N$)
that sets the main behavior of the gas w.r.t.\ fractional exclusion statistics.

Three classes of representations of special interest were
considered in some detail in \cite{LunQva-20}: Burau, Ising and Fibonacci.
Note that variations of these may arise by composing with abelian models.
Some important features have been obtained also for other reps,
such as scattering and dilute thermodynamics based on the nearest-neighbor 
pair exchange \cite{WilWu-90,Verlinde-91,ManTroMus-13a,ManTroMus-13b}.

\subsection{Burau} \label{sec:Burau}

The Burau rep \cite{Burau-35} is a simple deformation of the defining rep of the 
permutation group, $S_N \to \sU(N)$, that also must be reduced 
($D_N=N-1$ or $N-2$) 
and unitarized to fit in the above framework.
We are only aware of the simplest case 
$N=3$, $\cF=\C^2$ having 
been worked out explicitly so far \cite{Weinberger-15}, 
and then 
$$
	\spec U_{3,0} = \{1,-w^2\}, \qquad
	\spec U_{3,1} = \{w^3,-w^3\}
$$
for a parameter $w=e^{i\alpha\pi}$, $|\alpha|<1/3$, of the representation.
Here we find $\alpha_{3,2}(w) = 0$ for all such $w$
and indeed there are states that do not vanish 
on the diagonal because of a lack of exclusion there \cite{LunQva-20}.

\subsection{Ising} \label{sec:Ising}

Here $D_N = \dim \cF_N$ grows roughly as $(\sqrt{2})^N$, 
which is related to reps of Clifford algebras \cite{NayWil-96}
(and there are distinctions between even/odd numbers of such anyons
and of fermions which 
also appear in the models).
The spectra of all pair-exchange operators have been computed \cite{LunQva-20}:
$$
	\begin{array}{l}
	\spec U_{N,p} = \\
	\begin{cases}
		\{e^{-i\pi/8},e^{i\pi 3/8}\}, &\text{if $p=0$},\\
		\{e^{-i\pi/8},e^{i\pi 7/8}\}, &\text{if $p\ge 1$ is odd},\\
		\{e^{-i\pi 5/8},e^{-i\pi/8},e^{i\pi 3/8},e^{i\pi 7/8}\}, &\text{if $p\ge 2$ is even},
	\end{cases}
	\end{array}
$$
(in one rep, and their complex conjugates in another rep,
and with multiplicities that grow with $N$).
Thus,
$$
	\alpha_{N,n} = \beta_{N,0} = 1/8
	\qquad
	\text{for all $N,n \ge 2$,}
$$
so the Ising gas has an exclusion at the diagonals and a degeneracy pressure.
Since these $n$-fractionalities are independent of $n$ (and $N$),
we do not see any benefits of clustering to minimize statistical repulsion,
at this level of inquiry.

\subsection{Fibonacci} \label{sec:Fib}

Here $D_N = \dim \cF_N$ grows with the Fibonacci sequence, so that
$D_{N+1}/D_N \xrightarrow{N \to \infty} \phi$, the golden ratio.
Furthermore, 
the spectra of all pair-exchange operators have been computed
\cite{Qvarfordt-17}:
$$
	\spec U_{N,p} = \begin{cases}
		\{e^{-i\pi 3/5},e^{i\pi 4/5}\}, &\text{if $p=0$,}\\
		\{e^{-i\pi/5},e^{i\pi/5},e^{i\pi 4/5}\}, &\text{if $p=1$,}\\
		\{e^{-i\pi 3/5},e^{-i\pi/5},e^{i\pi/5},e^{i\pi 4/5}\}, &\text{if $p\ge 2$,}
	\end{cases}
$$
(in one rep, and their complex conjugates in another rep,
and with multiplicities that grow with $N$, also along the Fibonacci sequence).
Thus,
$$
	\alpha_{N,n} = \begin{cases}
		\beta_{N,0} = 3/5, &\text{if $n=2$,}\\
		\beta_{N,1} = 1/5, &\text{if $n \ge 3$,}\\
	\end{cases}
$$
and therefore also the Fibonacci gas is subject to exclusion 
at the diagonals and a degeneracy pressure.
Since $\alpha_{N,n} < \alpha_{N,2}$ for $n \ge 3$, 
this may be indicative of clustering in this
gas in order to minimize statistical repulsion over long ranges.

\section{Emergence of the anyon gas} \label{sec:emergence}

In the study of emergent quantum statistics, we typically consider two
distinct species of particle, one of which is numerous and numbered by 
e.g.\ $N \gg 1$,
which we shall refer to as the \keyword{bath}, 
and another species which is much less numerous,
$1 \le n \ll N$, which we refer to as \keyword{tracers} or \keyword{impurities}.
These two species could be physically the same type of particle but 
effectively distinguished by some quantum number, 
such as their Landau level index for example.
The bath forms a background and a collective coherence for
the tracers, so that its macroscopically well-defined state provides a
limiting reference frame with respect to which tracers may transmute their properties.

\subsection{The quantum Hall setting} \label{sec:QH}

In the quantum Hall (QH) setting we assume that all particles are confined to a plane
and subject to a large and approximately constant magnetic field 
$\bB = -b\be_z$, $b>0$,
i.e.\ the relevant one-body operator is again the Landau Hamiltonian 
\eqref{eq:H-Landau}
$$
	\hH_1^{\rm Lan} = \frac{1}{2m}\big( {-i}\hbar\nabla_\bx + qb\bx^\perp/2 \big)^2 
	= \hbar\omega_c (\ha^\dagger \ha + 1/2),
$$
$\omega_c = |qb|/m$,
and particles then distribute into the Landau levels (LLs) of this field.
The lowest Landau level (LLL) is the kernel of $\ha$
(let us now fix $\hbar=1$ and $q=-1$):
\begin{multline} \label{eq:LLL}
	LLL := \cH_0 \\ 
	= \Big\{ \psi \in L^2(\R^2) : \psi(\bx) = f(z) e^{-\frac{b}{4}|z|^2}, \ f \ \text{analytic} \Big\}
\end{multline}
and spanned by the orthonormal basis of Fock--Darwin states
$$
	\psi_l(z) = c_l z^l e^{-\frac{b}{4}|z|^2},
	\qquad l = 0,1,2,\ldots,
$$
for normalization constants $c_l>0$.
We assume electron spin is out of the picture due to the strong band separation,
and also that there is a radial trapping potential centered at $z=0$,
penalizing higher angular momenta $l$.

The first concrete prediction of emergent anyonic statistics
was made in the context of the \emph{fractional} quantum Hall effect (FQHE)
for electrons \cite{TsuStoGos-82},
i.e. fractional filling of the LLL,
for which Laughlin proposed the following trial states 
in the electrons' coordinates $z_j \in \C$, $\sz = (z_j)_{j=1,\ldots,N}$,
to explain the effect \cite{Laughlin-83,Laughlin-99}:\footnote{%
Note that these also 
coincide with the anyonic trial states \eqref{eq:trial-even}-\eqref{eq:trial-odd} 
if extended to $\alpha = \mu$ with trivial Jastrow factors \cite{Lundholm-16}.} 
$$
	\PsiLau_\mu(\sz) \propto \prod_{j<k} z_{jk}^\mu e^{-\frac{b}{4}|\sz|^2}.
$$
For electrons, $\mu$ is an odd integer, so that $\PsiLau_\mu \in L^2_\asym$.
In the case that $\mu=1$, suitable for an \emph{integer} QH effect, 
the state corresponds to a Fermi sea filling of LLL with the least angular momentum:
$\PsiLau_{\mu=1}(\sz)$ is 
proportional to
$$
	\bigwedge_{l=0}^{N-1} \psi_l(\sz) \\
	\propto 
	\det \mat{
			1 & 1 & \ldots & 1\\
			z_1 & z_2 & \ldots & z_N\\
			\vdots & \vdots & & \vdots \\
			z_1^{N-1} & z_2^{N-1} & \ldots & z_N^{N-1}
			} e^{-\frac{b}{4}|\sz|^2}, 
$$
by Vandermonde's formula for the determinant.
For fractional filling $1/\mu$, and to reduce 
Coulomb interaction energy among electrons, 
Laughlin thus proposed taking the determinant above to a power $\mu>1$.
On top of such states, Laughlin also considered states
of quasiholes pinned to positions $w_k \in \C$, $\sw = (w_k)_{k=1,\ldots,n}$:
$$
	\PsiQH_\mu(\sw;\sz) \propto \prod_{k=1}^n \prod_{j=1}^N (w_k-z_j) \PsiLau_\mu(\sz)
$$
(all states with appropriate normalization, and 
$\by_k \leftrightarrow w_k$, $\bx_j \leftrightarrow z_j$ 
respective real and complex coordinates).

\subsubsection{Berry phases} \label{sec:Berry}

Halperin \cite{Halperin-84} suggested that if the quasiholes 
are treated as quantum particles
then they ought to have fractional anyonic statistics,
and Arovas, Schrieffer and Wilczek \cite{AroSchWil-84}
subsequently proposed a \keyword{Berry-phase approach} to determine their statistics.
Namely, if we view $\PsiQH_\mu$ as a map 
$$
	\C^n \to L^2_\asym(\R^{2N}), \qquad
	\sw \mapsto \PsiQH_\mu(\sw;\cdot)
$$
from the parameters $\sw$ into the (fiber) Hilbert space of electrons,
then Berry associates to such a geometry a canonical connection
\cite{Berry-84,Simon-83}:
\begin{multline} \label{eq:Berry-conn}
	\tilde{\bA}_j(\sy) := 
	\langle -i\nabla_{\by_j} \rangle_{\PsiQH(\sy;\cdot)} \\
	= -i\int_{\C^N} \overline{\PsiQH_\mu(\sy;\sz)}\nabla_{\by_j}\PsiQH_\mu(\sy;\sz) \,d\sz.
\end{multline}
The holonomies of the connection w.r.t.\ loops on the parameters,
$\sigma\colon [0,1] \to \C^n$,
\begin{equation} \label{eq:Berry-holonomy}
	\rho(\sigma) = e^{i\Phi_\sigma}, \qquad
	\Phi_\sigma = \oint_\sigma \sum_{j=1}^n \tilde{\bA}_j \cdot d\by_j,
\end{equation}
are then interpreted to be the exchange phases \eqref{eq:Psi-equivariance}
(now including an ``external'' field of constant curvature) 
effectively associated to the motion of the quasiholes in the bath of electrons.
Thus, under the \keyword{adiabatic assumption} that the electrons move much faster 
than the holes, one can take statistical averages over $\sz \in \C^N$ 
and use an analogy with a classical charged plasma
to compute $|\PsiQH(\sy;\cdot)|^2$ and $\Phi_\sigma$, and then indeed
conclude that there is, on top of geometric holonomies corresponding 
to a constant field
$B/\mu = -b/\mu$ (i.e.\ effectively coupled to a fractional charge 
$+1/\mu$ that \emph{reduces} the field),
also topological exchange phases corresponding to 
the localized (but somewhat extended) attachment of fractional flux 
$-2\pi/\mu$ to the quasiholes.
For a relatively recent review of the approach, including possible extensions to 
nonabelian scenarios 
(for which the quasihole map $\PsiQH$ is multidimensional), 
see \cite{BonGurNay-11}.

\subsubsection{Transmuted Hamiltonian} \label{sec:trans-ham}

Forte \cite{Forte-91} argued that the above approach is not 
necessarily sufficient to 
determine the anyonic statistics because, as we have noted above 
with the freedom of statistics transmutation,
the knowledge of the specific Hamiltonian for the quasiholes 
is essential, since that is what
ultimately sets the scale of validity of any adiabatic treatment.
Thus, a different approach was suggested in \cite{LunRou-16,LamLunRou-22}
to remedy this situation and supply the missing dynamical information, 
by considering quasiholes pinned to already
present quantum particles. 

\begin{figure}
	\begin{tikzpicture}[>=stealth',scale=0.8]
		\draw [thick,fill=black!5!white] (0,0) -- (2,2) -- (10,2) -- (8,0) -- (0,0);
		
		\draw [green!60!white] (7.0,0.9) ellipse (0.14 and 0.07);
		\draw [green!60!white] (2.0,1.0) ellipse (0.14 and 0.07);
		\draw [green!60!white] (5.0,1.5) ellipse (0.14 and 0.07);
		\draw [black!50!white] (2.0,1.0) [xscale=2.4,->] arc(250:100:.20);
		\draw [black!50!white] (5.0,1.5) [xscale=2.1,->] arc(100:210:.30);
		\draw [black!50!white] (7.0,0.9) [xscale=2.1,->] arc(-90:50:.30);

		\draw [thick,blue,->] (0.5,1.5) -- (0.5,1.0);
		\draw [thick,blue,->] (0.75,1.7) -- (0.75,1.2);

		\draw [fill] (7.0,0.9) 		ellipse (0.06 and 0.03);
		\draw [fill] (2.0,1.0) 		ellipse (0.06 and 0.03);
		\draw [fill] (5.0,1.5) 		ellipse (0.06 and 0.03);
		
		\draw [fill,red] (2.1,1.8) 	ellipse (0.06 and 0.03);
		\draw [fill,red] (5.3,1.2)	ellipse (0.06 and 0.03);
		\draw [fill,red] (4.7,0.3)	ellipse (0.06 and 0.03);
		\draw [fill,red] (2.3,1.1)	ellipse (0.06 and 0.03);
		\draw [fill,red] (7.3,0.4)	ellipse (0.06 and 0.03);
		\draw [fill,red] (7.1,1.7)	ellipse (0.06 and 0.03);
		\draw [fill,red] (8.1,1.6)	ellipse (0.06 and 0.03);
		\draw [fill,red] (0.9,0.4)	ellipse (0.06 and 0.03);
		\draw [fill,red] (1.7,0.3)	ellipse (0.06 and 0.03);
		\draw [fill,red] (3.7,0.5)	ellipse (0.06 and 0.03);
		\draw [fill,red] (3.6,1.7)	ellipse (0.06 and 0.03);
		\draw [fill,red] (6.2,0.8)	ellipse (0.06 and 0.03);
		\draw [fill,red] (8.3,0.7)	ellipse (0.06 and 0.03);
		\draw [fill,red] (2.8,0.6)	ellipse (0.06 and 0.03);
		\draw [fill,red] (2.8,1.3)	ellipse (0.06 and 0.03);
		\draw [fill,red] (3.3,1.0)	ellipse (0.06 and 0.03);
		\draw [fill,red] (4.1,1.2)	ellipse (0.06 and 0.03);
		\draw [fill,red] (4.8,0.7)	ellipse (0.06 and 0.03);
		\draw [fill,red] (1.0,0.6)	ellipse (0.06 and 0.03);
		\draw [fill,red] (5.4,0.2)	ellipse (0.06 and 0.03);
		\draw [fill,red] (6.5,1.6)	ellipse (0.06 and 0.03);
		\draw [fill,red] (7.2,1.3)	ellipse (0.06 and 0.03);
		\draw [fill,red] (7.0,0.25)	ellipse (0.06 and 0.03);
		\draw [fill,red] (8.7,1.8)	ellipse (0.06 and 0.03);
		\draw [fill,red] (8.8,1.4)	ellipse (0.06 and 0.03);
		\draw [fill,red] (4.5,1.9)	ellipse (0.06 and 0.03);
		\draw [fill,red] (1.2,1.0)	ellipse (0.06 and 0.03);
		\draw [fill,red] (3.5,0.35)	ellipse (0.06 and 0.03);
		\draw [fill,red] (5.1,1.75)	ellipse (0.06 and 0.03);
		
		\node [above right] at (9.0,0.2) {$N \gg n$};
	\end{tikzpicture}
	
	\vspace{0.3cm}
	\begin{tikzpicture}[>=stealth',scale=0.8]
		\draw [thick,fill=red!50!white] (0,0) -- (2,2) -- (10,2) -- (8,0) -- (0,0);
		\draw [fill=black!5!white,white] (7.0,0.9) 	ellipse (0.14 and 0.07);
		\draw [fill=black!5!white,white] (2.0,1.0) 	ellipse (0.14 and 0.07);
		\draw [fill=black!5!white,white] (5.0,1.5) 	ellipse (0.14 and 0.07);
		
		\draw [red] (2.0,1.0) [xscale=2.4,->] arc(250:100:.20);
		\draw [red] (5.0,1.5) [xscale=2.1,->] arc(100:210:.30);
		\draw [red] (7.0,0.9) [xscale=2.1,->] arc(-90:50:.30);

		\draw [thick,blue,->] (0.5,1.5) -- (0.5,1.0);
		\draw [thick,blue,->] (0.75,1.7) -- (0.75,1.2);

		\draw [fill] (7.0,0.9) 		ellipse (0.06 and 0.03);
		\draw [fill] (2.0,1.0) 		ellipse (0.06 and 0.03);
		\draw [fill] (5.0,1.5) 		ellipse (0.06 and 0.03);

		\node [above right] at (9.0,0.2) {$N \to \infty$};
	\end{tikzpicture}
	
	\vspace{0.3cm}
	\begin{tikzpicture}[>=stealth',scale=0.8]
		\draw [thick,fill=black!5!white] (0,0) -- (2,2) -- (10,2) -- (8,0) -- (0,0);
		
		\draw [blue] (2.0,1.0) [xscale=2.4,->] arc(250:100:.20);
		\draw [blue] (5.0,1.5) [xscale=2.1,->] arc(100:210:.30);
		\draw [blue] (7.0,0.9) [xscale=2.1,->] arc(-90:50:.30);

		\draw [fill,blue!60!white] (7.0,0.9) 	ellipse (0.14 and 0.07);
		\draw [fill,blue!60!white] (2.0,1.0) 	ellipse (0.14 and 0.07);
		\draw [fill,blue!60!white] (5.0,1.5) 	ellipse (0.14 and 0.07);

		\draw [fill] (7.0,0.9) 		ellipse (0.06 and 0.03);
		\draw [fill] (2.0,1.0) 		ellipse (0.06 and 0.03);
		\draw [fill] (5.0,1.5) 		ellipse (0.06 and 0.03);

		\node [above right] at (9.0,0.2) {transm.};
	\end{tikzpicture}
	\caption{Statistics transmutation in the QH setting.
	Bath particles (red) are repelled from tracers (black) by short-range 
	interactions (green), and all particles move in an external homogeneous 
	magnetic field (blue arrows). 
	Subtracting the almost-homogeneous effective field generated by the bath
	as $b \sim N \gg 1$ then leaves a reduced/zero field 
	plus remnant field (blue disks) 
	at the holes punched in the bath by the tracers.}
	\label{fig:QH}
\end{figure}
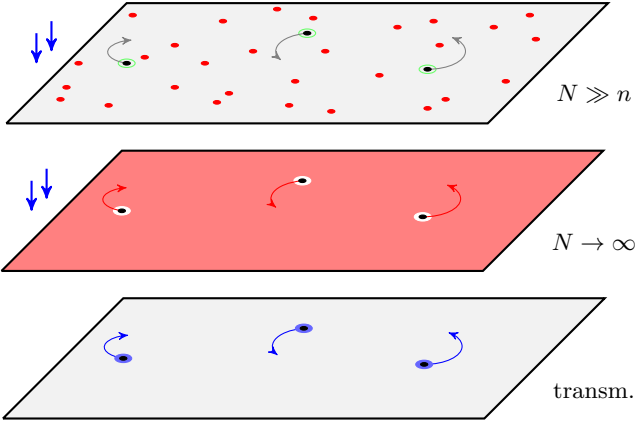

In brief, one may consider the following idealized situation 
(see Fig.~\ref{fig:QH}):
\begin{enumerate} 
\item large 2D bath of noninteracting \emph{fermions}, $N \gg 1$,
\item $n$ tracers/impurities (initially can be either bosons or fermions), $1 \le n \ll N$,
\item strong external transverse magnetic field, $b \to \infty$,
\item strong short-range repulsive bath-tracer interaction, at coupling $g \to \infty$.
\end{enumerate}
Taking the bath particles $\bx_j$ to have unit mass and unit negative charge,
the tracers $\by_k$ to have mass $m$ and charge $-q$,
and allowing for interaction and trapping of tracers in a joint potential $W$,
the Hamiltonian of the joint system is:
\begin{multline*}
	H_{n \oplus N} := \frac{1}{2} \sum_{j=1}^N \bigl(-i\nabla_{\bx_j} - b\bx_j^\perp/2\bigr)^2 \\
	+ \ \frac{1}{2m} \sum_{k=1}^n \bigl(-i\nabla_{\by_k} - qb\by_k^\perp/2\bigr)^2 \\
	+ \ g \sum_{j=1}^N \sum_{k=1}^n \delta(\bx_j-\by_k) \ 
	+ \ W(\by_1,\ldots,\by_n),
\end{multline*}
acting on states
$$
	\Psi(\by_1,\ldots,\by_n;\bx_1,\ldots,\bx_N) 
		\in L^2(\R^{2n}) \otimes L^2_\asym(\R^{2N}).
$$

If one takes $b \sim N$ and $N \gg n$, and assumes the bath then sits entirely
in the LLL, thus forming a disk-shaped QH droplet of size $\sim 1$,
the Hilbert space is reduced to
$$
	\cH^{n \oplus N}_{\sym/\asym} := L^2_{\sym/\asym}(\R^{2n}) \otimes \bigwedge\nolimits^N LLL
$$
(allowing for either bosonic or fermionic tracers),
with g.s.e.
$$
	E_{n \oplus N} := \inf_{0 \neq \Psi \in \cH^{n \oplus N}_{\sym/\asym}} \langle H_{n \oplus N}\rangle_\Psi.
$$
For $g \to \infty$, we then take as our ansatz for the ground state 
of the system a state in the kernel of the interaction:
\begin{equation} \label{eq:transm-ansatz}
	\Psi_\Phi(\sy;\sx) := \Phi(\sw) \cQH(\sw) \PsiQH(\sw;\sz),
\end{equation}
with $\Phi \in L^2_{\sym/\asym}(\C^n)$, $\cQH>0$ and
$$
	\int_{\C^n} |\Phi(\sw)|^2 \,d\sw = 1, \ \ 
	\cQH(\sw)^{-2} := \int_{\C^N} |\PsiQH(\sw;\sz)|^2 d\sz,
$$
ensuring the correct normalization 
$\int_{\C^{n+N}} |\Psi_\Phi|^2 d\sw d\sz = 1$.

The bath in LLL forces an \emph{analytic} vanishing as $z_j \to w_k$, 
i.e. $\bx_j \to \by_k$.
It suggests to take $\PsiQH = \PsiQH_\mu$ as above, 
corresponding to fractional filling $1/\mu$ of the LLL,
or
$$
	\PsiQH_{\mu,p}(\sw;\sz) := \prod_{k=1}^n \prod_{j=1}^N (w_k-z_j)^p \,\PsiLau_\mu(\sz),
$$
thus also allowing for non-simple quasiholes if $p>1$.

It was conjectured in \cite{LunRou-16,LamLunRou-22} that in this ansatz
\begin{equation} \label{eq:stat-trans}
	\langle H_{n \oplus N} \rangle_{\Psi_\Phi}
	= \frac{bN}{2} + \langle H_n^{\rm eff} \rangle_\Phi
	+ \text{error}_\Phi(n/N),
\end{equation}
for an effective \keyword{statistics-transmuted Hamiltonian} for the tracers:
\begin{multline*}
	H_n^{\rm eff} := \frac{bn}{2m}\frac{p}{\mu}
		+ \frac{1}{2m} \sum_{j=1}^n \Bigl(-i\nabla_{\by_j} 
		- \bigl(q-\frac{p}{\mu}\bigr) \frac{b}{2}\by_j^\perp 
		- \frac{p^2}{\mu} \bA_j\Bigr)^2 \\
		+ W(\sy),
\end{multline*}
thus implying an effective renormalized charge $-q+p/\mu$ and shifted 
statistics parameter by $-p^2/\mu$.

In \cite{LamLunRou-22} it was proved mathematically that \eqref{eq:stat-trans} is
indeed a correct conclusion, with an error term of order $o(N)$ at fixed $n$
and reasonable $\Phi$,
\emph{if} $\mu=1$ and $p=1$, 
corresponding to transmutation of bosons into fermions and vice versa
(cf.\ Sec.~\ref{sec:stat-transm}).
Many crucial simplifications occur for these specific parameters,
having to do with the Vandermonde structure of the state,
and in the proof it is also required that tracers have sufficient extra repulsion 
to not cluster too strongly at the diagonal set $\bDelta_n$,
which is ensured either by them being initially bosons 
and then eventually (free, if $q=1$) fermions,
or that $W$ includes an additional and 
sufficiently strong intra-species repulsion.

\subsection{Impurities in the plane and polarons} \label{sec:polarons}

Another route to statistics transmutation is to place tracers/impurities 
in a bath of bosons, such as the phonon vibrational modes of a crystal lattice,
or photonic optical modes.
Such systems are well known to enable transmutation of the impurities 
into \keyword{polarons}, i.e.\ bound states of impurity particles
and coherent states of phonons/photons
\cite{Landau-33,Pekar-46,Frohlich-54}.
In \cite{YakLem-18,Yakaboylu-etal-19} it was observed that, 
subject to suitable external fields and rotation of such systems, 
the polarons can also transmute their statistics.

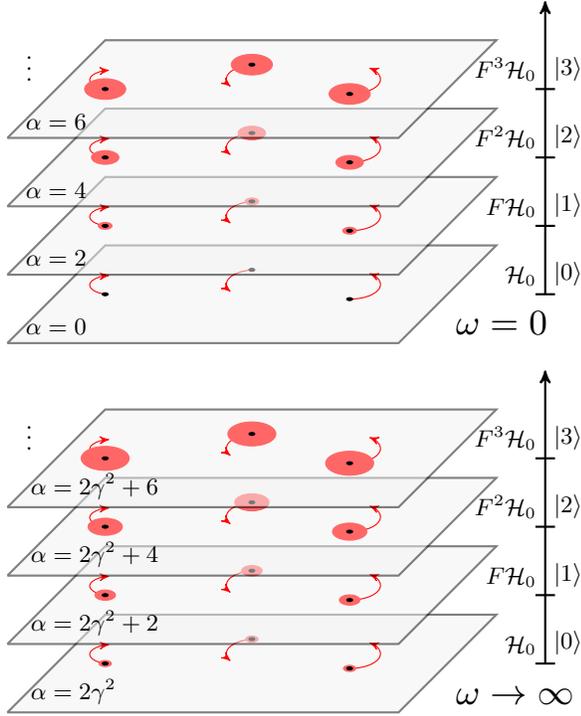
\begin{figure}
	\begin{tikzpicture}[>=stealth',scale=0.65]
		\fluxplane{0}{0}{0.1}
		\fluxplane{0}{1.4}{1}
		\fluxplane{0}{2.8}{2}
		\fluxplane{0}{4.2}{3}
		\draw [thick,->] (11,1) -- (11,7);
		\draw [thick] (10.8,1.0) -- (11.2,1.0);
		\draw [thick] (10.8,2.4) -- (11.2,2.4);
		\draw [thick] (10.8,3.8) -- (11.2,3.8);
		\draw [thick] (10.8,5.2) -- (11.2,5.2);
		\node [above left] at (11,1.0) {\scalebox{1.0}{$\cH_0$}};
		\node [above left] at (11,2.4) {\scalebox{1.0}{$F\cH_0$}};
		\node [above left] at (11,3.8) {\scalebox{1.0}{$F^2\cH_0$}};
		\node [above left] at (11,5.2) {\scalebox{1.0}{$F^3\cH_0$}};
		\node [above right] at (11,1.0) {\scalebox{1.0}{$|0\rangle$}};
		\node [above right] at (11,2.4) {\scalebox{1.0}{$|1\rangle$}};
		\node [above right] at (11,3.8) {\scalebox{1.0}{$|2\rangle$}};
		\node [above right] at (11,5.2) {\scalebox{1.0}{$|3\rangle$}};
		\node [above right] at (0.2,0) {\scalebox{1.0}{$\alpha=0$}};
		\node [above right] at (0.2,1.4) {\scalebox{1.0}{$\alpha=2$}};
		\node [above right] at (0.2,2.8) {\scalebox{1.0}{$\alpha=4$}};
		\node [above right] at (0.2,4.2) {\scalebox{1.0}{$\alpha=6$}};
		\node [above right] at (0.2,5.2) {\scalebox{1.0}{$\vdots$}};
		\node [above right] at (9,0) {\scalebox{1.5}{$\omega = 0$}};
	\end{tikzpicture}
	
	\vspace{0.3cm}
	\begin{tikzpicture}[>=stealth',scale=0.65]
		\fluxplane{0}{0}{0.9}
		\fluxplane{0}{1.4}{1.5}
		\fluxplane{0}{2.8}{2.5}
		\fluxplane{0}{4.2}{3.5}
		\draw [thick,->] (11,1) -- (11,7);
		\draw [thick] (10.8,1.0) -- (11.2,1.0);
		\draw [thick] (10.8,2.4) -- (11.2,2.4);
		\draw [thick] (10.8,3.8) -- (11.2,3.8);
		\draw [thick] (10.8,5.2) -- (11.2,5.2);
		\node [above left] at (11,1.0) {\scalebox{1.0}{$\cH_0$}};
		\node [above left] at (11,2.4) {\scalebox{1.0}{$F\cH_0$}};
		\node [above left] at (11,3.8) {\scalebox{1.0}{$F^2\cH_0$}};
		\node [above left] at (11,5.2) {\scalebox{1.0}{$F^3\cH_0$}};
		\node [above right] at (11,1.0) {\scalebox{1.0}{$|0\rangle$}};
		\node [above right] at (11,2.4) {\scalebox{1.0}{$|1\rangle$}};
		\node [above right] at (11,3.8) {\scalebox{1.0}{$|2\rangle$}};
		\node [above right] at (11,5.2) {\scalebox{1.0}{$|3\rangle$}};
		\node [above right] at (0.3,0.0) {\scalebox{1.0}{$\alpha=2\gamma^2$}};
		\node [above right] at (0.3,1.4) {\scalebox{1.0}{$\alpha=2\gamma^2+2$}};
		\node [above right] at (0.3,2.8) {\scalebox{1.0}{$\alpha=2\gamma^2+4$}};
		\node [above right] at (0.3,4.2) {\scalebox{1.0}{$\alpha=2\gamma^2+6$}};
		\node [above right] at (0.2,5.2) {\scalebox{1.0}{$\vdots$}};
		\node [above right] at (9,0) {\scalebox{1.5}{$\omega \to \infty$}};
	\end{tikzpicture}
	\caption{Statistics transmutation in the polaron setting.
	An initial ladder of even-integer fluxes/vortices 
	(composite bosons or fermions) 
	is in the adiabatic limit 
	transmuted into a ladder of fractional fluxes/vortices (anyons).}
	\label{fig:flux-ladder}
\end{figure}

As an illustrative toy model, 
we consider a single bosonic collective degree of freedom:
$$
	[\ha,\ha^\dagger]=1, \quad 
	\hat{N} := \ha^\dagger \ha, \ \ 
	\text{with eigenstates}\ |N\rangle, \ N\ge 0,
$$
and an initial $n$-particle Hamiltonian 
$$
	\hH_0 = \sum_{j=1}^n (-i\nabla_{\bx_j})^2 + W(\sx)
$$
for bosons or fermions on $\cH_0 = L^2_{\sym/\asym}(\R^{2n})$.
Let us then define the interaction Hamiltonian
\begin{equation} \label{eq:toy-Hamiltonian}
	\hH_{\omega,\gamma} := \hH_0 + \omega \ha^\dagger\ha 
		+ \gamma\omega (F\ha^\dagger + F^{-1}\ha) + \gamma^2\omega,
\end{equation}
with two parameters $\omega > 0$ and $\gamma \in \R$.
The interpretation of the terms added to $\hH_0$ is that 
$\omega$ is the energy associated to an 
attachment or detachment of two units of flux or vorticity 
to each of the particles,
which is done by the respective hopping terms with coupling $\gamma\omega$ 
and such that $\gamma$ defines the ratio between these two effects.
The last term is only a constant added for convenience.

We now denote $Z = \prod_{j<k} z_{jk}$, and
consider two possible choices 
for $F$ (or their complex conjugates, 
depending on which way the orientation symmetry is to be broken):
\begin{enumerate}[wide, labelwidth=!, labelindent=0pt]
\item $F = (Z/|Z|)^2 = U^2$ corresponding to \keyword{flux attachment},
\item $F = Z^2$ corresponding to \keyword{vortex attachment}.
\end{enumerate}
By acting with $Fa^\dagger$ on the reference system $\cH^0 = \cH_0|0\rangle$ one
obtains a ladder of Hilbert spaces of \keyword{composite bosons/fermions}
(see Fig.~\ref{fig:flux-ladder}):
$$
	\cH^N := F^N\cH_0|N\rangle,
	\qquad N=0,1,2,\ldots
$$

Upon taking the ``adiabatic limit''
$\omega \to \infty$ at fixed $\gamma$, 
we claim that in the bottom of the spectrum of $\hH_{\omega,\gamma}$ we then obtain
a ladder of transmuted anyons with statistics parameter 
$\alpha=2\gamma^2 + 2N$ \cite{Yakaboylu-etal-19}.
These anyons are either \emph{interacting} by means of a scalar potential 
$\sum_j \bA_j^2$
(for the more realistic, unitary choice $F=U^2$),
or \emph{free} (for the more algebraic, non-unitary choice $F=Z^2$,
yielding a non-hermitian Hamiltonian with $i\bA_j = \nabla_{\bx_j} \log Z$).
Namely, the transformations 
\begin{equation} \label{eq:polaron-transf}
	\hat{S} = F^{\hat{N}} = e^{\hat{N} \log F}, \qquad
	\hat{U} = e^{-\gamma(\ha^\dagger - \ha)},
\end{equation}
which diagonalize $\hH_{\omega,\gamma}$, 
\begin{multline*}
	\hat{U}^{-1}\hat{S}^{-1} \hH_{\omega,\gamma} \hat{S}\hat{U}\\
	= \omega \ha^\dagger \ha + \sum_{j=1}^N \left[ 
		-i\nabla_{\bx_j} + 2\bA_j(\ha^\dagger-\gamma)(\ha-\gamma)
		\right]^2 + W(\sx),
\end{multline*}
create coherent-state mixtures 
\begin{equation} \label{eq:coherent-state}
	\hat{S}\hat{U}|0\rangle = e^{-\gamma^2/2}\sum_{N=0}^\infty \frac{(-\gamma)^N}{\sqrt{N!}} F^N|N\rangle
\end{equation}
over the initial ladder of integer composites and enable a shift of
the expected amount of flux/vorticity by $\alpha_0 = 2\gamma^2 \ge 0$, 
possibly to non-integer 
values.
This transmutation procedure has indeed been used (with the choice $F=Z^2$)
to compute the 
numerical spectra seen in Fig.~\ref{fig:spectrum-2-3}.

\begin{figure}
	\centering
	\includegraphics[scale=0.85]{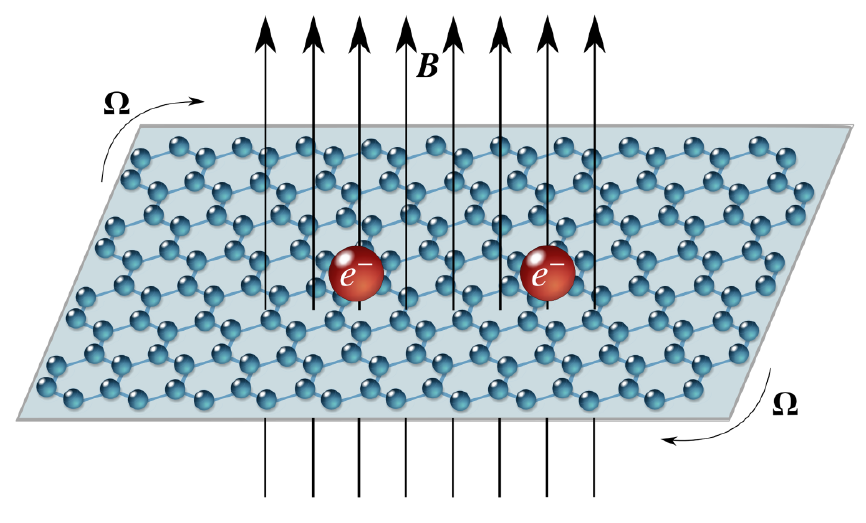}\\
	\includegraphics[scale=0.85]{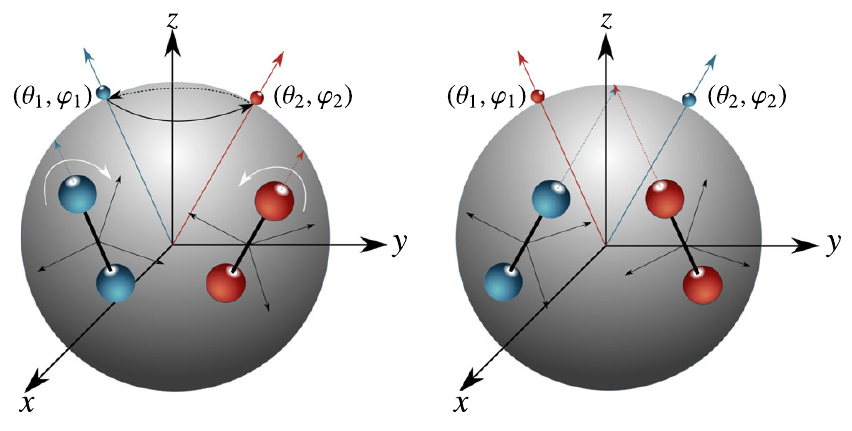}
	\caption{Two possible experimental probes for emergent statistics transmutation: 
	rotating plane in which 
	Fr\"ohlich polarons turn into anyons (upper, from \cite{Yakaboylu-etal-19})
	and the exchange of linear molecules (lower, from \cite{Yakaboylu-etal-20}).}
	\label{fig:experiments}
\end{figure}

A possible plane polaron model with the above form is obtained
upon applying a constant magnetic field $\bB$ and shifting to a
counter-rotating frame at half the cyclotron frequency $\Omega = B/(2m)$
(cf.\ Fig.~\ref{fig:experiments}, upper):
\begin{multline*}
	\hH_N = \frac{1}{2m}\sum_{j=1}^n \hat{\bp}^2_j + W^\Omega(\sx) 
	+ \sum_\bk \omega_\bk^\Omega \hb_\bk^\dagger \hb_\bk \\
	+ \sum_\bk \lambda_\bk(\sx) \left( e^{-i\beta_\bk(\sx)} b^\dagger_\bk + e^{i\beta_\bk(\sx)} b_\bk \right),
\end{multline*}
for phonon vibration modes $b_\bk$, $\bk \in \R^2$ with dispersion $\omega_\bk^\Omega$
(which is their bare dispersion shifted by their angular momentum coupled to $\Omega$), 
and phonon-impurity interaction parameters $\lambda_\bk$, $\beta_\bk$.
For $n=2$ and a suitable interaction, 
realizable within the Fr\"ohlich plane polaron model with quasi-2D Coulomb forces
\cite{Yakaboylu-etal-19},
the emergent gauge field
\begin{equation} \label{eq:polaron-gauge-field}
	\alpha\bA_j(\sx) \approx \langle -i\nabla_{\bx_j} \rangle_{\hat{S}\hat{U}|0\rangle}
	\approx -\sum_{\bk} (\lambda_\bk(\sx)/\omega_\bk^\Omega)^2 \nabla_{\bx_j}\beta_\bk(\sx)
\end{equation}
is then approximately of the correct Aharonov-Bohm type 
and such as to transmute the statistics by 
$\alpha \approx \alpha(\Omega)$, 
ascertained by the system's correlations to the collective reference rotation/field.

\subsection{Impurities on the sphere and angulons} \label{sec:angulons}

A similar approach \cite{Yakaboylu-etal-20,Brooks-etal-21}
can be taken for particles moving effectively on a sphere $\bbS^2$,
namely one may e.g.\ replace the planar kinetic energy \eqref{eq:T-free} 
by the rotation energy
\begin{equation} \label{eq:T-rot}
	T_{\rm rot} = \sum_{j=1}^n \bL^2_j
\end{equation}
(with units normalized suitably)
of $n$ identical linear molecules (cf.\ Fig.~\ref{fig:experiments}, lower).
The resulting model for the Hamiltonian 
coupled to a bath of rotational angular momenta
is called an \keyword{angulon} model \cite{SchLem-15}.
This transmutation method has also been used to compute 
the energy spectrum of two anyons
on a sphere \cite{Yakaboylu-etal-20}; 
cf. \cite{OuvPol-19,PolOuv-20},
and see \cite{Einarsson-92} concerning anyons in various topologies.

\subsection{Other emergent models for anyons} \label{sec:other-emerge}

Other approaches to investigate statistics transmutation in the
QH setting include effects of anyonic shifted angular momentum 
on the energy spectrum \cite{CooSim-15},
and on pair correlation and other aspects of the density 
\cite{ZhaSreGemJai-14,MorTurPolWil-17,Dubcek-etal-18,UmuMacComCar-18,Grass-etal-20},
including nonabelian scenarios \cite{ZhaSreJai-15,Baldelli-etal-21}.
Emergent anyons can also be expected on spin networks and lattices 
\cite{Bachmann-17,Bachmann-etal-20,Bachmann-etal-23}.
Certain classes of nonabelions have been proposed to emerge in the physics of 
neutrons and neutron stars \cite{MasMizNit-23}.
Even in the quantum gravity context might we expect nonabelions, 
as they can capture
the relevant degrees of freedom on the black-hole event horizon \cite{PitRui-15}.

\bigskip

\section{Conclusions and outlook} \label{sec:conclusions}

In light of recent experiments concerning anyons 
\cite{Nakamura-etal-20,Bartolomei-etal-20,Google-22,Fan-etal-22},
it is increasingly important to gain a firm 
theoretical understanding of the basic properties of the anyon gas.
We have here focused on its precise definition in typical ideal and nonideal contexts, 
its most basic physics 
as regards the 
connection between exchange and exclusion statistics,
as well as its inevitable emergence in a few conceptually simple scenarios 
involving relatively well-understood bosonic 
and fermionic systems 
with manifest breaking of orientation symmetry,
such as the rotated polaron and the quantum Hall setup.

In preparation for future experiments and theoretical research, 
one could make a few observations
and highlight some potential 
obstacles. 
First, it is hoped that accurate density functional theories 
for large systems of anyons 
will provide more robust, detectable signatures, 
such as spatial and momentum density profiles, 
as compared to the typically extremely 
fragile phases probed by interferometry 
(cf. \cite{CamZhoGol-05,Nakamura-etal-20}).
Second, the development of impurity models 
and the pinning of quasiholes, fluxes and vortices to already well-understood
quantum particles promises to 
eliminate some of the 
ambiguities inherent in conventional Berry phase and adiabatic treatments
(cf. \cite{Forte-91,Myrheim-99,Jain-07,LunRou-16,LamLunRou-22}).
Third, with the steadily growing literature on anyons 
and an increasing interest in their fascinating physics, 
even from the general public, 
it is important to be aware of
potential confusion
on how the term ``anyon'' is used. 
Hopefully,
this overview has helped to clarify both the nontrivial relationship between
exchange and exclusion statistics, as well as the
crucial differences between: 
a) the \keyword{choice or construction} of 
a concrete braid group representation
$\rho_N\colon B_N \to \sU(\cF_N)$
(algebra, kinematics, or simply an action of a permutation on a system); 
b) its \keyword{practical realization} such as by means of motion of classical parameters 
(magnetics, or computation using a classical or quantum computer; 
cf.\ \cite{Noh-etal-20});
and 
c) \keyword{actual anyons} (or bosons or fermions) endowed with 
quantum kinetic energy $\hT_{\rho_N}$ (geometrodynamics), uncertainty and exclusion.

\begin{ack}
	The study of the magnetic TF functional for anyons was done in 
	collaboration with Dinh Thi Nguyen, following helpful discussions with 
	Nicolas Rougerie and Th\'eotime Girardot.
	Further, I thank the editor Tapash Chakraborty for the timely invitation to
	write this overview, as well as 
	Vincent Cavalier, Ask Ellingsen, Gerald Goldin and Wolfgang Staubach
	for useful comments on the manuscript.
	I am thankful to numerous other colleagues for insightful discussions
	on this topic over the last 13 years, 
	in particular Jan Philip Solovej, who raised my interest in it, 
	as well as
	Eddy Ardonne,
	Morris Brooks, 
	Michele Correggi, 
	Romain Duboscq, 
	Luca Fresta, 
	J\"urg Fr\"ohlich, 
	Thors Hans Hansson,
	Gaultier Lambert, 
	Simon Larson, 
	Jon Magne Leinaas,
	Tomasz Maciazek, 
	Per Moosavi, 
	Luca Oddis, 
	St\'ephane Ouvry, 
	Viktor Qvarfordt, 
	Robert Seiringer, 
	Andrea Trombettoni,
	Susanne Viefers, and 
	Enderalp Yakaboylu.
	The research project was funded by the
	Swedish Research Council (grant no. 2021-05328, ``Mathematics of anyons and intermediate quantum statistics'').
\end{ack}

\def\MR#1{} 

\begin{thebibliography}{203}
\providecommand{\natexlab}[1]{#1}

\bibitem[Abramsky and Brandenburger(2011)]{AbrBra-11}
S.~Abramsky and A.~Brandenburger, \emph{The sheaf-theoretic structure of
  non-locality and contextuality}, New Journal of Physics \textbf{13} (2011),
  no.~11, 113036,
  \href{http://dx.doi.org/10.1088/1367-2630/13/11/113036}{\path{doi}}.

\bibitem[Aftalion et~al.(2006{\natexlab{a}})Aftalion, Blanc, and
  Nier]{AftBlaNie-06a}
A.~Aftalion, X.~Blanc, and F.~Nier, \emph{{Vortex distribution in the lowest
  Landau level}}, Phys. Rev. A \textbf{73} (2006), 011601(R),
  \href{http://dx.doi.org/10.1103/PhysRevA.73.011601}{\path{doi}}.

\bibitem[Aftalion et~al.(2006{\natexlab{b}})Aftalion, Blanc, and
  Nier]{AftBlaNie-06b}
A.~Aftalion, X.~Blanc, and F.~Nier, \emph{Lowest {L}andau level functional and
  {B}argmann spaces for {B}ose-{E}instein condensates}, J. Funct. Anal.
  \textbf{241} (2006), no.~2, 661--702,
  \href{http://dx.doi.org/10.1016/j.jfa.2006.04.027}{\path{doi}}. \MR{MR2271933
  (2008c:82052)}

\bibitem[Aharonov and Bohm(1959)]{AhaBoh-59}
Y.~Aharonov and D.~Bohm, \emph{Significance of electromagnetic potentials in
  the quantum theory}, Phys. Rev. \textbf{115} (1959), 485--491,
  \href{http://dx.doi.org/10.1103/PhysRev.115.485}{\path{doi}}.

\bibitem[Albeverio et~al.(2005)Albeverio, Gesztesy, H{\o}egh-Krohn, and
  Holden]{Albeverio-etal-05}
S.~Albeverio, F.~Gesztesy, R.~H{\o}egh-Krohn, and H.~Holden, \emph{Solvable
  models in quantum mechanics}, second ed., AMS Chelsea Publishing, Providence,
  RI, 2005, With an appendix by Pavel Exner. \MR{2105735}

\bibitem[Alonso and Simon(1980)]{AloSim-80}
A.~Alonso and B.~Simon, \emph{The {Birman-Kre\v{\i}n-Vishik} theory of
  self-adjoint extensions of semibounded operators}, Journal of Operator Theory
  (1980), 251--270.

\bibitem[Arovas et~al.(1984)Arovas, Schrieffer, and Wilczek]{AroSchWil-84}
D.~Arovas, J.~R. Schrieffer, and F.~Wilczek, \emph{Fractional statistics and
  the quantum {H}all effect}, Phys. Rev. Lett. \textbf{53} (1984), 722--723,
  \href{http://dx.doi.org/10.1103/PhysRevLett.53.722}{\path{doi}}.

\bibitem[Arovas et~al.(1985)Arovas, Schrieffer, Wilczek, and
  Zee]{AroSchWilZee-85}
D.~P. Arovas, R.~Schrieffer, F.~Wilczek, and A.~Zee, \emph{Statistical
  mechanics of anyons}, Nuclear Physics B \textbf{251} (1985), 117 -- 126,
  \href{http://dx.doi.org/10.1016/0550-3213(85)90252-4}{\path{doi}}.

\bibitem[Artin(1947)]{Artin-47}
E.~Artin, \emph{Theory of braids}, Ann. of Math. (2) \textbf{48} (1947),
  101--126, \href{http://dx.doi.org/10.2307/1969218}{\path{doi}}. \MR{19087}

\bibitem[Artin(1925)]{Artin-25}
E.~Artin, \emph{Theorie der {Z}\"{o}pfe}, Abh. Math. Sem. Univ. Hamburg
  \textbf{4} (1925), no.~1, 47--72,
  \href{http://dx.doi.org/10.1007/BF02950718}{\path{doi}}. \MR{3069440}

\bibitem[Bachmann(2017)]{Bachmann-17}
S.~Bachmann, \emph{Local disorder, topological ground state degeneracy and
  entanglement entropy, and discrete anyons}, Rev. Math. Phys. \textbf{29}
  (2017), no.~06, 1750018,
  \href{http://dx.doi.org/10.1142/S0129055X17500180}{\path{doi}}.

\bibitem[Bachmann et~al.(2020)Bachmann, Bols, De~Roeck, and
  Fraas]{Bachmann-etal-20}
S.~Bachmann, A.~Bols, W.~De~Roeck, and M.~Fraas, \emph{Many-body {F}redholm
  index for ground-state spaces and abelian anyons}, Phys. Rev. B \textbf{101}
  (2020), 085138,
  \href{http://dx.doi.org/10.1103/PhysRevB.101.085138}{\path{doi}}.

\bibitem[Bachmann et~al.(2023)Bachmann, Nachtergaele, and
  Vadnerkar]{Bachmann-etal-23}
S.~Bachmann, B.~Nachtergaele, and S.~Vadnerkar, \emph{Dynamical abelian anyons
  with bound states and scattering states}, arXiv e-prints, 2023,
  \href{http://arxiv.org/abs/2303.07379}{\path{arXiv:2303.07379}}.

\bibitem[Baldelli et~al.(2021)Baldelli, Juli{\'a}-D{\'{\i}}az, Bhattacharya,
  Lewenstein, and Gra{\ss{}}]{Baldelli-etal-21}
N.~Baldelli, B.~Juli{\'a}-D{\'{\i}}az, U.~Bhattacharya, M.~Lewenstein, and
  T.~Gra{\ss{}}, \emph{Tracing non-abelian anyons via impurity particles},
  Phys. Rev. B \textbf{104} (2021), 035133,
  \href{http://dx.doi.org/10.1103/PhysRevB.104.035133}{\path{doi}}.

\bibitem[Bartolomei et~al.(2020)Bartolomei, Kumar, Bisognin, Marguerite,
  Berroir, Bocquillon, Pla{\c c}ais, Cavanna, Dong, Gennser, Jin, and
  F{\`e}ve]{Bartolomei-etal-20}
H.~Bartolomei, M.~Kumar, R.~Bisognin, A.~Marguerite, J.-M. Berroir,
  E.~Bocquillon, B.~Pla{\c c}ais, A.~Cavanna, Q.~Dong, U.~Gennser, Y.~Jin, and
  G.~F{\`e}ve, \emph{Fractional statistics in anyon collisions}, Science
  \textbf{368} (2020), no.~6487, 173--177,
  \href{http://dx.doi.org/10.1126/science.aaz5601}{\path{doi}}.

\bibitem[Beer et~al.(2018)Beer, Bondarenko, Hahn, Kalabakov, Knust, Niermann,
  Osborne, Schridde, Seckmeyer, Stiegemann, et~al.]{Beer-etal-18}
K.~Beer, D.~Bondarenko, A.~Hahn, M.~Kalabakov, N.~Knust, L.~Niermann, T.~J.
  Osborne, C.~Schridde, S.~Seckmeyer, D.~E. Stiegemann, et~al., \emph{From
  categories to anyons: a travelogue}, arXiv e-prints, 2018,
  \href{http://arxiv.org/abs/1811.06670}{\path{arXiv:1811.06670}}.

\bibitem[Berry(1984)]{Berry-84}
M.~V. Berry, \emph{Quantal phase factors accompanying adiabatic changes}, Proc.
  Roy. Soc. London, Ser. A \textbf{392} (1984), 45--57,
  \href{http://dx.doi.org/10.1098/rspa.1984.0023}{\path{doi}}.

\bibitem[Biedenharn et~al.(1990)Biedenharn, Lieb, Simon, and
  Wilczek]{BieLieSimWil-90}
L.~Biedenharn, E.~Lieb, B.~Simon, and F.~Wilczek, \emph{The ancestry of the
  {`Anyon'}}, Physics Today \textbf{43} (1990), no.~8, 90,
  \href{http://dx.doi.org/10.1063/1.2810672}{\path{doi}}.

\bibitem[Birman(1974)]{Birman-74}
J.~S. Birman, \emph{Braids, links, and mapping class groups}, Princeton
  University Press, Princeton, N.J.; University of Tokyo Press, Tokyo, 1974,
  Annals of Mathematics Studies, No. 82. \MR{0375281}

\bibitem[Bonderson et~al.(2011)Bonderson, Gurarie, and Nayak]{BonGurNay-11}
P.~Bonderson, V.~Gurarie, and C.~Nayak, \emph{Plasma analogy and non-abelian
  statistics for {I}sing-type quantum {H}all states}, Phys. Rev. B \textbf{83}
  (2011), 075303,
  \href{http://dx.doi.org/10.1103/PhysRevB.83.075303}{\path{doi}}.

\bibitem[Bourdeau and Sorkin(1992)]{BorSor-92}
M.~Bourdeau and R.~D. Sorkin, \emph{When can identical particles collide?},
  Phys. Rev. D \textbf{45} (1992), 687--696,
  \href{http://dx.doi.org/10.1103/PhysRevD.45.687}{\path{doi}}.

\bibitem[Brooks et~al.(2021{\natexlab{a}})Brooks, Lemeshko, Lundholm, and
  Yakaboylu]{Yakaboylu-etal-20}
M.~Brooks, M.~Lemeshko, D.~Lundholm, and E.~Yakaboylu, \emph{Molecular
  impurities as a realization of anyons on the two-sphere}, Phys. Rev. Lett.
  \textbf{126} (2021), 015301,
  \href{http://dx.doi.org/10.1103/PhysRevLett.126.015301}{\path{doi}}.

\bibitem[Brooks et~al.(2021{\natexlab{b}})Brooks, Lemeshko, Lundholm, and
  Yakaboylu]{Brooks-etal-21}
M.~Brooks, M.~Lemeshko, D.~Lundholm, and E.~Yakaboylu, \emph{Emergence of
  anyons on the two-sphere in molecular impurities}, Atoms \textbf{9} (2021),
  no.~4, 106, \href{http://dx.doi.org/10.3390/atoms9040106}{\path{doi}}.

\bibitem[Burau(1935)]{Burau-35}
W.~Burau, \emph{\"{U}ber {Z}opfgruppen und gleichsinnig verdrillte
  {V}erkettungen}, Abh. Math. Sem. Univ. Hamburg \textbf{11} (1935), no.~1,
  179--186, \href{http://dx.doi.org/10.1007/BF02940722}{\path{doi}}.
  \MR{3069652}

\bibitem[Caenepeel and MacKenzie(1994)]{CaeMac-94}
D.~Caenepeel and R.~MacKenzie, \emph{Parity violation, anyon scattering, and
  the mean field approximation}, Phys. Rev. D \textbf{50} (1994), 5418--5424,
  \href{http://dx.doi.org/10.1103/PhysRevD.50.5418}{\path{doi}}.

\bibitem[Calogero(1969{\natexlab{a}})]{Calogero-69b}
F.~Calogero, \emph{Ground state of a one-dimensional {N}-body system}, J. Math.
  Phys. \textbf{10} (1969), no.~12, 2197--2200,
  \href{http://dx.doi.org/10.1063/1.1664821}{\path{doi}}.

\bibitem[Calogero(1969{\natexlab{b}})]{Calogero-69a}
F.~Calogero, \emph{Solution of a three-body problem in one dimension}, J. Math.
  Phys. \textbf{10} (1969), no.~12, 2191--2196,
  \href{http://dx.doi.org/10.1063/1.1664820}{\path{doi}}.

\bibitem[Camino et~al.(2005)Camino, Zhou, and Goldman]{CamZhoGol-05}
F.~E. Camino, W.~Zhou, and V.~J. Goldman, \emph{{Realization of a Laughlin
  quasiparticle interferometer: Observation of fractional statistics}}, Phys.
  Rev. B \textbf{72} (2005), 075342,
  \href{http://dx.doi.org/10.1103/PhysRevB.72.075342}{\path{doi}}.

\bibitem[Canright and Johnson(1994)]{CanJoh-94}
G.~S. Canright and M.~D. Johnson, \emph{Fractional statistics: alpha to beta},
  J. Phys. A: Math. Gen. \textbf{27} (1994), no.~11, 3579,
  \href{http://dx.doi.org/10.1088/0305-4470/27/11/009}{\path{doi}}.

\bibitem[Cappelli et~al.(2001)Cappelli, Georgiev, and Todorov]{CapGeoTod-01}
A.~Cappelli, L.~S. Georgiev, and I.~T. Todorov, \emph{Parafermion {H}all states
  from coset projections of abelian conformal theories}, Nucl. Phys. B
  \textbf{599} (2001), no.~3, 499--530,
  \href{http://dx.doi.org/10.1016/S0550-3213(00)00774-4}{\path{doi}}.

\bibitem[Chen et~al.(1989)Chen, Wilczek, Witten, and Halperin]{CheWilWitHal-89}
Y.~H. Chen, F.~Wilczek, E.~Witten, and B.~I. Halperin, \emph{On anyon
  superconductivity}, Int. J. Mod. Phys. B \textbf{03} (1989), 1001--1067,
  \href{http://dx.doi.org/10.1142/S0217979289000725}{\path{doi}}.

\bibitem[Chitra and Sen(1992)]{ChiSen-92}
R.~Chitra and D.~Sen, \emph{{Ground state of many anyons in a harmonic
  potential}}, Phys. Rev. B \textbf{46} (1992), 10923--10930,
  \href{http://dx.doi.org/10.1103/PhysRevB.46.10923}{\path{doi}}.

\bibitem[Chou(1991)]{Chou-91a}
C.~Chou, \emph{Multianyon spectra and wave functions}, Phys. Rev. D \textbf{44}
  (1991), 2533--2547,
  \href{http://dx.doi.org/10.1103/PhysRevD.44.2533}{\path{doi}}.

\bibitem[Cooper and Simon(2015)]{CooSim-15}
N.~R. Cooper and S.~H. Simon, \emph{Signatures of fractional exclusion
  statistics in the spectroscopy of quantum {H}all droplets}, Phys. Rev. Lett.
  \textbf{114} (2015), 106802,
  \href{http://dx.doi.org/10.1103/PhysRevLett.114.106802}{\path{doi}}.

\bibitem[Correggi and Fermi(2021)]{CorFer-21}
M.~Correggi and D.~Fermi, \emph{Magnetic perturbations of anyonic and
  {A}haronov-{B}ohm {S}chr{\"o}dinger operators}, J. Math. Phys. \textbf{62}
  (2021), 032101, \href{http://dx.doi.org/10.1063/5.0018933}{\path{doi}}.

\bibitem[Correggi and Oddis(2018)]{CorOdd-18}
M.~Correggi and L.~Oddis, \emph{Hamiltonians for two-anyon systems}, Rend. Mat.
  Appl. \textbf{39} (2018), 277--292,
  \url{http://www1.mat.uniroma1.it/ricerca/rendiconti/39_2_(2018)_277-292.html}.

\bibitem[Correggi et~al.(2017)Correggi, Lundholm, and Rougerie]{CorLunRou-16}
M.~Correggi, D.~Lundholm, and N.~Rougerie, \emph{Local density approximation
  for the almost-bosonic anyon gas}, Analysis \& PDE \textbf{10} (2017),
  1169--1200, \href{http://dx.doi.org/10.2140/apde.2017.10.1169}{\path{doi}}.

\bibitem[Correggi et~al.(2018)Correggi, Lundholm, and
  Rougerie]{CorLunRou-proc-17}
M.~Correggi, D.~Lundholm, and N.~Rougerie, \emph{Local density approximation
  for almost-bosonic anyons}, Proceedings of QMath13, Atlanta, October 8--11,
  2016, Mathematical problems in quantum physics (F.~Bonetto, D.~Borthwick,
  E.~Harrell, and M.~Loss, eds.), Contemp. Math., vol. 717, 2018, pp.~77--92,
  \href{http://dx.doi.org/10.1090/conm/717}{\path{doi}}.

\bibitem[Correggi et~al.(2019)Correggi, Duboscq, Lundholm, and
  Rougerie]{CorDubLunRou-19}
M.~Correggi, R.~Duboscq, D.~Lundholm, and N.~Rougerie, \emph{Vortex patterns in
  the almost-bosonic anyon gas}, EPL (Europhysics Letters) \textbf{126} (2019),
  20005, \href{http://dx.doi.org/10.1209/0295-5075/126/20005}{\path{doi}}.

\bibitem[Coutinho et~al.(1992)Coutinho, Nogami, and
  Fernando~Perez]{CouNogPer-92}
F.~A.~B. Coutinho, Y.~Nogami, and J.~Fernando~Perez, \emph{Self-adjoint
  extensions of the {H}amiltonian for a charged particle in the presence of a
  thread of magnetic flux}, Phys. Rev. A \textbf{46} (1992), 6052--6055,
  \href{http://dx.doi.org/10.1103/PhysRevA.46.6052}{\path{doi}}.

\bibitem[Date and Murthy(1993)]{DatMur-93}
G.~Date and M.~V.~N. Murthy, \emph{Classical dynamics of anyons and the quantum
  spectrum}, Phys. Rev. A \textbf{48} (1993), 105--110,
  \href{http://dx.doi.org/10.1103/PhysRevA.48.105}{\path{doi}}.

\bibitem[Date et~al.(2003)Date, Murthy, and Vathsan]{DatMurVat-03}
G.~Date, M.~V.~N. Murthy, and R.~Vathsan, \emph{Classical and quantum mechanics
  of anyons}, arXiv e-prints, 2003,
  \href{http://arxiv.org/abs/cond-mat/0302019}{\path{arXiv:cond-mat/0302019}}.

\bibitem[Delaney et~al.(2016)Delaney, Rowell, and Wang]{DelRowWan-16}
C.~Delaney, E.~C. Rowell, and Z.~Wang, \emph{Local unitary representations of
  the braid group and their applications to quantum computing}, Rev. Colombiana
  Mat. \textbf{50} (2016), no.~2, 207--272,
  \href{http://dx.doi.org/10.15446/recolma.v50n2.62211}{\path{doi}}.
  \MR{3605648}

\bibitem[Dell'Antonio et~al.(1997)Dell'Antonio, Figari, and Teta]{DelFigTet-97}
G.~Dell'Antonio, R.~Figari, and A.~Teta, \emph{Statistics in space dimension
  two}, Lett. Math. Phys. \textbf{40} (1997), no.~3, 235--256,
  \href{http://dx.doi.org/10.1023/A:1007361832622}{\path{doi}}. \MR{1453763}

\bibitem[Dirac(1967)]{Dirac-64}
P.~A.~M. Dirac, \emph{Lectures on quantum mechanics}, Belfer Graduate School of
  Science Monographs Series, vol.~2, Belfer Graduate School of Science, New
  York; produced and distributed by Academic Press, Inc., New York, 1967,
  Second printing of the 1964 original. \MR{2220894}

\bibitem[Dowker(1972)]{Dowker-72}
J.~S. Dowker, \emph{Quantum mechanics and field theory on multiply connected
  and on homogeneous spaces}, J. Phys. A \textbf{5} (1972), no.~7, 936,
  \href{http://dx.doi.org/10.1088/0305-4470/5/7/004}{\path{doi}}.

\bibitem[Dowker(1985)]{Dowker-85}
J.~S. Dowker, \emph{Remarks on nonstandard statistics}, J. Phys. A \textbf{18}
  (1985), no.~18, 3521--3530,
  \href{http://dx.doi.org/10.1088/0305-4470/18/18/015}{\path{doi}}. \MR{822912}

\bibitem[Dub{\v{c}}ek et~al.(2018)Dub{\v{c}}ek, Klajn, Pezer, Buljan, and
  Juki{\'{c}}]{Dubcek-etal-18}
T.~Dub{\v{c}}ek, B.~Klajn, R.~Pezer, H.~Buljan, and D.~Juki{\'{c}},
  \emph{Quasimomentum distribution and expansion of an anyonic gas}, Phys. Rev.
  A \textbf{97} (2018), 011601,
  \href{http://dx.doi.org/10.1103/PhysRevA.97.011601}{\path{doi}}.

\bibitem[Dyson(1967)]{Dyson-67}
F.~J. Dyson, \emph{Ground-state energy of a finite system of charged
  particles}, J. Math. Phys. \textbf{8} (1967), no.~8, 1538--1545,
  \href{http://dx.doi.org/10.1063/1.1705389}{\path{doi}}.

\bibitem[Dyson and Lenard(1967)]{DysLen-67}
F.~J. Dyson and A.~Lenard, \emph{Stability of matter. {I}}, J. Math. Phys.
  \textbf{8} (1967), no.~3, 423--434,
  \href{http://dx.doi.org/10.1063/1.1705209}{\path{doi}}.

\bibitem[Ehrenberg and Siday(1949)]{EhrSid-49}
W.~Ehrenberg and R.~E. Siday, \emph{The refractive index in electron optics and
  the principles of dynamics}, Proceedings of the Physical Society. Section B
  \textbf{62} (1949), no.~1, 8.

\bibitem[Einarsson(1992)]{Einarsson-92}
T.~Einarsson, \emph{Anyons and antiferromagnets: Two two-dimensional topics},
  Ph.D. thesis, Institute of Theoretical Physics, Chalmers University of
  Technology, Gothenburg, 1992,
  \url{https://research.chalmers.se/en/publication/176319}.

\bibitem[Fan et~al.(2022)Fan, Li, Hu, Li, Long, Liu, Yang, Nie, Li, Xin,
  et~al.]{Fan-etal-22}
Y.-a. Fan, Y.~Li, Y.~Hu, Y.~Li, X.~Long, H.~Liu, X.~Yang, X.~Nie, J.~Li,
  T.~Xin, et~al., \emph{Experimental realization of a topologically protected
  {H}adamard gate via braiding {F}ibonacci anyons}, arXiv e-prints, 2022,
  \href{http://arxiv.org/abs/2210.12145}{\path{arXiv:2210.12145}}.

\bibitem[Fermi(1927)]{Fermi-27}
E.~Fermi, \emph{Un metodo statistico per la determinazione di alcune priorieta
  dell'atome}, Rend. Accad. Naz. Lincei \textbf{6} (1927), 602--607.

\bibitem[Fetter et~al.(1989)Fetter, Hanna, and Laughlin]{FetHanLau-89}
A.~L. Fetter, C.~B. Hanna, and R.~B. Laughlin, \emph{Random-phase approximation
  in the fractional-statistics gas}, Phys. Rev. B \textbf{39} (1989),
  9679--9681, \href{http://dx.doi.org/10.1103/PhysRevB.39.9679}{\path{doi}}.

\bibitem[Formanek(1996)]{Formanek-96}
E.~Formanek, \emph{Braid group representations of low degree}, Proc. London
  Math. Soc. (3) \textbf{73} (1996), no.~2, 279--322,
  \href{http://dx.doi.org/10.1112/plms/s3-73.2.279}{\path{doi}}. \MR{1397691}

\bibitem[Forte(1991)]{Forte-91}
S.~Forte, \emph{{Berry's phase, fractional statistics and the Laughlin wave
  function}}, Mod. Phys. Lett. A \textbf{6} (1991), no.~34, 3152--3162,
  \href{http://dx.doi.org/10.1142/S021773239100364X}{\path{doi}}.

\bibitem[Forte(1992)]{Forte-92}
S.~Forte, \emph{Quantum mechanics and field theory with fractional spin and
  statistics}, Rev. Mod. Phys. \textbf{64} (1992), 193--236,
  \href{http://dx.doi.org/10.1103/RevModPhys.64.193}{\path{doi}}.

\bibitem[Fredenhagen et~al.(1989)Fredenhagen, Rehren, and
  Schroer]{FreRehSch-89}
K.~Fredenhagen, K.-H. Rehren, and B.~Schroer, \emph{Superselection sectors with
  braid group statistics and exchange algebras. {I}. {G}eneral theory}, Comm.
  Math. Phys. \textbf{125} (1989), no.~2, 201--226,
  \url{http://projecteuclid.org/euclid.cmp/1104179464}. \MR{1016869}

\bibitem[Freedman et~al.(2003)Freedman, Kitaev, Larsen, and
  Wang]{FreKitLarWan-03}
M.~H. Freedman, A.~Kitaev, M.~J. Larsen, and Z.~Wang, \emph{Topological quantum
  computation}, Bull. Amer. Math. Soc. (N.S.) \textbf{40} (2003), no.~1,
  31--38, Mathematical challenges of the 21st century (Los Angeles, CA, 2000),
  \href{http://dx.doi.org/10.1090/S0273-0979-02-00964-3}{\path{doi}}.
  \MR{1943131}

\bibitem[Fr{\"o}hlich(1954)]{Frohlich-54}
H.~Fr{\"o}hlich, \emph{{Electrons in lattice fields}}, Advances in Physics
  \textbf{3} (1954), 325--361,
  \href{http://dx.doi.org/10.1080/00018735400101213}{\path{doi}}.

\bibitem[Fr\"ohlich and Gabbiani(1990)]{FroGab-90}
J.~Fr\"ohlich and F.~Gabbiani, \emph{Braid statistics in local quantum theory},
  Rev. Math. Phys. \textbf{2} (1990), no.~3, 251--353,
  \href{http://dx.doi.org/10.1142/S0129055X90000107}{\path{doi}}. \MR{1104414}

\bibitem[Fr\"{o}hlich and Marchetti(1988)]{FroMar-88}
J.~Fr\"{o}hlich and P.~A. Marchetti, \emph{Quantum field theory of anyons},
  Lett. Math. Phys. \textbf{16} (1988), no.~4, 347--358,
  \href{http://dx.doi.org/10.1007/BF00402043}{\path{doi}}. \MR{974482}

\bibitem[Fr\"{o}hlich and Marchetti(1989)]{FroMar-89}
J.~Fr\"{o}hlich and P.~A. Marchetti, \emph{Quantum field theories of vortices
  and anyons}, Commun. Math. Phys. \textbf{121} (1989), no.~2, 177--223,
  \href{http://dx.doi.org/10.1007/BF01217803}{\path{doi}}. \MR{985396}

\bibitem[Fr{\"o}hlich(1976)]{Froehlich-76}
J.~Fr{\"o}hlich, \emph{New super-selection sectors {(``soliton-states'')} in
  two dimensional {B}ose quantum field models}, Commun. Math. Phys. \textbf{47}
  (1976), no.~3, 269--310.

\bibitem[Fr{\"o}hlich(1988)]{Froehlich-88}
J.~Fr{\"o}hlich, \emph{Statistics of fields, the {Y}ang-{B}axter equation, and
  the theory of knots and links}, Nonperturbative quantum field theory
  ({Carg\`ese}, 1987), NATO Adv. Sci. Inst. Ser. B Phys., vol. 185, Plenum, New
  York, 1988, pp.~71--100. \MR{1008276}

\bibitem[Fr{\"o}hlich(1990)]{Froehlich-90}
J.~Fr{\"o}hlich, \emph{Quantum statistics and locality}, Proceedings of the
  {G}ibbs {S}ymposium ({N}ew {H}aven, {CT}, 1989), Amer. Math. Soc.,
  Providence, RI, 1990, pp.~89--142. \MR{1095329}

\bibitem[Fr{\"o}hlich(2009)]{Froehlich-09}
J.~Fr{\"o}hlich, \emph{Spin, or actually: Spin and quantum statistics}, The
  Spin: Poincar{\'e} Seminar 2007 (B.~Duplantier, J.-M. Raimond, and
  V.~Rivasseau, eds.), Birkh{\"a}user Basel, Basel, 2009, pp.~1--60,
  \href{http://dx.doi.org/10.1007/978-3-7643-8799-0_1}{\path{doi}}.

\bibitem[Fr{\"o}hlich and Kerler(1993)]{FroKer-93}
J.~Fr{\"o}hlich and T.~Kerler, \emph{Quantum groups, quantum categories and
  quantum field theory}, Berlin-Heidelberg-New York: Springer-Verlag, 1993,
  \href{http://dx.doi.org/10.1007/BFb0084244}{\path{doi}}.

\bibitem[Gentile(1940)]{Gentile-40}
G.~Gentile, \emph{Osservazioni sopra le statistiche intermedie}, Il Nuovo
  Cimento \textbf{17} (1940), no.~10, 493--497,
  \href{http://dx.doi.org/10.1007/BF02960187}{\path{doi}}.

\bibitem[Gentile(1942)]{Gentile-42}
G.~Gentile, \emph{Le statistiche intermedie e le propriet{\`a} dell'elio
  liquido}, Il Nuovo Cimento \textbf{19} (1942), no.~4, 109--125,
  \href{http://dx.doi.org/10.1007/BF02960192}{\path{doi}}.

\bibitem[Girardeau(1960)]{Girardeau-60}
M.~Girardeau, \emph{Relationship between systems of impenetrable bosons and
  fermions in one dimension}, J. Math. Phys. \textbf{1} (1960), 516--523,
  \href{http://dx.doi.org/10.1063/1.1703687}{\path{doi}}. \MR{0128913 (23
  {\#}B1950)}

\bibitem[Girardeau(1965)]{Girardeau-65}
M.~D. Girardeau, \emph{Permutation symmetry of many-particle wave functions},
  Phys. Rev. \textbf{139} (1965), B500--B508,
  \href{http://dx.doi.org/10.1103/PhysRev.139.B500}{\path{doi}}.

\bibitem[Girardot(2020)]{Girardot-20}
T.~Girardot, \emph{Average field approximation for almost bosonic anyons in a
  magnetic field}, J. Math. Phys. \textbf{61} (2020), no.~7, 071901, 23,
  \href{http://dx.doi.org/10.1063/1.5143205}{\path{doi}}. \MR{4123638}

\bibitem[Girardot(2021)]{Girardot-21}
T.~Girardot, \emph{Mean-field approximation for the anyon gas}, Ph.D. thesis,
  Universit{\'e} Grenoble Alpes {\&} CNRS, l'{\'E}cole Doctorale
  Math{\'e}matiques, Sciences et technologies de l'information, Informatique,
  2021, \url{https://www.theses.fr/2021GRALM031}.

\bibitem[Girardot and Rougerie(2021)]{GirRou-21}
T.~Girardot and N.~Rougerie, \emph{Semiclassical limit for almost fermionic
  anyons}, Commun. Math. Phys. \textbf{387} (2021), no.~1, 427--480,
  \href{http://dx.doi.org/10.1007/s00220-021-04164-1}{\path{doi}}.

\bibitem[Girardot and Rougerie(2022)]{GirRou-22}
T.~Girardot and N.~Rougerie, \emph{A {L}ieb-{T}hirring inequality for extended
  anyons}, arXiv e-prints, 2022,
  \href{http://arxiv.org/abs/2209.02543}{\path{arXiv:2209.02543}}.

\bibitem[Goldin et~al.(1981)Goldin, Menikoff, and Sharp]{GolMenSha-81}
G.~A. Goldin, R.~Menikoff, and D.~H. Sharp, \emph{Representations of a local
  current algebra in nonsimply connected space and the {A}haronov-{B}ohm
  effect}, J. Math. Phys. \textbf{22} (1981), no.~8, 1664--1668,
  \href{http://dx.doi.org/10.1063/1.525110}{\path{doi}}.

\bibitem[Goldin et~al.(1980)Goldin, Menikoff, and Sharp]{GolMenSha-80}
G.~Goldin, R.~Menikoff, and D.~Sharp, \emph{Particle statistics from induced
  representations of a local current group}, J. Math. Phys. \textbf{21} (1980),
  no.~4, 650--664, \href{http://dx.doi.org/10.1063/1.524510}{\path{doi}}.

\bibitem[Goldin(2022)]{Goldin-22}
G.~A. Goldin, \emph{The prediction of anyons: Its history and wider
  implications}, arXiv e-prints, 2022,
  \href{http://arxiv.org/abs/2212.12632}{\path{arXiv:2212.12632}}.

\bibitem[Goldin and Majid(2004)]{GolMaj-04}
G.~A. Goldin and S.~Majid, \emph{On the {F}ock space for nonrelativistic anyon
  fields and braided tensor products}, J. Math. Phys. \textbf{45} (2004),
  no.~10, 3770--3787, \href{http://dx.doi.org/10.1063/1.1787620}{\path{doi}}.
  \MR{2095672}

\bibitem[Goldin and Sharp(1996)]{GolSha-96}
G.~A. Goldin and D.~H. Sharp, \emph{Diffeomorphism groups, anyon fields, and
  {$q$} commutators}, Phys. Rev. Lett. \textbf{76} (1996), 1183--1187,
  \href{http://dx.doi.org/10.1103/PhysRevLett.76.1183}{\path{doi}}.

\bibitem[Goldin et~al.(1983)Goldin, Menikoff, and Sharp]{GolMenSha-83}
G.~A. Goldin, R.~Menikoff, and D.~H. Sharp, \emph{Diffeomorphism groups, gauge
  groups, and quantum theory}, Phys. Rev. Lett. \textbf{51} (1983), 2246--2249,
  \href{http://dx.doi.org/10.1103/PhysRevLett.51.2246}{\path{doi}}.

\bibitem[Goldin et~al.(1985)Goldin, Menikoff, and Sharp]{GolMenSha-85}
G.~A. Goldin, R.~Menikoff, and D.~H. Sharp, \emph{Comments on {"General Theory
  for Quantum Statistics in Two Dimensions"}}, Phys. Rev. Lett. \textbf{54}
  (1985), 603--603,
  \href{http://dx.doi.org/10.1103/PhysRevLett.54.603}{\path{doi}}.

\bibitem[{Google Quantum AI and Collaborators}(2022)]{Google-22}
{Google Quantum AI and Collaborators}, \emph{Observation of non-abelian
  exchange statistics on a superconducting processor}, arXiv e-prints, 2022,
  \href{http://arxiv.org/abs/2210.10255}{\path{arXiv:2210.10255}}.

\bibitem[Gra{\ss{}} et~al.(2020)Gra{\ss{}}, Juli{\'a}-D{\'{\i}}az, Baldelli,
  Bhattacharya, and Lewenstein]{Grass-etal-20}
T.~Gra{\ss{}}, B.~Juli{\'a}-D{\'{\i}}az, N.~Baldelli, U.~Bhattacharya, and
  M.~Lewenstein, \emph{Fractional angular momentum and anyon statistics of
  impurities in {L}aughlin liquids}, Phys. Rev. Lett. \textbf{125} (2020),
  136801, \href{http://dx.doi.org/10.1103/PhysRevLett.125.136801}{\path{doi}}.

\bibitem[Green(1953)]{Green-53}
H.~S. Green, \emph{A generalized method of field quantization}, Phys. Rev.
  \textbf{90} (1953), 270--273,
  \href{http://dx.doi.org/10.1103/PhysRev.90.270}{\path{doi}}.

\bibitem[Gross(1961)]{Gross-61}
E.~Gross, \emph{Structure of a quantized vortex in boson systems}, Nuovo
  Cimento \textbf{20} (1961), no.~3, 454--477.

\bibitem[Grundberg et~al.(1991)Grundberg, Hansson, Karlhede, and
  Leinaas]{Grundberg-etal-91}
J.~Grundberg, T.~Hansson, A.~Karlhede, and J.~Leinaas, \emph{On singular anyon
  wavefunctions}, Mod. Phys. Lett. B \textbf{05} (1991), no.~07, 539--546,
  \href{http://dx.doi.org/10.1142/S0217984991000642}{\path{doi}}.

\bibitem[Haldane(1991)]{Haldane-91}
F.~D.~M. Haldane, \emph{{``Fractional statistics'' in arbitrary dimensions: A
  generalization of the Pauli principle}}, Phys. Rev. Lett. \textbf{67} (1991),
  937--940, \href{http://dx.doi.org/10.1103/PhysRevLett.67.937}{\path{doi}}.

\bibitem[Halperin(1984)]{Halperin-84}
B.~I. Halperin, \emph{Statistics of quasiparticles and the hierarchy of
  fractional quantized {H}all states}, Phys. Rev. Lett. \textbf{52} (1984),
  1583--1586, \href{http://dx.doi.org/10.1103/PhysRevLett.52.1583}{\path{doi}}.

\bibitem[Hoffmann-Ostenhof et~al.(2008)Hoffmann-Ostenhof, Hoffmann-Ostenhof,
  Laptev, and Tidblom]{HofLapTid-08}
M.~Hoffmann-Ostenhof, T.~Hoffmann-Ostenhof, A.~Laptev, and J.~Tidblom,
  \emph{{Many-particle Hardy Inequalities}}, J. London Math. Soc. \textbf{77}
  (2008), 99--114, \href{http://dx.doi.org/10.1112/jlms/jdm091}{\path{doi}}.

\bibitem[Hu et~al.(2021)Hu, Murthy, Rao, and Jain]{Hu-etal-21}
Y.~Hu, G.~Murthy, S.~Rao, and J.~K. Jain, \emph{{K}ohn-{S}ham density
  functional theory of abelian anyons}, Phys. Rev. B \textbf{103} (2021),
  035124, \href{http://dx.doi.org/10.1103/PhysRevB.103.035124}{\path{doi}}.

\bibitem[Iengo and Lechner(1992)]{IenLec-92}
R.~Iengo and K.~Lechner, \emph{Anyon quantum mechanics and {C}hern-{S}imons
  theory}, Phys. Rep. \textbf{213} (1992), 179--269,
  \href{http://dx.doi.org/10.1016/0370-1573(92)90039-3}{\path{doi}}.

\bibitem[Jackiw(1990)]{Jackiw-90}
R.~Jackiw, \emph{Topics in planar physics}, Nuclear Phys. B Proc. Suppl.
  \textbf{18A} (1990), 107--170, Integrability and quantization (Jaca, 1989),
  \href{http://dx.doi.org/10.1016/0920-5632(90)90648-E}{\path{doi}}.
  \MR{1128794}

\bibitem[Jain(2007)]{Jain-07}
J.~K. Jain, \emph{{Composite fermions}}, Cambridge Univ. Press, 2007.

\bibitem[Khare(2005)]{Khare-05}
A.~Khare, \emph{{Fractional Statistics and Quantum Theory}}, 2nd ed., World
  Scientific, Singapore, 2005.

\bibitem[Kitaev(2003)]{Kitaev-03}
A.~Y. Kitaev, \emph{Fault-tolerant quantum computation by anyons}, Ann. Physics
  \textbf{303} (2003), no.~1, 2--30,
  \href{http://dx.doi.org/10.1016/S0003-4916(02)00018-0}{\path{doi}}.
  \MR{1951039}

\bibitem[Kitaev(2006)]{Kitaev-06}
A.~Kitaev, \emph{Anyons in an exactly solved model and beyond}, Ann. Physics
  \textbf{321} (2006), no.~1, 2--111,
  \href{http://dx.doi.org/10.1016/j.aop.2005.10.005}{\path{doi}}. \MR{2200691}

\bibitem[Klaiber(1968)]{Klaiber-68}
B.~Klaiber, \emph{The {T}hirring model}, Lectures in Theoretical Physics, Vol.
  X-A: Quantum Theory and Statistical Physics (A.~O. Barut and W.~E. Brittin,
  eds.), Gordon and Breach, New York, 1968, Presented at the Theoretical
  Physics Institute, University of Colorado, Summer 1967, pp.~141--176.

\bibitem[Kohno(1987)]{Kohno-87}
T.~Kohno, \emph{Monodromy representations of braid groups and {Y}ang-{B}axter
  equations}, Ann. Inst. Fourier (Grenoble) \textbf{37} (1987), no.~4,
  139--160, \url{http://www.numdam.org/item?id=AIF_1987__37_4_139_0}.
  \MR{927394}

\bibitem[Korff et~al.(1999)Korff, Lang, and Schrader]{KorLanSch-99}
C.~Korff, G.~Lang, and R.~Schrader, \emph{Two-particle scattering theory for
  anyons}, J. Math. Phys. \textbf{40} (1999), no.~4, 1831--1869,
  \href{http://dx.doi.org/10.1063/1.532837}{\path{doi}}. \MR{1683162}

\bibitem[Kundu(1999)]{Kundu-99}
A.~Kundu, \emph{Exact solution of double {$\ensuremath{\delta}$} function bose
  gas through an interacting anyon gas}, Phys. Rev. Lett. \textbf{83} (1999),
  1275--1278, \href{http://dx.doi.org/10.1103/PhysRevLett.83.1275}{\path{doi}}.

\bibitem[Lahtinen and Pachos(2017)]{LahPac-17}
V.~Lahtinen and J.~Pachos, \emph{A short introduction to topological quantum
  computation}, SciPost Physics \textbf{3} (2017), no.~3, 021,
  \href{http://dx.doi.org/10.21468/SciPostPhys.3.3.021}{\path{doi}}.

\bibitem[Laidlaw and DeWitt(1971)]{LaiDeW-71}
M.~G.~G. Laidlaw and C.~M. DeWitt, \emph{Feynman functional integrals for
  systems of indistinguishable particles}, Phys. Rev. D \textbf{3} (1971),
  1375--1378, \href{http://dx.doi.org/10.1103/PhysRevD.3.1375}{\path{doi}}.

\bibitem[Lambert et~al.(2023)Lambert, Lundholm, and Rougerie]{LamLunRou-22}
G.~Lambert, D.~Lundholm, and N.~Rougerie, \emph{Quantum statistics
  transmutation via magnetic flux attachment}, to appear in Prob. Math. Phys.,
  2023, \href{http://arxiv.org/abs/2201.03518}{\path{arXiv:2201.03518}}.

\bibitem[Landau(1933)]{Landau-33}
L.~Landau, \emph{{\"U}ber die {B}ewegung der {E}lektronen in {K}ristallgitter},
  Physik Z. Sowjetunion \textbf{3} (1933), 644--645.

\bibitem[Lankhorst et~al.(2018)Lankhorst, Brinkman, Hilgenkamp, Poccia, and
  Golubov]{Lankhorst-etal-18}
M.~Lankhorst, A.~Brinkman, H.~Hilgenkamp, N.~Poccia, and A.~Golubov,
  \emph{{Annealed Low Energy States in Frustrated Large Square Josephson
  Junction Arrays}}, Condens. Matter \textbf{3} (2018), 19,
  \href{http://dx.doi.org/10.3390/condmat3020019}{\path{doi}}.

\bibitem[Larson and Lundholm(2018)]{LarLun-16}
S.~Larson and D.~Lundholm, \emph{Exclusion bounds for extended anyons}, Arch.
  Ration. Mech. Anal. \textbf{227} (2018), 309--365,
  \href{http://dx.doi.org/10.1007/s00205-017-1161-9}{\path{doi}}.

\bibitem[Larson et~al.(2021)Larson, Lundholm, and Nam]{LarLunNam-19}
S.~Larson, D.~Lundholm, and P.~T. Nam, \emph{{Lieb--Thirring} inequalities for
  wave functions vanishing on the diagonal set}, Ann. Henri Lebesgue \textbf{4}
  (2021), 251--282, \href{http://dx.doi.org/10.5802/ahl.72}{\path{doi}}.

\bibitem[Laughlin(1983)]{Laughlin-83}
R.~B. Laughlin, \emph{Anomalous quantum {H}all effect: An incompressible
  quantum fluid with fractionally charged excitations}, Phys. Rev. Lett.
  \textbf{50} (1983), no.~18, 1395--1398,
  \href{http://dx.doi.org/10.1103/PhysRevLett.50.1395}{\path{doi}}.

\bibitem[Laughlin(1999)]{Laughlin-99}
R.~B. Laughlin, \emph{Nobel lecture: Fractional quantization}, Rev. Mod. Phys.
  \textbf{71} (1999), 863--874,
  \href{http://dx.doi.org/10.1103/RevModPhys.71.863}{\path{doi}}.

\bibitem[Lee and Oh(1994)]{LeeOh-94}
T.~Lee and P.~Oh, \emph{Non-abelian {C}hern-{S}imons quantum mechanics and
  non-abelian {A}haronov-{B}ohm effect}, Ann. Physics \textbf{235} (1994),
  no.~2, 413--434, \href{http://dx.doi.org/10.1006/aphy.1994.1103}{\path{doi}}.
  \MR{1297823}

\bibitem[{Leinaas} and {Myrheim}(1977)]{LeiMyr-77}
J.~M. {Leinaas} and J.~{Myrheim}, \emph{{On the theory of identical
  particles}}, Nuovo Cimento B \textbf{37} (1977), 1--23,
  \href{http://dx.doi.org/10.1007/BF02727953}{\path{doi}}.

\bibitem[Leinaas and Myrheim(1993)]{LeiMyr-93}
J.~M. Leinaas and J.~Myrheim, \emph{{H}eisenberg quantization for systems of
  identical particles}, International Journal of Modern Physics A \textbf{8}
  (1993), no.~21, 3649--3695,
  \href{http://dx.doi.org/10.1142/S0217751X93001491}{\path{doi}}.

\bibitem[Leinaas and Myrheim(2022)]{LeiMyr-22}
J.~M. Leinaas and J.~Myrheim, \emph{Fractional statistics in low-dimensional
  systems}, For the Encyclopedia of Condensed Matter Physics, 2e., 2022.

\bibitem[Lerda(1992)]{Lerda-92}
A.~Lerda, \emph{{Anyons}}, Springer-Verlag, Berlin--Heidelberg, 1992.

\bibitem[Li et~al.(1992)Li, Bhaduri, and Murthy]{LiBhaMur-92}
S.~Li, R.~K. Bhaduri, and M.~V.~N. Murthy, \emph{Thomas-fermi approximation for
  confined anyons}, Phys. Rev. B \textbf{46} (1992), 1228--1231,
  \href{http://dx.doi.org/10.1103/PhysRevB.46.1228}{\path{doi}}.

\bibitem[Lieb and Liniger(1963)]{LieLin-63}
E.~H. Lieb and W.~Liniger, \emph{Exact analysis of an interacting {B}ose gas.
  {I}. {T}he general solution and the ground state}, Phys. Rev. (2)
  \textbf{130} (1963), 1605--1616,
  \href{http://dx.doi.org/10.1103/PhysRev.130.1605}{\path{doi}}. \MR{0156630
  (27 \#6551)}

\bibitem[Lieb and Seiringer(2010)]{LieSei-09}
E.~H. Lieb and R.~Seiringer, \emph{The {S}tability of {M}atter in {Q}uantum
  {M}echanics}, Cambridge Univ. Press, 2010.

\bibitem[Lieb and Thirring(1975)]{LieThi-75}
E.~H. Lieb and W.~E. Thirring, \emph{Bound for the kinetic energy of fermions
  which proves the stability of matter}, Phys. Rev. Lett. \textbf{35} (1975),
  687--689, \href{http://dx.doi.org/10.1103/PhysRevLett.35.687}{\path{doi}}.

\bibitem[Lieb and Thirring(1976)]{LieThi-76}
E.~H. Lieb and W.~E. Thirring, \emph{Inequalities for the moments of the
  eigenvalues of the {S}chr{\"o}dinger hamiltonian and their relation to
  {S}obolev inequalities}, Studies in Mathematical Physics, pp.~269--303,
  Princeton Univ. Press, 1976.

\bibitem[Lieb and Yngvason(2001)]{LieYng-01}
E.~H. Lieb and J.~Yngvason, \emph{The ground state energy of a dilute
  two-dimensional {B}ose gas}, J. Statist. Phys. \textbf{103} (2001), no.~3-4,
  509--526, Special issue dedicated to the memory of Joaquin M. Luttinger,
  \href{http://dx.doi.org/10.1023/A:1010337215241}{\path{doi}}. \MR{1827922}

\bibitem[Lieb et~al.(1995)Lieb, Solovej, and Yngvason]{LieSolYng-95}
E.~H. Lieb, J.~P. Solovej, and J.~Yngvason, \emph{Ground states of large
  quantum dots in magnetic fields}, Phys. Rev. B \textbf{51} (1995),
  10646--10665, \href{http://dx.doi.org/10.1103/PhysRevB.51.10646}{\path{doi}}.

\bibitem[Lieb et~al.(2005)Lieb, Seiringer, Solovej, and
  Yngvason]{LieSeiSolYng-05}
E.~H. Lieb, R.~Seiringer, J.~P. Solovej, and J.~Yngvason, \emph{The mathematics
  of the {B}ose gas and its condensation}, Oberwolfach {S}eminars,
  Birkh{\"a}user, 2005,
  \href{http://arxiv.org/abs/cond-mat/0610117}{\path{arXiv:cond-mat/0610117}}.

\bibitem[Lundholm(2017)]{Lundholm-16}
D.~Lundholm, \emph{Many-anyon trial states}, Phys. Rev. A \textbf{96} (2017),
  012116, \href{http://dx.doi.org/10.1103/PhysRevA.96.012116}{\path{doi}}.

\bibitem[Lundholm(2019)]{Lundholm-17}
D.~Lundholm, \emph{Methods of modern mathematical physics: Uncertainty and
  exclusion principles in quantum mechanics}, KTH and LMU graduate course
  textbook (latest version at
  \url{http://www.mathematik.uni-muenchen.de/~lundholm/methmmp.pdf}), 2019,
  \href{http://arxiv.org/abs/1805.03063}{\path{arXiv:1805.03063}}.

\bibitem[Lundholm and Qvarfordt(2020)]{LunQva-20}
D.~Lundholm and V.~Qvarfordt, \emph{Exchange and exclusion in the non-abelian
  anyon gas}, arXiv e-prints, 2020,
  \href{http://arxiv.org/abs/2009.12709}{\path{arXiv:2009.12709}}.

\bibitem[Lundholm and Rougerie(2015)]{LunRou-15}
D.~Lundholm and N.~Rougerie, \emph{{The average field approximation for almost
  bosonic extended anyons}}, J. Stat. Phys. \textbf{161} (2015), no.~5,
  1236--1267, \href{http://dx.doi.org/10.1007/s10955-015-1382-y}{\path{doi}}.

\bibitem[Lundholm and Rougerie(2016)]{LunRou-16}
D.~Lundholm and N.~Rougerie, \emph{Emergence of fractional statistics for
  tracer particles in a {L}aughlin liquid}, Phys. Rev. Lett. \textbf{116}
  (2016), 170401,
  \href{http://dx.doi.org/10.1103/PhysRevLett.116.170401}{\path{doi}}.

\bibitem[Lundholm and Seiringer(2018)]{LunSei-17}
D.~Lundholm and R.~Seiringer, \emph{Fermionic behavior of ideal anyons}, Lett.
  Math. Phys. \textbf{108} (2018), 2523--2541,
  \href{http://dx.doi.org/10.1007/s11005-018-1091-y}{\path{doi}}.

\bibitem[Lundholm and Solovej(2013{\natexlab{a}})]{LunSol-13a}
D.~Lundholm and J.~P. Solovej, \emph{{Hardy and Lieb-Thirring inequalities for
  anyons}}, Comm. Math. Phys. \textbf{322} (2013), 883--908,
  \href{http://dx.doi.org/10.1007/s00220-013-1748-4}{\path{doi}}.

\bibitem[Lundholm and Solovej(2013{\natexlab{b}})]{LunSol-13b}
D.~Lundholm and J.~P. Solovej, \emph{Local exclusion principle for identical
  particles obeying intermediate and fractional statistics}, Phys. Rev. A
  \textbf{88} (2013), 062106,
  \href{http://dx.doi.org/10.1103/PhysRevA.88.062106}{\path{doi}}.

\bibitem[Lundholm and Solovej(2014)]{LunSol-14}
D.~Lundholm and J.~P. Solovej, \emph{{Local exclusion and Lieb-Thirring
  inequalities for intermediate and fractional statistics}}, Ann. Henri
  Poincar\'e \textbf{15} (2014), 1061--1107,
  \href{http://dx.doi.org/10.1007/s00023-013-0273-5}{\path{doi}}.

\bibitem[Lundholm et~al.(2015)Lundholm, Portmann, and Solovej]{LunPorSol-15}
D.~Lundholm, F.~Portmann, and J.~P. Solovej, \emph{Lieb-{T}hirring bounds for
  interacting {B}ose gases}, Comm. Math. Phys. \textbf{335} (2015), no.~2,
  1019--1056, \href{http://dx.doi.org/10.1007/s00220-014-2278-4}{\path{doi}}.

\bibitem[Lundholm et~al.(2016)Lundholm, Nam, and Portmann]{LunNamPor-16}
D.~Lundholm, P.~T. Nam, and F.~Portmann, \emph{Fractional
  {H}ardy-{L}ieb-{T}hirring and related inequalities for interacting systems},
  Arch. Ration. Mech. Anal. \textbf{219} (2016), no.~3, 1343--1382,
  \href{http://dx.doi.org/10.1007/s00205-015-0923-5}{\path{doi}}.

\bibitem[Maciazek and Sawicki(2019)]{MacSaw-19}
T.~Maciazek and A.~Sawicki, \emph{Non-abelian quantum statistics on graphs},
  Commun. Math. Phys. \textbf{371} (2019), 921--973,
  \href{http://dx.doi.org/10.1007/s00220-019-03583-5}{\path{doi}}.

\bibitem[Mancarella et~al.(2013{\natexlab{a}})Mancarella, Trombettoni, and
  Mussardo]{ManTroMus-13a}
F.~Mancarella, A.~Trombettoni, and G.~Mussardo, \emph{Statistical mechanics of
  an ideal gas of non-abelian anyons}, Nucl. Phys. B \textbf{867} (2013),
  no.~3, 950--976,
  \href{http://dx.doi.org/10.1016/j.nuclphysb.2012.10.020}{\path{doi}}.

\bibitem[Mancarella et~al.(2013{\natexlab{b}})Mancarella, Trombettoni, and
  Mussardo]{ManTroMus-13b}
F.~Mancarella, A.~Trombettoni, and G.~Mussardo, \emph{Statistical interparticle
  potential of an ideal gas of non-abelian anyons}, J. Phys. A: Math. Theor.
  \textbf{46} (2013), no.~27, 275001,
  \href{http://dx.doi.org/10.1088/1751-8113/46/27/275001}{\path{doi}}.

\bibitem[Manuel and Tarrach(1991)]{ManTar-91}
C.~Manuel and R.~Tarrach, \emph{Contact interactions of anyons}, Phys. Lett. B
  \textbf{268} (1991), no.~2, 222--226,
  \href{http://dx.doi.org/10.1016/0370-2693(91)90807-3}{\path{doi}}.
  \MR{1131514}

\bibitem[Masaki et~al.(2023)Masaki, Mizushima, and Nitta]{MasMizNit-23}
Y.~Masaki, T.~Mizushima, and M.~Nitta, \emph{Non-abelian anyons and non-abelian
  vortices in topological superconductors}, For the Encyclopedia of Condensed
  Matter Physics, 2e., 2023,
  \href{http://arxiv.org/abs/2301.11614}{\path{arXiv:2301.11614}}.

\bibitem[McEuen et~al.(1992)McEuen, Foxman, Kinaret, Meirav, Kastner, Wingreen,
  and Wind]{McEuen-etal-92}
P.~L. McEuen, E.~B. Foxman, J.~Kinaret, U.~Meirav, M.~A. Kastner, N.~S.
  Wingreen, and S.~J. Wind, \emph{Self-consistent addition spectrum of a
  {C}oulomb island in the quantum {H}all regime}, Phys. Rev. B \textbf{45}
  (1992), 11419--11422,
  \href{http://dx.doi.org/10.1103/PhysRevB.45.11419}{\path{doi}}.

\bibitem[Messiah and Greenberg(1964)]{MesGre-64}
A.~M.~L. Messiah and O.~W. Greenberg, \emph{Symmetrization postulate and its
  experimental foundation}, Phys. Rev. \textbf{136} (1964), B248--B267,
  \href{http://dx.doi.org/10.1103/PhysRev.136.B248}{\path{doi}}.

\bibitem[Moore and Read(1991)]{MooRea-91}
G.~Moore and N.~Read, \emph{Nonabelions in the fractional quantum {H}all
  effect}, Nucl. Phys. B \textbf{360} (1991), no.~2, 362 -- 396,
  \href{http://dx.doi.org/10.1016/0550-3213(91)90407-O}{\path{doi}}.

\bibitem[Morampudi et~al.(2017)Morampudi, Turner, Pollmann, and
  Wilczek]{MorTurPolWil-17}
S.~C. Morampudi, A.~M. Turner, F.~Pollmann, and F.~Wilczek, \emph{Statistics of
  fractionalized excitations through threshold spectroscopy}, Phys. Rev. Lett.
  \textbf{118} (2017), 227201,
  \href{http://dx.doi.org/10.1103/PhysRevLett.118.227201}{\path{doi}}.

\bibitem[Mouchet(2021)]{Mouchet-21}
A.~Mouchet, \emph{Path integrals in a multiply-connected configuration space
  (50 years after)}, Foundations of Physics \textbf{51} (2021), no.~6, 107,
  \href{http://dx.doi.org/10.1007/s10701-021-00497-y}{\path{doi}}.

\bibitem[Mund and Schrader(1995)]{MunSch-95}
J.~Mund and R.~Schrader, \emph{Hilbert spaces for nonrelativistic and
  relativistic ``free'' plektons (particles with braid group statistics)},
  Advances in dynamical systems and quantum physics ({C}apri, 1993), World Sci.
  Publ., River Edge, NJ, 1995, pp.~235--259,
  \href{http://arxiv.org/abs/hep-th/9310054}{\path{arXiv:hep-th/9310054}}.
  \MR{1414702}

\bibitem[Mund(2009)]{Mund-09}
J.~Mund, \emph{The spin-statistics theorem for anyons and plektons in d= 2+ 1},
  Commun. Math. Phys. \textbf{286} (2009), no.~3, 1159--1180,
  \href{http://dx.doi.org/10.1007/s00220-008-0628-9}{\path{doi}}.

\bibitem[Murthy et~al.(1991)Murthy, Law, Brack, and Bhaduri]{MurLawBraBha-91}
M.~V.~N. Murthy, J.~Law, M.~Brack, and R.~K. Bhaduri, \emph{Quantum spectrum of
  three anyons in an oscillator potential}, Phys. Rev. Lett. \textbf{67}
  (1991), 1817--1820,
  \href{http://dx.doi.org/10.1103/PhysRevLett.67.1817}{\path{doi}}.

\bibitem[Murthy et~al.(1992)Murthy, Law, Bhaduri, and Date]{MurLawBhaDat-92}
M.~V.~N. Murthy, J.~Law, R.~K. Bhaduri, and G.~Date, \emph{On a class of
  noninterpolating solutions of the many-anyon problem}, J. Phys. A: Math. Gen.
  \textbf{25} (1992), no.~23, 6163,
  \href{http://dx.doi.org/10.1088/0305-4470/25/23/013}{\path{doi}}.

\bibitem[Myrheim(1999)]{Myrheim-99}
J.~Myrheim, \emph{Anyons}, Topological aspects of low dimensional systems
  (A.~Comtet, T.~Jolic{\oe}ur, S.~Ouvry, and F.~David, eds.), Les Houches -
  Ecole d'Ete de Physique Theorique, vol.~69, (Springer-Verlag, Berlin,
  Germany), 1999, pp.~265--413,
  \href{http://dx.doi.org/10.1007/3-540-46637-1_4}{\path{doi}}.

\bibitem[Nakamura et~al.(2020)Nakamura, Liang, Gardner, and
  Manfra]{Nakamura-etal-20}
J.~Nakamura, S.~Liang, G.~C. Gardner, and M.~J. Manfra, \emph{Direct
  observation of anyonic braiding statistics}, Nature Phys. \textbf{16} (2020),
  931--936, \href{http://dx.doi.org/10.1038/s41567-020-1019-1}{\path{doi}}.

\bibitem[Nayak and Wilczek(1996)]{NayWil-96}
C.~Nayak and F.~Wilczek, \emph{{$2n$}-quasihole states realize
  {$2^{n-1}$}-dimensional spinor braiding statistics in paired quantum {H}all
  states}, Nuclear Phys. B \textbf{479} (1996), no.~3, 529--553,
  \href{http://dx.doi.org/10.1016/0550-3213(96)00430-0}{\path{doi}}.
  \MR{1418835}

\bibitem[Nayak et~al.(2008)Nayak, Simon, Stern, Freedman, and
  Das~Sarma]{Nayak-etal-08}
C.~Nayak, S.~H. Simon, A.~Stern, M.~Freedman, and S.~Das~Sarma,
  \emph{Non-abelian anyons and topological quantum computation}, Rev. Mod.
  Phys. \textbf{80} (2008), 1083--1159,
  \href{http://dx.doi.org/10.1103/RevModPhys.80.1083}{\path{doi}}.

\bibitem[Noh et~al.(2020)Noh, Schuster, Iadecola, Huang, Wang, Chen, Chamon,
  and Rechtsman]{Noh-etal-20}
J.~Noh, T.~Schuster, T.~Iadecola, S.~Huang, M.~Wang, K.~P. Chen, C.~Chamon, and
  M.~C. Rechtsman, \emph{Braiding photonic topological zero modes}, Nature
  Physics \textbf{16} (2020), no.~9, 989--993,
  \href{http://dx.doi.org/10.1038/s41567-020-1007-5}{\path{doi}}.

\bibitem[Oddis(2020)]{Oddis-20}
L.~Oddis, \emph{Two-anyon schr{\"o}dinger operators}, Ph.D. thesis,
  Universit{\`a} di Roma, Sapienza, Facolt{\`a} di Scienze Matematiche, Fisiche
  e Naturali, 2020,
  \url{https://phd.uniroma1.it/web/LUCA-ODDIS_nT1669369_EN.aspx}.

\bibitem[Ouvry(2007)]{Ouvry-07}
S.~Ouvry, \emph{{Anyons and lowest Landau level anyons}}, S\'eminaire
  Poincar\'e \textbf{11} (2007), 77--107,
  \href{http://dx.doi.org/10.1007/978-3-7643-8799-0_3}{\path{doi}}.

\bibitem[Ouvry and Polychronakos(2019)]{OuvPol-19}
S.~Ouvry and A.~P. Polychronakos, \emph{Anyons on the sphere: Analytic states
  and spectrum}, Nuclear Physics B \textbf{949} (2019), 114797,
  \href{http://dx.doi.org/https://doi.org/10.1016/j.nuclphysb.2019.114797}{\path{doi}}.

\bibitem[Pancharatnam(1956)]{Pancharatnam-56}
S.~Pancharatnam, \emph{Generalized theory of interference, and its
  applications: Part i. coherent pencils}, Proceedings of the Indian Academy of
  Sciences-Section A, vol.~44, Springer India New Delhi, 1956, pp.~247--262.

\bibitem[Pekar(1946)]{Pekar-46}
S.~Pekar, \emph{Local quantum states of electrons in an ideal ion crystal},
  Zhurnal Eksperimentalnoi I Teoreticheskoi Fiziki \textbf{16} (1946), no.~4,
  341--348.

\bibitem[Pitaevskii(1961)]{Pitaevskii-61}
L.~P. Pitaevskii, \emph{Vortex lines in an imperfect bose gas}, Zh. Eksper.
  Teor. fiz. \textbf{40} (1961), no.~40, 646--651.

\bibitem[Pithis and Ruiz~Euler(2015)]{PitRui-15}
A.~G.~A. Pithis and H.-C. Ruiz~Euler, \emph{Anyonic statistics and large
  horizon diffeomorphisms for loop quantum gravity black holes}, Phys. Rev. D
  \textbf{91} (2015), 064053,
  \href{http://dx.doi.org/10.1103/PhysRevD.91.064053}{\path{doi}}.

\bibitem[Polychronakos(1989)]{Polychronakos-89}
A.~P. Polychronakos, \emph{Non-relativistic bosonization and fractional
  statistics}, Nuclear Physics B \textbf{324} (1989), no.~3, 597--622,
  \href{http://dx.doi.org/10.1016/0550-3213(89)90522-1}{\path{doi}}.

\bibitem[Polychronakos and Ouvry(2020)]{PolOuv-20}
A.~P. Polychronakos and S.~Ouvry, \emph{Two anyons on the sphere: Nonlinear
  states and spectrum}, Nuclear Physics B \textbf{951} (2020), 114906,
  \href{http://dx.doi.org/https://doi.org/10.1016/j.nuclphysb.2019.114906}{\path{doi}}.

\bibitem[Qvarfordt(2017)]{Qvarfordt-17}
V.~Qvarfordt, \emph{Non-abelian anyons: Statistical repulsion and topological
  quantum computation}, MSc thesis, KTH, 2017,
  \url{http://urn.kb.se/resolve?urn=urn%3Anbn%3Ase%3Akth%3Adiva-207177}.

\bibitem[Read and Rezayi(1999)]{ReaRez-99}
N.~Read and E.~Rezayi, \emph{{Beyond paired quantum Hall states: Parafermions
  and incompressible states in the first excited Landau level}}, Phys. Rev. B
  \textbf{59} (1999), 8084--8092,
  \href{http://dx.doi.org/10.1103/PhysRevB.59.8084}{\path{doi}}.

\bibitem[Regnault et~al.(2008)Regnault, Goerbig, and Jolicoeur]{RegGoeJol-08}
N.~Regnault, M.~O. Goerbig, and T.~Jolicoeur, \emph{Bridge between abelian and
  non-abelian fractional quantum {H}all states}, Phys. Rev. Lett. \textbf{101}
  (2008), 066803,
  \href{http://dx.doi.org/10.1103/PhysRevLett.101.066803}{\path{doi}}.

\bibitem[Rougerie and Yang(2023)]{RouYan-23b}
N.~Rougerie and Q.~Yang, \emph{Dimensional reduction for a system of {2D}
  anyons}, arXiv e-prints, 2023,
  \href{http://arxiv.org/abs/2305.06670}{\path{arXiv:2305.06670}}.

\bibitem[Rowell and Wang(2012)]{RowWan-12}
E.~C. Rowell and Z.~Wang, \emph{Localization of unitary braid group
  representations}, Comm. Math. Phys. \textbf{311} (2012), no.~3, 595--615,
  \href{http://dx.doi.org/10.1007/s00220-011-1386-7}{\path{doi}}. \MR{2909757}

\bibitem[Schick(1971)]{Schick-71}
M.~Schick, \emph{Two-dimensional system of hard-core bosons}, Phys. Rev. A
  \textbf{3} (1971), 1067--1073,
  \href{http://dx.doi.org/10.1103/PhysRevA.3.1067}{\path{doi}}.

\bibitem[Schmidt and Lemeshko(2015)]{SchLem-15}
R.~Schmidt and M.~Lemeshko, \emph{Rotation of quantum impurities in the
  presence of a many-body environment}, Phys. Rev. Lett. \textbf{114} (2015),
  203001, \href{http://dx.doi.org/10.1103/PhysRevLett.114.203001}{\path{doi}}.

\bibitem[Schulman(1968)]{Schulman-68}
L.~Schulman, \emph{A path integral for spin}, Phys. Rev. \textbf{176} (1968),
  1558--1569, \href{http://dx.doi.org/10.1103/PhysRev.176.1558}{\path{doi}}.

\bibitem[Seiringer and Solovej(2023)]{SeiSol-23}
R.~Seiringer and J.~P. Solovej, \emph{A simple approach to {Lieb--Thirring}
  type inequalities}, arXiv e-prints, 2023,
  \href{http://arxiv.org/abs/2303.04504}{\path{arXiv:2303.04504}}.

\bibitem[Simon(1983)]{Simon-83}
B.~Simon, \emph{Holonomy, the quantum adiabatic theorem, and {B}erry's phase},
  Phys. Rev. Lett. \textbf{51} (1983), 2167--2170,
  \href{http://dx.doi.org/10.1103/PhysRevLett.51.2167}{\path{doi}}.

\bibitem[Souriau(1970)]{Souriau-70}
J.-M. Souriau, \emph{Structure des syst\`emes dynamiques}, Ma{\^{i}}trises de
  math\'ematiques, Dunod, Paris, 1970, English translation by R. H. Cushman and
  G. M. Tuynman, Progress in Mathematics, 149, Birkh\"auser Boston Inc.,
  Boston, MA, 1997,
  \url{http://www.jmsouriau.com/structure_des_systemes_dynamiques.htm}.
  \MR{0260238}

\bibitem[Sporre et~al.(1991)Sporre, Verbaarschot, and Zahed]{SpoVerZah-91}
M.~Sporre, J.~J.~M. Verbaarschot, and I.~Zahed, \emph{Numerical solution of the
  three-anyon problem}, Phys. Rev. Lett. \textbf{67} (1991), 1813--1816,
  \href{http://dx.doi.org/10.1103/PhysRevLett.67.1813}{\path{doi}}.

\bibitem[Sporre et~al.(1992)Sporre, Verbaarschot, and Zahed]{SpoVerZah-92}
M.~Sporre, J.~J.~M. Verbaarschot, and I.~Zahed, \emph{Four anyons in a harmonic
  well}, Phys. Rev. B \textbf{46} (1992), 5738--5741,
  \href{http://dx.doi.org/10.1103/PhysRevB.46.5738}{\path{doi}}.

\bibitem[Stern(2008)]{Stern-08}
A.~Stern, \emph{{Anyons and the quantum Hall effect -- A pedagogical review}},
  Ann. Phys. \textbf{323} (2008), no.~1, 204--249, January Special Issue 2008,
  \href{http://dx.doi.org/10.1016/j.aop.2007.10.008}{\path{doi}}.

\bibitem[Stormer(1999)]{Stormer-99}
H.~L. Stormer, \emph{Nobel lecture: The fractional quantum {H}all effect}, Rev.
  Mod. Phys. \textbf{71} (1999), 875--889,
  \href{http://dx.doi.org/10.1103/RevModPhys.71.875}{\path{doi}}.

\bibitem[Streater and Wilde(1970)]{StrWil-70}
R.~F. Streater and I.~F. Wilde, \emph{Fermion states of a boson field}, Nuclear
  Physics B \textbf{24} (1970), no.~3, 561--575,
  \href{http://dx.doi.org/10.1016/0550-3213(70)90445-1}{\path{doi}}.

\bibitem[{Sutherland}(1971)]{Sutherland-71}
B.~{Sutherland}, \emph{{Quantum Many-Body Problem in One Dimension: Ground
  State}}, J. Math. Phys. \textbf{12} (1971), 246--250,
  \href{http://dx.doi.org/10.1063/1.1665584}{\path{doi}}.

\bibitem[Thomas(1927)]{Thomas-27}
L.~H. Thomas, \emph{The calculation of atomic fields}, Proc. Camb. Philos. Soc.
  \textbf{23} (1927), no.~05, 542--548,
  \href{http://dx.doi.org/10.1017/S0305004100011683}{\path{doi}}.

\bibitem[Tonks(1936)]{Tonks-36}
L.~Tonks, \emph{The complete equation of state of one, two and
  three-dimensional gases of hard elastic spheres}, Phys. Rev. \textbf{50}
  (1936), 955--963,
  \href{http://dx.doi.org/10.1103/PhysRev.50.955}{\path{doi}}.

\bibitem[Trugenberger(1992{\natexlab{a}})]{Trugenberger-92}
C.~Trugenberger, \emph{{The anyon fluid in the Bogoliubov approximation}},
  Phys. Rev. D \textbf{45} (1992), 3807--3817,
  \href{http://dx.doi.org/10.1103/PhysRevD.45.3807}{\path{doi}}.

\bibitem[Trugenberger(1992{\natexlab{b}})]{Trugenberger-92b}
C.~Trugenberger, \emph{{Ground state and collective excitations of extended
  anyons}}, Phys. Lett. B \textbf{288} (1992), 121--128,
  \href{http://dx.doi.org/10.1016/0370-2693(92)91965-C}{\path{doi}}.

\bibitem[Tsuchiya and Kanie(1987)]{TsuKan-87}
A.~Tsuchiya and Y.~Kanie, \emph{Vertex operators in the conformal field theory
  on p1 and monodromy representations of the braid group}, Lett. Math. Phys.
  \textbf{13} (1987), 303--312,
  \href{http://dx.doi.org/10.1007/BF00401159}{\path{doi}}.

\bibitem[Tsuchiya and Kanie(1988)]{TsuKan-88}
A.~Tsuchiya and Y.~Kanie, \emph{Vertex operators in conformal field theory on
  p1 and monodromy representations of braid group}, Conformal Field Theory and
  Solvable Lattice Models, Advanced Studies in Pure Math \textbf{16} (1988),
  297--372, \href{http://dx.doi.org/10.1142/9789812798329_0034}{\path{doi}}.

\bibitem[Tsui et~al.(1982)Tsui, Stormer, and Gossard]{TsuStoGos-82}
D.~C. Tsui, H.~L. Stormer, and A.~C. Gossard, \emph{Two-dimensional
  magnetotransport in the extreme quantum limit}, Phys. Rev. Lett. \textbf{48}
  (1982), 1559--1562,
  \href{http://dx.doi.org/10.1103/PhysRevLett.48.1559}{\path{doi}}.

\bibitem[Tsui(1999)]{Tsui-99}
D.~C. Tsui, \emph{Nobel lecture: Interplay of disorder and interaction in
  two-dimensional electron gas in intense magnetic fields}, Rev. Mod. Phys.
  \textbf{71} (1999), 891--895,
  \href{http://dx.doi.org/10.1103/RevModPhys.71.891}{\path{doi}}.

\bibitem[Umucal{\i}lar et~al.(2018)Umucal{\i}lar, Macaluso, Comparin, and
  Carusotto]{UmuMacComCar-18}
R.~O. Umucal{\i}lar, E.~Macaluso, T.~Comparin, and I.~Carusotto,
  \emph{Time-of-flight measurements as a possible method to observe anyonic
  statistics}, Phys. Rev. Lett. \textbf{120} (2018), 230403,
  \href{http://dx.doi.org/10.1103/PhysRevLett.120.230403}{\path{doi}}.

\bibitem[Verlinde(1991)]{Verlinde-91}
E.~P. Verlinde, \emph{A note on braid statistics and the non-{A}belian
  {A}haronov-{B}ohm effect}, Modern Quantum Field Theory: proceedings. (S.~Das,
  A.~Dhar, S.~Mukhi, A.~Raina, and A.~Sen., eds.), World Scientific, 1991,
  Proc. of Conference: International Colloquium on Modern Quantum Field Theory,
  Bombay, 1990, pp.~450--461,
  \url{https://lib-extopc.kek.jp/preprints/PDF/1991/9106/9106160.pdf}.

\bibitem[Weinberger(2015)]{Weinberger-15}
O.~Weinberger, \emph{The braid group, representations and non-abelian anyons},
  BSc thesis, KTH, 2015,
  \url{http://urn.kb.se/resolve?urn=urn%3Anbn%3Ase%3Akth%3Adiva-167993}.

\bibitem[Wen(1991)]{Wen-91}
X.~G. Wen, \emph{Non-abelian statistics in the fractional quantum {H}all
  states}, Phys. Rev. Lett. \textbf{66} (1991), 802--805,
  \href{http://dx.doi.org/10.1103/PhysRevLett.66.802}{\path{doi}}.

\bibitem[Wilczek(1982{\natexlab{a}})]{Wilczek-82a}
F.~Wilczek, \emph{Magnetic flux, angular momentum, and statistics}, Phys. Rev.
  Lett. \textbf{48} (1982), 1144--1146,
  \href{http://dx.doi.org/10.1103/PhysRevLett.48.1144}{\path{doi}}.

\bibitem[Wilczek(1982{\natexlab{b}})]{Wilczek-82b}
F.~Wilczek, \emph{Quantum mechanics of fractional-spin particles}, Phys. Rev.
  Lett. \textbf{49} (1982), 957--959,
  \href{http://dx.doi.org/10.1103/PhysRevLett.49.957}{\path{doi}}.

\bibitem[Wilczek(1990)]{Wilczek-90}
F.~Wilczek, \emph{{Fractional Statistics and Anyon Superconductivity}}, World
  Scientific, Singapore, 1990.

\bibitem[Wilczek and Wu(1990)]{WilWu-90}
F.~Wilczek and Y.-S. Wu, \emph{Space-time approach to holonomy scattering},
  Phys. Rev. Lett. \textbf{65} (1990), 13--16,
  \href{http://dx.doi.org/10.1103/PhysRevLett.65.13}{\path{doi}}.

\bibitem[Wu(1984{\natexlab{a}})]{Wu-84a}
Y.-S. Wu, \emph{General theory for quantum statistics in two dimensions}, Phys.
  Rev. Lett. \textbf{52} (1984), 2103--2106,
  \href{http://dx.doi.org/10.1103/PhysRevLett.52.2103}{\path{doi}}.

\bibitem[Wu(1984{\natexlab{b}})]{Wu-84b}
Y.-S. Wu, \emph{Multiparticle quantum mechanics obeying fractional statistics},
  Phys. Rev. Lett. \textbf{53} (1984), 111--114,
  \href{http://dx.doi.org/10.1103/PhysRevLett.53.111}{\path{doi}}.

\bibitem[Yakaboylu and Lemeshko(2018)]{YakLem-18}
E.~Yakaboylu and M.~Lemeshko, \emph{Anyonic statistics of quantum impurities in
  two dimensions}, Phys. Rev. B \textbf{98} (2018), 045402,
  \href{http://dx.doi.org/10.1103/PhysRevB.98.045402}{\path{doi}}.

\bibitem[Yakaboylu et~al.(2020)Yakaboylu, Ghazaryan, Lundholm, Rougerie,
  Lemeshko, and Seiringer]{Yakaboylu-etal-19}
E.~Yakaboylu, A.~Ghazaryan, D.~Lundholm, N.~Rougerie, M.~Lemeshko, and
  R.~Seiringer, \emph{Quantum impurity model for anyons}, Phys. Rev. B
  \textbf{102} (2020), 144109,
  \href{http://dx.doi.org/10.1103/PhysRevB.102.144109}{\path{doi}}.

\bibitem[Zhang et~al.(2014)Zhang, Sreejith, Gemelke, and Jain]{ZhaSreGemJai-14}
Y.~Zhang, G.~J. Sreejith, N.~D. Gemelke, and J.~K. Jain, \emph{{Fractional
  angular momentum in cold atom systems}}, Phys. Rev. Lett. \textbf{113}
  (2014), 160404,
  \href{http://dx.doi.org/10.1103/PhysRevLett.113.160404}{\path{doi}}.

\bibitem[Zhang et~al.(2015)Zhang, Sreejith, and Jain]{ZhaSreJai-15}
Y.~Zhang, G.~J. Sreejith, and J.~K. Jain, \emph{{Creating and manipulating
  non-Abelian anyons in cold atom systems using auxiliary bosons}}, Phys. Rev.
  B. \textbf{92} (2015), 075116,
  \href{http://dx.doi.org/10.1103/PhysRevB.92.075116}{\path{doi}}.

\end{thebibliography}

\end{document}